\documentclass[prd,amsfonts,amsmath,amssymb,showpacs,nofootinbib]{revtex4-1}

\usepackage{bm}
\usepackage{pst-plot}
\usepackage{pst-node}
\usepackage{graphicx}

\newcommand{\stM}{\mathcal{M}}
\newcommand{\J}{\mathcal{J}}
\newcommand{\reals}{\mathbb{R}}
\newcommand{\complex}{\mathbb{C}}
\newcommand{\sign}{\mbox{sign}}

\renewcommand\Re{\operatorname{Re}}
\renewcommand\Im{\operatorname{Im}}

\begin{document}

\title{Particle creation and energy conditions for a quantized
scalar field in the\\ presence of an external, time-dependent,
Mamev-Trunov potential.}
\author{Michael J. Pfenning}
\email{Michael.Pfenning@usma.edu}
\affiliation{Department of Physics, United States Military Academy,%
 West Point, New York, 10996-1790, USA}

\date{\today}

\begin{abstract}
We study the behavior of a massless, quantized, scalar field on a two-dimensional
cylinder spacetime as it responds to the time-dependent evolution of a Mamev-Trunov
potential of the form $V(x,t) = 2 \xi \delta(x) \theta(-t)$.  We begin by
constructing mode solutions to the classical  Klein-Gordon-Fock equation
with potential on the whole spacetime.  For a given eigen-mode solution of the
IN region of the spacetime ($t<0$), we determine its evolution into the
OUT region ($t>0$) through the use of a Fourier decomposition in terms of the
OUT region eigen-modes.  The classical system is then second quantized in the
canonical quantization scheme.  On the OUT region, there is a unitarily
equivalent representation of the quantized field in terms of the OUT region
eigen-modes, including zero-frequency modes which we also quantize
in a manner which allows for their interpretation as particles in the typical
sense.  After determining the Bogolubov coefficients
between the two representations, we study the production of quanta out of
the vacuum when the potential turns off. We find that the number of ``particles''
created on the OUT region is finite for the standard modes, and with the usual
ambiguity in the number of particles created in the zero frequency modes.
We then look at the renormalized expectation value of the stress-energy-tensor
on the IN and OUT regions for the IN vacuum state.  We find that
the resulting stress-tensor can violate the null, weak, strong, and dominant energy
conditions because the standard Casimir energy-density of the cylinder spacetime
is negative.  Finally, we show that the same stress-tensor satisfies a quantum
inequality on the OUT region.
\end{abstract}

\maketitle

\section{Introduction}

\subsection{Quantum Inequalities}
In quantum field theory (QFT), it is well known that the renormalized expectation
value of the energy-density operator for a free, quantized field can be negative.
Epstein, Glaser, and Jaffe \cite{Epstein} demonstrated this to be
a generic property of QFTs under relatively weak assumptions. Also,
negative energies seem to be a generic property for the vacuum state of a QFT
in many curved spacetimes, and additionally, for the vacuum state in
both flat and curved spacetimes with boundaries.  The effect of a nonzero value for
the renormalized expectation value of the vacuum state is often referred to as
vacuum polarization, vacuum energy, zero-point energy, or equivalently, as
the Casimir energy \cite{Casimir}.  The study of this phenomenon has driven
extensive research throughout the later half of the twentieth century which has
continued into the twenty-first.

Negative energy densities also occur for multi-particle states
where interference terms arise in the expectation value of the
stress-energy tensor which have sufficient magnitude to
overpower any positive constant positive terms.  (See
\cite{Pfen98b} for an easy example and \cite{Ford78} for a
thorough discussion.) It was noted by Ford \cite{Ford78}, that
unrestrained negative energies can be used to violate the
second law of thermodynamics. In the same paper, he argues that
no such breakdown would occur in two dimensions if a negative-%
energy flux $F$ obeys an inequality of the form $|F|\lesssim
\tau^{-2}$, where $\tau$ is the duration over which the flux
occurs.  In a subsequent paper \cite{Ford91}, Ford was able to
derive such an inequality constraining negative-energy fluxes
directly from QFT which applies to all
possible quantum states for the massless Klein-Gordon scalar
field in flat spacetimes. In particular, if
the flux is smeared in time against a normalized Lorentzian
sampling function of characteristic width $t_0$, then, in two
dimensions
\begin{equation}
\hat{F}\equiv\frac{t_0}{\pi}\int_{-\infty}^{\infty} \frac{F(t)}{t^2+t_0^2} dt \geq
-\frac{1}{16\pi t_0^2},
\end{equation}
and in four dimensions
\begin{equation}
\hat{F} \geq -\frac{3}{32\pi^2 t_0^4}.
\end{equation}

A few years later, Ford and Roman \cite{F&Ro95} extended this analysis to the
energy-density observed along the worldline of a geodesic.  Their analysis
begins with with the derivation of a {\em difference quantum inequality} on the two-dimensional,
spatially-compactified, cylinder spacetime ($\reals\times S^1$).  Consider a timelike
geodesic $\gamma(\tau)$ parameterized by proper time $\tau$,  whose tangent vector is
denoted by $u^\mu(\tau)$. Letting $|\psi\rangle$ be an arbitrary quantum state
and $|0_C\rangle$ be the Casimir vacuum state on the cylinder spacetime, they define
the difference in the expectation value of the energy-density between these states as
\begin{equation}
D\langle \bm{T}_{\mu\nu} u^\mu u^\nu \rangle\equiv \langle\psi| \bm{T}_{\mu\nu} u^\mu u^\nu
|\psi\rangle -\langle 0_C| \bm{T}_{\mu\nu} u^\mu u^\nu |0_C \rangle.
\end{equation}
On their own, each of the two terms in the difference are formally divergent,
but both have the same singular structure, thus the difference is finite.
(This is a typical ``regularization'' process employed in QFT.)

In the specific case of an inertial observer, and again using a Lorentzian
weighting function with characteristic width $\tau_0$, they derive the
lower bound
\begin{equation}
\hat{D}\langle \bm{T}_{\mu\nu} u^\mu u^\nu \rangle \equiv \frac{\tau_0}{\pi}
\int_{-\infty}^{\infty} \frac{D\langle \bm{T}_{\mu\nu}u^\mu u^\nu\rangle}
{\tau^2 +\tau_0^2} d\tau\geq-\frac{1}{8\pi\tau_0^2}.
\end{equation}
The important thing to note, which is true for most all forms of inequalities,
is that the lower bound is on the difference between the expectation values
between two different states.   The difference
quantum inequality can be converted to bounds on the renormalized value of
the energy-density by noting that
\begin{equation}
D\langle \bm{T}_{\mu\nu} u^\mu u^\nu \rangle = \langle\psi| \bm{T}_{\mu\nu} u^\mu u^\nu
|\psi\rangle_{\rm Ren.} -\langle 0_C|  \bm{T}_{\mu\nu} u^\mu u^\nu |0_C \rangle_{\rm Ren.}.
\end{equation}
Thus, an {\em absolute quantum inequality} takes the form
\begin{equation}
\frac{\tau_0}{\pi}
\int_{-\infty}^{\infty} \frac{\langle\psi| \bm{T}_{\mu\nu} u^\mu u^\nu|\psi\rangle_{\rm Ren.}}
{\tau^2 +\tau_0^2} d\tau\geq
\frac{1+v^2}{1-v^2}\langle 0_C| \bm{T}_{tt} | 0_C\rangle_{\rm Ren.}-\frac{1}{8\pi\tau_0^2},
\end{equation}
where we have made use of the time independence and symmetry properties of the renormalized
stress-tensor in the Casimir vacuum state on the cylinder spacetime.

In the same paper, Ford and Roman go on to derive a {\em quantum inequality}
in four-dimensional Minkowski spacetime;
\begin{equation}
\hat{\rho} \equiv \frac{\tau_0}{\pi}\int_{-\infty}^{\infty} \frac{\langle : \bm{T}_{\mu\nu}u^\mu
u^\nu:\rangle}{\tau^2 +\tau_0^2} d\tau\geq -\frac{3}{32\pi^2\tau_0^4}.
\end{equation}
Here, the colons denote normal ordering with respect to the standard
Minkowski space vacuum; in other words, it is again a lower bound on
the difference between expectation values between two states.

Since their initial discovery, quantum inequalities have been developed for
an assortment of QFTs, both massless and massive, in a
variety of spacetimes, both flat and curved.  Additionally, they have been
proven for a large class of weighting function beyond just the Lorenzian;
first by Flanagan \cite{Flan97} for the scalar field in two-dimensional
Minkowski spacetime, and followed by  Fewster and Eveson \cite{Fe&E98}
for the massive scalar field in $(n+1)$-dimensional Minkowski spacetime.
For example, let $f(\xi)$ be a smooth, strictly positive function on the
real line with unit area under the curve (one can relax the unit area condition),
then Flanagan's lower bound states
\begin{equation}
\int_{-\infty}^{\infty} \langle : \bm{T}_{tt} (0,t) : \rangle\, f(t)\,dt \geq -\frac{1}{24\pi}
\int_{-\infty}^{\infty} \frac{[f'(\nu)]^2}{f(\nu)} d\nu.
\label{eq:Flan_timelike}
\end{equation}
In the same paper, Flanagan also derives a spatial quantum inequality with nearly
identical form,
\begin{equation}
\int_{-\infty}^{\infty} \langle : \bm{T}_{tt} (x,0) : \rangle\, f(x)\,dx \geq -\frac{1}{24\pi}
\int_{-\infty}^{\infty} \frac{[f'(\nu)]^2}{f(\nu)} d\nu.
\label{eq:Flan_spacelike}
\end{equation}
Both bounds are derived in the rest frame of an inertial observer and are the optimal
lower bound over all possible states. The colons again means normal ordering with
respect to the Minkowski vacuum state.

Significant improvements in the mathematical rigor for the derivation of quantum
inequalities were made by Fewster \cite{Fews00} by employing microlocal analysis
in the context of algebraic QFT in curved spacetime.  The
pairing { quantum inequality} now serves as an umbrella term, of which the
the most frequently studied type is the {\em quantum weak energy inequality}
(QWEI), which typically takes the form \cite{Fews00,Pfen03}
\begin{equation}
\int\langle\omega| :\bm{T}_{\mu\nu}u^\mu u^\nu :_{\omega_0} |\omega\rangle f(\tau) d\tau
\geq -\mathcal{Q}(\omega_0,\gamma,f).
\end{equation}
Here, $f$ is a smooth compactly-supported test function, $\omega$ and $\omega_0$ are
Hadamard states, and the colon with the subscript denotes normal ordering with respect
to $\omega_0$, thus these are again a form of difference inequality, with $\omega_0$ serving
as the reference state.  Finally, the functional $\mathcal Q$ is independent of the
state $\omega$, and microlocal analysis is used to prove that it is finite. These can
again be recast in terms of the renormalized expectations values, resulting in
\begin{equation}
\int\langle\omega| \bm{T}_{\mu\nu}u^\mu u^\nu  |\omega\rangle_{\rm Ren.} f(\tau) d\tau
\geq \int\langle\omega_0| \bm{T}_{\mu\nu}u^\mu u^\nu  |\omega_0\rangle_{\rm Ren.} f(\tau)
d\tau-\mathcal{Q}(\omega_0,\gamma,f).
\end{equation}
In applications, it is commonplace to take the reference state $\omega_0$ to be
the Casimir vacuum state, although this is by no means a requirement.

\subsection{Claims of Violations of Quantum Inequalities}
In two recent papers, Solomon \cite{Solo11,Solo12} puts forth models of
a massless, quantized, scalar field in two-dimensional Minkowski spacetime
with the
presence of an external, time-dependent potential of the form $V(x,t)
= \theta(-t)V_\xi(x)$. Here $\theta$ is the standard Heaviside unit-step-%
function and $\xi$ is the coupling constant between the potential to the field.
The scalar field obeys the Klein-Gordon-Fock wave equation
\begin{equation}
\Box\bm{\Phi}(x,t) + V(x,t)\bm{\Phi}(x,t) = 0.
\label{eq:KGF}
\end{equation}
Such models can be interpreted as a quantum field which transitions from
a field interacting with the potential to being a free field at the $t=0$
Cauchy surface. We call the causal past and future of this Cauchy
surface the IN and OUT regions, respectively.

The classical wave equation associated with the equation above can be solved
independently in both regions using standard PDE techniques for the potentials
chosen by Solomon.
For the IN region, one assumes harmonic time dependence, such that
positive-frequency modes are given by
\begin{equation}
\phi^{\rm IN}(x,t) = \frac{1}{\sqrt{2\omega_{\xi,j}}}\, \chi_{\xi,j}(x)\, e^{-i\omega_{\xi,j} t},
\end{equation}
where the $\chi_{\xi,j}(x)$'s are a complete set of orthonormalized eigenfunctions
to the equation
\begin{equation}
\left[-\partial_x^2 + V_\xi(x)\right]\chi_{\xi,j}(x) = \omega_{\xi,j}^2\, \chi_{\xi,j}(x)
\label{eq:schrodinger_like}
\end{equation}
and $j$ is a label for uniquely identifying an eigenfunction.

The transition across the $t=0$ Cauchy surface is then
handled by assuming $C^1$ continuity conditions in time, i.e.,
\begin{equation}
\phi^{\rm IN}(x,0) = \phi^{\rm OUT}(x,0) \qquad\mbox{ and }\qquad
\partial_t \phi^{\rm IN}(x,0) = \partial_t\phi^{\rm OUT}(x,0).
\end{equation}
From a physical standpoint, this is reasonable; one evolves a
solution to the wave equation with potential up to the $t=0$
Cauchy surface, at which point $\phi^{IN}(x,0)$ and $\partial_t
\phi^{IN}(x,0)$ serve as the Cauchy data for the continued
evolution of the wave into the causal future of the $t=0$
Cauchy surface. In two-dimensional Minkowski spacetime, where
Solomon is working, the future evolution is easily determined
using d'Alembert's solution to the wave equation,
\begin{equation}
\phi^{\rm OUT}(x,t)= \frac{1}{2}\left[\phi^{\rm IN}(x+t,0)+\phi^{\rm IN}(x-t,0)
\right]+\frac{1}{2}\int_{x-t}^{x+t} \partial_t \phi^{\rm IN}(z,0) \,dz.
\end{equation}
Thus, we have mode solutions to the classical wave equation on
the whole spacetime of the form
\begin{equation}
\phi_j(x,t) = \left\{ \begin{array}{cl}
\phi^{\rm IN}(x,t)&\mbox{ for } t\leq 0,\\[3pt]
\phi^{\rm OUT}(x,t)&\mbox{ for }t\geq 0.
\end{array}\right.
\end{equation}
Using canonical quantization, one can then {\it lift} the general solution to
the classical wave equation to a self-adjoint operator,
\begin{equation}
\bm{\Phi}(x,t) = \int d\mu(j) \left[ \bm{a}_j \phi_j(x,t)
+\bm{a}_j^\dagger \overline{\phi_j(x,t)} \right],
\end{equation}
where $d\mu(j)$ is an appropriate measure for the labeling set of the $j$'s,
$\bm{a}_j^\dagger$ and $\bm{a}_j$ are the standard creation and
annihilation operators, respectively, with the usual commutation relations,
and we use the standard QFT Fock space on which these operators act.
In particular, the IN vacuum state $|0\rangle$ is defined such that
$\bm{a}_j |0\rangle = 0$ for all $j$.

The stress-tensor operator associated with the quantized scalar
field for the Klein-Gordon-Fock equation can be separated into two parts,
\begin{equation}
\bm{T}_{\mu\nu} = \bm{K}_{\mu\nu} + \bm{U}_{\mu\nu},
\end{equation}
where, using the terminology of Solomon, the {\em kinetic-tensor}
is defined as the portion of the stress-tensor that is explicitly
free of the potential, i.e.,
\begin{equation}
\bm{K}_{\mu\nu} \equiv (\nabla_\mu\bm{\Phi})(\nabla_\nu \bm{\Phi}) - \frac{1}{2}
g_{\mu\nu} (\nabla^\alpha\bm{\Phi})(\nabla_\alpha \bm{\Phi}),
\end{equation}
while the {\em potential}-tensor is everything in the stress-tensor explicitly
involving the potential, i.e.,
\begin{equation}
\bm{U}_{\mu\nu} \equiv \frac{1}{2}g_{\mu\nu} V(x,t)\bm{\Phi}^2.
\end{equation}
It is important to note that the support of the kinetic-tensor is
the whole spacetime, while the support of the potential-tensor is
restricted to the support of the potential.  Thus, if the support
of the potential is closed or compact, then the support of the
potential-tensor will be closed or compact.
Solomon's {\em kinetic energy-density} is just the $K_{00}$ component
of the kinetic-tensor\footnote{Solomon uses the letter $T$ to
represent both the stress-tensor and the kinetic-tensor.  We choose
the alternate notation of $T$ and $K$ to avoid any unintended confusion
between them.}.  In regions of space where the potential vanishes,
the stress-tensor is equal to the kinetic-tensor. Because of the potential,
all three of the tensors defined above have nontrivial traces and
nontrivial divergences%
\footnote{The traces are given by
${K^\mu}_\mu = \left(1-\frac{n}{2}\right) \left(\nabla^\alpha \phi\right)
\left(\nabla_\alpha \phi\right)$
, ${U^\mu}_\mu = \frac{n}{2} V(x,t)\phi^2$, and
${T^\mu}_\mu = \left(1-\frac{n}{2}\right) \left(\nabla^\alpha \phi\right)\left(\nabla_\alpha \phi\right)
+\frac{n}{2} V(x,t)\phi^2$,
where $n$ is the dimension of the spacetime. The divergences are
$\nabla^\mu K_{\mu\nu} = -V(x,t)\phi\left(\nabla_\nu \phi\right)$,
$\nabla^\mu U_{\mu\nu} = \frac{1}{2}\left(\nabla_\nu V(x,t)\right)\phi^2
+V(x,t)\phi\left(\nabla_\nu \phi\right)$
and
$\nabla^\mu T_{\mu\nu} = \frac{1}{2}\left(\nabla_\nu V(x,t)\right)\phi^2$.
}.

In his papers, Solomon calculates the expectation value of the
kinetic energy-density for the IN vacuum state $|0_\xi\rangle$
on the IN region, finding
\begin{equation}
\langle 0_\xi | \bm{K}_{tt}^{\rm IN} | 0_\xi \rangle (x) = \int \frac{d\mu(j)}{4\omega_{\xi,j}}
\left( \omega_{\xi,j}^2 |\chi_{\xi,j}(x)|^2 + |\chi'_{\xi,j}(x) |^2\right),
\end{equation}
where the prime denotes differentiation of the function with respect
to the argument. After a lengthy calculation, the expectation
value of the energy-density for the IN vacuum state on the OUT
region is \begin{equation}
\langle 0_\xi | \bm{T}_{tt}^{\rm OUT} | 0_\xi \rangle (x,t) = \frac{1}{2}\left[
\langle 0_\xi | \bm{K}_{tt}^{\rm IN} | 0_\xi \rangle (x+t) +
\langle 0_\xi | \bm{K}_{tt}^{\rm IN} | 0_\xi \rangle (x-t)\right],
\end{equation}
where the basis eigenfunctions are chosen to be real valued%
\footnote{If the basis of eigenfunctions is not real valued, then the expression for
the energy-density in the OUT region would be
$ \langle 0 |T_{tt}^{\rm OUT}%
| 0 \rangle (x,t) = \frac{1}{2}\left[ \langle 0 | K_{tt}^{\rm IN} | 0 \rangle
(x+t) + \langle 0 | K_{tt}^{\rm IN} | 0 \rangle (x-t)\right]
-\frac{1}{4}\int d\mu(j) \Im\left[ \chi'_{\xi,j}(x+t)\overline{\chi_{\xi,j}(x+t)}\right]%
+\frac{1}{4}\int d\mu(j) \Im\left[ \chi'_{\xi,j}(x-t)\overline{\chi_{\xi,j}(x-t)}\right].%
$}.
In the regions of the spacetime where the potential is zero,
Solomon conjectures that we may use any of the standard
renormalizations schemes to determine the renormalized values
of both of these expressions. Thus, if there is a stationary
Casimir effect due to the potential in any portion of the IN
region of the
spacetime, this will become a left and right moving pulse of
energy on the OUT region of the spacetime.
For example, one model that Solomon presents is that of a
double-delta-function potential of the form
\begin{equation}
V_\xi(x) = \xi\left[ \delta(x-a/2)+\delta(x+a/2)\right],
\end{equation}
for which Mamev and Trunov \cite{Ma&Tr81} have shown that there
is a constant, negative-valued, Casimir effect for the vacuum
expectation value of the energy-density in the region of space
between the two delta-functions and vanishing outside. Explicitly,
\begin{equation}
\langle 0_\xi | \bm{T}_{tt} | 0_\xi \rangle_{Ren.}(x) =
\langle 0_\xi | \bm{K}_{tt} | 0_\xi \rangle_{Ren.}(x) =
\left\{ \begin{array}{cl} -\eta & \mbox{ for } |x|<a/2,\\
0 & \mbox{ for } |x|>a/2,\end{array}\right.
\label{Eq;Mamev_Trunov}
\end{equation}
where $\eta$ is a positive function of the coupling constant $\xi$
and separation $a$ given by
\begin{equation}
\eta = \eta(\xi,a) = \frac{\xi}{2\pi a} \left(\int_0^\infty \frac{y e^{-y}}%
{y e^{y} + \frac{\xi a}{2}\sinh y}dy - \int_0^\infty \frac{y e^{-y}}%
{y e^{y} + \frac{\xi a}{2}\cosh y}dy
\right).
\end{equation}
Mamev and Trunov are silent
on what the renormalized expectation value of the energy-density
is at the locations of the delta-function potentials ($x =\pm a/2$) . They
do state in \cite{Ma&Tr82} that additional renormalization terms
are required that depend on the potential and its derivatives to
determine $\langle 0_\xi | \bm{T}_{tt}^{IN} | 0_\xi \rangle_{Ren.}(\pm a/2)$.

Solomon uses the Mamev and Trunov double-delta-function potential
on the IN region of his spacetime, which he rewrites as
\begin{equation}
\langle 0_\xi | \bm{K}_{tt}^{\rm IN} | 0_\xi \rangle_{\rm Ren.}(x)
= -\eta\left[ \theta(x+a/2) - \theta(x-a/2)\right].
\label{eq:Solomon_posits1}
\end{equation}
For the OUT region of the spacetime, Solomon
then posits
\begin{equation}
\langle 0_\xi | \bm{T}_{tt}^{\rm OUT} | 0_\xi \rangle_{\rm Ren.} =
-\frac{\eta}{2}\left[ \theta(x+t+a/2)- \theta(x+t-a/2)\right]
- \frac{\eta}{2}\left[ \theta(x-t+a/2)- \theta(x-t-a/2)\right].
\label{eq:Solomon_posits2}
\end{equation}
As was the case with Mamev and Trunov, Solomon is silent about
the value of the renormalized kinetic-tensor on
the IN region at $x=\pm a/2$, and consequently for the renormalized
stress-tensor at points along the future-pointing null rays emanating from
the points $(t,x) = (0,\pm a/2)$ on the OUT region. Solomon then goes
on to show that this particular expression for the vacuum expectation value
of the energy-density would indeed violate the quantum inequalities
of Flannagan \cite{Flan97} on the OUT region of the spacetime.

Unfortunately, Solomon's conclusions are incorrect, as
Eqs.~(\ref{eq:Solomon_posits1}) and~(\ref{eq:Solomon_posits2}) are
incomplete expressions for both the IN-region kinetic energy-density
and the OUT-region energy-density, respectively.  In the case of
static Minkowski spacetime with a time-independent double-delta-function
potential, it has been shown by Graham and colleagues \cite{Gr&Ja02},
in the context of a massive scalar field, that the renormalized
energy-density has nonzero contributions at the points $x=\pm a/2$.
Unfortunately, there is no straightforward way to take the $m\rightarrow 0$
limit of the Graham et. al. results and then separate the renormalized
kinetic energy-density out of the expression for the renormalized
energy-density.

However, for the massless field we can conjecture that the renormalized
IN-region kinetic-tensor will have a $tt$-component of the form
\begin{equation}
\langle 0 | \bm{K}_{tt}^{\rm IN} | 0 \rangle_{\rm Ren.}(x)
= -\eta \left[ \theta(x+a/2) - \theta(x-a/2)\right]
+ q\, \delta(|x|-a/2),
\end{equation}
where $q=q(\xi,a)$ is another function of the coupling constant $\xi$
and separation $a$. This yields an OUT-region energy-density of the form
\begin{eqnarray}
\langle 0 | \bm{T}_{tt}^{\rm OUT} | 0 \rangle_{\rm Ren.}(x)
&=& -\frac{\eta}{2} \left[ \theta(x+t+a/2)- \theta(x+t-a/2)\right] +
\frac{q}{2}\delta(|x+t|-a/2)\nonumber\\
&& -\frac{\eta}{2} \left[ \theta(x-t+a/2) - \theta(x-t-a/2)\right] +
\frac{q}{2}\,\delta(|x-t|-a/2).
\end{eqnarray}
Physically, this describes two square-wave pulses of negative energy
with amplitude $-\eta/2$ traveling outward  at the speed of light
from the initial location of the potential; one moving
to the left and one moving to the right. Additionally, on the leading
and trailing edges of the square-wave pulses are delta-function spikes
of energy, with magnitude $q/2$ which, as we will see below for a
related model, are positive.
The positive energy comes from the creation of particles out of the vacuum
by the quantum field in response to the shutting off of the potential.

Using this new expression for the renormalized energy-density, we can
again consider Flanagan's quantum inequality on the OUT region. To do this,
we use unit-area test functions with the constraint that they only
have support on the OUT region of the spacetime.  Then, substituting the above
energy-density into the quantum inequality, and  using a geodesic
parameterized by $\gamma^\mu(\tau) = (\tau, x_0)$, where $x_0>a/2$ and
$\tau\in[0,\infty)$, results in
\begin{equation}
\frac{q}{2}\left[f(x_0-a/2) + f(x_0+a/2)\right]-\frac{\eta}{2}
\int_{x_0-a/2}^{x_0+a/2} f(\tau)\, d\tau \geq -\frac{1}{24\pi}
\int_0^\infty \frac{|f'(\nu)|^2}{f(\nu)}\, d\nu.
\end{equation}
To determine if the quantum inequality is violated will depend on
the relative strength of the delta-function contributions
to the negative-energy contribution of the square wave
part of the energy-density.

We will put off definitively settling whether or not Flanagan's
quantum inequality is violated for a follow-up paper.  Instead,
for the remainder of this paper, we determine the renormalized
kinetic-tensor on the IN region and the renormalized stress-tensor
on the OUT region of a the two-dimensional cylinder spacetime with
a single delta-function potential that is abruptly shut off at $t=0$.
We find that particle creation in our model causes a left- and right-%
moving delta-function of positive energy in the OUT region stress-tensor.
We also show that all of the classical point-wise energy conditions
fail on this spacetime because of a negative-energy Casimir effect,
but that the positive-energy pulses are sufficiently large to ensure
that the quantum inequality for this spacetime is satisfied
for all inertial observers on the OUT region of the spacetime,
and for all values of the coupling constant $\xi$.

\subsection{Outline}

We begin by considering a massless, quantized, scalar field coupled to a scalar
potential on the spatially-compact, two-dimensional, cylinder spacetime
$\reals\times S^1$. We use the standard $(t,x)$
coordinates with the identification
of points such that $(t,x) = (t,x+L)$.  Here, $L$ is the circumference of the
spatial sections of the universe and we use the standard metric $g_{\mu\nu}
= {\rm diag}(1,-1)$.
The choice of spacetime is made such that the mathematics which follows
is tractable.  Similar calculations could be performed in other spacetimes.

The quantized scalar field obeys the Klein-Gordon-Fock wave equation,
Eq.~(\ref{eq:KGF}), with a Mamev-Trunov-type potential of the form
\begin{equation}
V(x,t) = 2 \xi\,\delta(x)\,\theta(-t),
\label{eq:M-T-potential}
\end{equation}
where $\xi$ is a positive coupling constant and $\delta(x)$ is the Dirac-delta-function. The factor of 2 is included solely for convenience. The potential is
a delta-function of strength $2\xi$ that is abruptly turned off at time $t=0$.

The Mamev-Trunov-type potential breaks the spacetime into two regions: a static
IN region for $t<0$ where the scalar field is coupled to a non-zero delta-function
potential, and a static OUT region for $t>0$ where the scalar field is free
from the potential. A graphical representation of this spacetime with the potential
is presented in Fig.\ 1.

In Sect.~\ref{sec:classical}, we determine the mode solutions to
the Klein-Gordon-Fock wave equation for both regions.  It is
advantageous to separate the modes based on their spatial
symmetry/antisymmetry properties about  $x=0$.  On the whole
spacetime, both the IN and OUT regions, there exists a complete set
of antisymmetric, orthonormal, positive-frequency,  modes solutions of the
form
\begin{equation}
\Phi^{\rm odd}(n,x,t) = (k_n L)^{-1/2} \, \sin(k_n x) \, e^{-i k_n t},
\label{eq:phi_odd}
\end{equation}
with $k_n = 2\pi n/L$ and $n=1,2,3, \dots$.  There are also negative-frequency
antisymmetric mode solutions given by the complex conjugate.  Because these modes
vanish at the origin, they do not experience or interact with the potential.

There also exist symmetric, orthonormal, positive-frequency, mode solutions to
the wave equation, which are sensitive to the potential, of the form
\begin{equation}
\Phi^{\rm even}(j,x,t) = \left\{ \begin{array}{cc}
\phi^{\rm even}(j,x,t)\quad & \mbox{for } t\le 0,\\[4pt]
\phi^{\rm even}_{\rm OUT}(j,x,t)\quad & \mbox{for } t\ge0.
\end{array}\right.
\end{equation}
The IN portion of this mode solution is given by the
expression
\begin{equation}
\phi^{\rm even}(j,x,t) = (\kappa_j L)^{-1/2}A_j \left[ \cos(\kappa_j x) + \frac{\xi}{\kappa_j}
\sin(\kappa_j |x|)\right] \, e^{-i \kappa_j t},
\end{equation}
where $\kappa_j = {2 Z_j}/{L}$, $Z_j$ is the $j$-th positive root of the
transcendental equation
\begin{equation}
Z = \frac{\xi L}{2}\cot(Z),
\end{equation}
and $A_j$ is a normalization constant defined in Eq.\ (\ref{eq:A_j_normalization_coeff})
below. The IN portion of the modes solutions have a corner at the location of the
delta-function potential, while the OUT portion have corners that propagate outward
from the origin of the spacetime at the speed of light.

The OUT portion of the symmetric mode solution is given by a Fourier series,
Eq.\ (\ref{eq:psi_odd_mode_decomposition}), in terms of the ``standard'' basis of
symmetric modes for the potential-free Klein-Gordon equation, of which there are
two kinds: a) an infinite family of time-oscillatory mode solutions,
with the positive-frequency solutions given by
\begin{equation}
\psi^{\rm even}(n,x,t) = (k_n L)^{-1/2} \, \cos(k_n x) \, e^{-i k_n t},
\end{equation}
where $k_n=2\pi n/L$, $n=1,2,3,\dots,$ and the negative-frequency solutions
given by the complex conjugate, and b) two topological, zero-frequency, mode
solutions given by
\begin{equation}
\psi^{\rm top.}(x,t) = \sqrt{\frac{\ell}{2L}}\left(1 - i \frac{t}{\ell}\right)
\label{eq_top_mode}
\end{equation}
and its complex conjugate. The topological modes exist because the
spatial sections of the cylinder spacetime
are compact.  Furthermore, they are necessary to have a complete basis set
to represent a solution to the Cauchy problem for all initial data.  The
antisymmetric, mode solutions, given by
Eq.~(\ref{eq:phi_odd}), do not appear in the Fourier representation
for the OUT portion of the symmetric mode solutions on the whole spacetime.

To determine the Fourier coefficients for the OUT portion of the symmetric
mode solution, we require, like Solommon, $C^1$ continuity (in time) of the complete
mode solution across the $t=0$ Cauchy surface. Essentially, we are using the known
behavior of the IN symmetric mode functions at time $t=0$ as the Cauchy data to
determine a unique solution of the wave equation in the OUT region.
The resulting Fourier series has non-zero Fourier coefficients,
Eqs.\ (\ref{Eq_Fourier_coeff_a}) and (\ref{Eq_Fourier_coeff_b}), for both the
topological modes and the positive and negative-frequency even mode solutions.
Thus, the initially positive-frequency even mode solution on the IN region
of the spacetime develops both positive and negative-frequency components at the
moment that the potential turns off which persist through the OUT region.

In Sec.~\ref{sec:Quantization}, we second-quantize our system, following
the standard canonical quantization scheme in literature (see, for example, Birrell
and Davies \cite{Brl&Dv}).  In this scheme, one promotes the real-valued classical
field $\Phi$ to a self-adjoint operator $ \bm{\Phi}$ on a Hilbert space of states.
The typical Hilbert space is usually given by a standard Fock space. For a Bosonic field
theory, the field operator and its conjugate momenta $\bm{\Pi}$ also satisfy a standard
set of equal time commutation relations. This process works well for our spacetime
because it has a convenient timelike Killing vector.

On the IN region of the spacetime, the Fock space associated with the field
algebra has the usual form, and we define the IN vacuum state $|0_L\rangle$
to be the state destroyed by all of the annihilation operators of the field
algebra, Eq.~(\ref{eq:defn_IN_vac_state}). The subscripted $L$ is included in the
notation to remind us that this is the ground state on a spatially-closed spacetime
of circumference $L$, and not the standard Minkowski-space vacuum state, which we
will denote by $|0\rangle$. States with higher particle content can be constructed
in the usual way by acting with the creation operators.

On the OUT region of the spacetime, there exists an unitarily equivalent
field algebra based upon the ``standard'' mode solutions to the potential-free
wave equation.  So we also present the second-quantization of this equivalent system.
However, we do make one modification to the standard quantization procedure; along
with the time-oscillatory modes, we also second-quantize the topological modes using
the method developed by Ford and Pathinayake \cite{F&Pa89}. At the classical
level, the topological modes given by Eq.~(\ref{eq_top_mode}) have nonzero conjugate
momenta, therefore they can be included in the classical symplectic form that gets
lifted to the commutator relation of the field algebra.  It is found that such a process
produces an algebra with a non-trivial center \cite{D&Lang}.

Because the OUT region had two equivalent field algebras and Fock spaces, we
determine the Bogolubov transformation between the elements of the algebras. Since
the OUT portion of the symmetric mode solutions is already given by a Fourier
series in terms of the ``standard'' modes, determining the explicit form of the
Bogolubov coefficients is simply a task of identifying the correct Fourier coefficient.

Working in the Heisenberg picture, we then calculate the number of ``standard''
quanta created on the OUT region of the spacetime for the IN vacuum state
$|0_L\rangle$. We find that (a) no quanta are created in the odd modes, (b) a finite,
non-zero number of quanta are created in the topological modes,
Eq.~(\ref{eq:number_topological_excited}), (c) a finite, non-zero number of quanta
are created in the time-oscillatory even modes, Eq.~(\ref{eq:number_oscillatory_excited}),
and (d) the total number of quanta created is finite.  All the quanta created in this
model come into existence at the moment the potential is shut off, i.e., at $t=0$.

In Sec.~\ref{sec:Stress-Tensor}, we determine the renormalized expectation value of
the stress-tensor for the IN ground state $|0_L\rangle$ on both the IN
and OUT regions of the spacetime.  For the IN region of the spacetime we find
\begin{equation}
\langle 0_L|\bm{T}_{\mu\nu}|0_L \rangle_{Ren.} = \left(-\frac{\pi}{6L^2}+
\frac{\mathcal{B}-\mathcal{C}}{L^2}\right) \delta_{\mu\nu},
\end{equation}
which holds everywhere except at the location of the delta-function potential.
The $-\pi/6L^2$ part of this expression is the standard Casimir energy-density
for the cylinder spacetime.  The $(\mathcal{B}-\mathcal{C})/L^2$ is the
correction to the ground state energy-density due to the presence of the
potential. Here, both coefficients $\mathcal{B}$ and
$\mathcal{C}$ are positive functions of $\chi\equiv\xi L/2$, given by infinite summations
over the transcendental eigenvalues, Eqs.~(\ref{eq_B_defn}) and (\ref{eq_C_defn})
respectively, and are plotted in Fig.~2.  We explicitly prove that both are convergent,
and we determine that the difference between them always satisfies
\begin{equation}
0\le(\mathcal{B}-\mathcal{C})\leq \frac{\pi}{6}.
\end{equation}

We also determine the renormalized expectation of the stress-tensor on
the OUT region for the same state;
\begin{eqnarray}
\langle 0_L|\bm{T}_{\mu\nu}| 0_L \rangle_{Ren.} &=& \left\{-\frac{\pi}{6 L^2}+\frac{\mathcal{B
- C}}{L^2} + \frac{\mathcal{C}}{2L^2}\sum_{n=-\infty}^\infty
\left[\delta\left(\frac{t+x}{L}-n\right) + \delta\left(\frac{t-x}{L}-n\right)\right]\right\}
\delta_{\mu\nu}\nonumber\\
 &&+ \frac{\mathcal{C}}{2L^2}\sum_{n=-\infty}^\infty\left[\delta\left(\frac{t+x}{L}-n\right)
- \delta\left(\frac{t-x}{L}-n\right)\right] \left( \begin{array}{cc}
0 & 1\\ 1 & 0 \end{array}\right),
\end{eqnarray}
which holds for all spacetime locations to the future of the $t=0$ Cauchy surface.
It is covariantly conserved, i.e., $\nabla^\mu\langle 0_L|\bm{T}_{\mu\nu}| 0_L
\rangle_{Ren.}=0$, and
we find the standard Casimir energy-density for the cylinder spacetime followed by a
correction to the ground state energy-density given by the
$(\mathcal{B}-\mathcal{C})/L^2$ term.
The remaining terms in the above expression are the contributions to the stress-tensor
due to the quanta excited (i.e. particle creation) from the shutting off of the potential.
The remarkably simple expression of two classical, point-like particles moving outward from
the origin to the left and right with equal amplitude $\mathcal{C}/2$ is the result of a
very detailed analysis of the properties of the Bogolubov coefficients and identities,
and their application to the very complicated expression for the ``moving'' parts of
the stress-tensor given by the Fourier series in Eq.~(\ref{eq:rho_defn}).

In Sec.~\ref{sec:Energy Cond}, we evaluate the energy conditions
from general relativity on the OUT region of the spacetime, using
the expression above for the renormalized stress-tensor.  For a
timelike geodesic worldline, the renormalized expectation value
of the energy-density is given by Eq.~(\ref{eq:Timelike_contracted}), and
for a null geodesic worldline by Eq.~(\ref{eq:Null_contracted}).
We find that the null energy condition (NEC), weak
energy condition (WEC), the strong energy condition (SEC), and the dominant
energy condition (DEC) all fail on some region of the space-time for the OUT-%
region stress-tensor because the difference $\mathcal{B}-\mathcal{C}\leq\pi/6$,
and is therefore insufficiently large to overcome the usual $-\pi/6L^2$
term of the Casimir energy.  We then calculate the total energy in a
constant-time Cauchy surface on the OUT region, finding
\begin{equation}
\mathcal{E} = -\frac{\pi}{6L}+\frac{\mathcal{B}}{L}.
\end{equation}
We note that the total energy is a constant, independent of time, further
indicating that the renormalized stress-tensor on the OUT region is conserved for
all time $t>0$.  Additionally, because of the dependance of $\mathcal{B}$ on the
value of $\chi$, the total energy in the Cauchy surface is negative for values of
$\chi\leq0.82$, positive for values of $\chi\geq0.83$, and it passes through
zero somewhere in the range $0.82<\chi<0.83$.

In the final part of Sec.~\ref{sec:Energy Cond}, we use our normal-ordered
expectation value of the energy-density for the IN vacuum state on the OUT
region in a QWEI for the two-dimensional cylinder
spacetime without potential, given by
\begin{eqnarray}
\int_\reals d\tau\, \langle\omega| \bm{\rho}  |\omega\rangle_{\rm Ren.} (\tau)
\left[g(\tau)\right]^2 &\geq&
\frac{1+v^2}{1-v^2}\left(-\frac{\pi}{6 L^2}\right) \int_\reals d\tau\, \left[g(\tau)\right]^2
-\frac{1}{2L}\sum_{n=1}^\infty k_n \left[\frac{1+v}{1-v} \int_0^\infty
\frac{d\alpha}{\pi} \left| \hat{g}\left(\alpha + k_n\sqrt{\frac{1+v}{1-v}}\right)\right|^2
\right.
\nonumber\\
&&\left. +\frac{1-v}{1+v} \int_0^\infty
\frac{d\alpha}{\pi} \left| \hat{g}\left(\alpha + k_n\sqrt{\frac{1-v}{1+v}}\right)\right|^2
\right],
\end{eqnarray}
where $|\omega\rangle$ is any Hadamard state on the cylinder spacetime, and
$g(\tau)$ is a smooth, real-valued, compactly-supported test function on
real line. The derivation of this QWEI, with the inclusion of the topological
modes, is contained in Appendix~\ref{sec:QWEI for cylinder}. We can use this
inequality on the OUT region of our spacetime if we restrict the set of test
functions to only those which have compact support to the future of the $t=0$
Cauchy surface.

Evaluating the left-hand side of the inequality for the state
$|0_L\rangle$ yields
\begin{eqnarray}
L.H.S &=& \int d\tau \, \langle 0_L | :\bm{\rho}:_{\tilde{0}_L}  | 0_L \rangle [g(\tau)]^2\nonumber\\
&=& \frac{1+v^2}{1-v^2}\left(-\frac{\pi}{6 L^2}\right) \int d\tau\, [g(\tau)]^2
+\frac{1+v^2}{1-v^2}\left(\frac{\mathcal{B - C}}{L^2}\right) \int d\tau [g(\tau)]^2\nonumber\\
&&+\frac{\mathcal{C}}{2L^2} \sum_{n=-\infty}^{\infty} \left[ \frac{1+v}{1-v} \int d\tau [g(\tau)]^2
\delta\left(\frac{t_0+x_0+(1+v)\gamma\tau}{L}-n\right)\right.\nonumber\\
&&\left.+\frac{1-v}{1+v} \int d\tau [g(\tau)]^2
\delta\left(\frac{t_0-x_0+(1-v)\gamma\tau}{L}-n\right)\right].
\end{eqnarray}
Notice that only the first of the four terms in the result for the
left-hand side is negative, and that it is identical to the first
term on the right-hand side of the QWEI. The remaining terms on the
right-hand side are negative.  Thus, the QWEI is satisfied by the
stress-energy tensor of the IN vacuum state on the OUT region of
the spacetime for all allowed test functions $g(t)$ with support
to the future of the $t=0$ Cauchy surface, and for all values of $\xi$.

The main body of the paper concludes with some comments and
conjectures in Sec.~\ref{sec:Conclusion}.  In addition to the main
body, there are five appendices containing technical information
necessary for the paper to be complete, and to which we refer
throughout the document.  The appendices include: a proof of
the equivalence of the IN and OUT region mode functions on a
bow-tie shaped domain surrounding the $t=0$ Cauchy surface; the
construction of the advanced-minus-retarded Green's function on
the cylinder spacetime when topological modes are included;
the convergence and properties of certain summations over the
eigenvalues of the transcendental equation; notes on an alternative
way to determine the IN vacuum stress-tensor on the IN region
and why it fails; and finally the derivation of the QWEI
on the cylinder spacetime.

\subsection{Mathematical Notation}

We use units in which $\hbar$, $c$ and $G$ are set to unity throughout
the paper.  The complex conjugate of a complex number $z\in\complex$ is
denoted by $\overline{z}$, and similarly for functions.  For complex-valued
functions $u(x)$ and $v(x)$, we use the standard L$^2$ inner-product,
\begin{equation}
(u,v)_{L^2} \equiv \int_{S^1} u(x) \overline{v(x)}\, dx.
\end{equation}
The normalization for mode solutions of the wave equation is chosen
such that the modes are pseudo-orthonormal with respect to the standard
bilinear product used in QFT \cite{Brl&Dv},
\begin{equation}
\left(\phi_1, \phi_2  \right)_{\rm QFT} \equiv -i \int_{-L/2}^{L/2} \left[
\phi_1(x,t) \left(\partial_t \overline{\phi_2}(x,t) \right) - \left(
\partial_t \phi_1(x,t) \right)\overline{\phi_2}(x,t)\right]dx.
\label{eq:normalization}
\end{equation}

Operators will be typeset in bold face to distinguish then from variables
and functions.  The Hermitian conjugate of an operator $\bm{a}$ will be
denoted by $\bm{a}^\dagger$.

We define the Fourier transform on a Schwartz class function $f\in
\mathcal{S}(\reals)$, the space of smooth functions that decay at infinity, as
\begin{equation}
\hat{f}(\alpha) = \int_{-\infty}^\infty f(x)\,e^{i\alpha x} dx.
\end{equation}
Since the Fourier transform is an automorphism on Schwartz class functions, we have
that the inverse Fourier transform is
\begin{equation}
f(x) = \frac{1}{2\pi} \int_{-\infty}^\infty \hat{f}(\alpha)\,e^{-i\alpha x} d\alpha.
\end{equation}
For this choice of definition of the Fourier transform, the convolution theorem states
\begin{equation}
\int_{-\infty}^\infty f(x) g(y-x) dx = \frac{1}{2\pi}\int_{-\infty}^{\infty} \hat{f}(\alpha)
\hat{g}(\alpha)\, e^{-i\alpha y} d\alpha,
\end{equation}
which has as a corollary Parseval's theorem,
\begin{equation}
\int_{-\infty}^{\infty} | f(x) |^2 dx = \frac{1}{2\pi} \int_{-\infty}^{\infty} | \hat{f}
(\alpha) |^2 d\alpha.
\end{equation}

\section{The Classical Formalism}\label{sec:classical}

Let $\stM$ be an $n$-dimensional, globally hyperbolic, Lorentzian
spacetime with smooth metric $\bm g$ of signature
$(+,-,\dots,-)$.  On this spacetime we have a real-valued scalar field
$\phi:\stM\rightarrow\reals$, which interacts with a scalar
potential $V(x)$.  This situation is described by the action
\begin{equation}
S_{\rm matter}(\phi,g^{\mu\nu})= \frac{1}{2} \int \left[ g^{\mu\nu}
\left(\partial_\mu \phi\right) \left(\partial_\nu \phi\right) - V(x)
\phi^2 \right]\sqrt{-g}\, d^nx,
\end{equation}
where $g_{\mu\nu}$ is the spacetime metric,
$g=\det{g_{\mu\nu}}$, $g^{\mu\nu}$ is the inverse of the
metric, and $\partial_\mu$ is the partial derivative. Variation
of the action with respect to the scalar field yields the
standard Klein-Gordon-Fock wave equation
\begin{equation}
\frac{1}{\sqrt{-g}} \partial_\mu\sqrt{-g} g^{\mu\nu}\partial_\nu \phi + V\phi=0,
\end{equation}
or, more succinctly, $\Box\phi + V\phi = 0$.  Similarly, the
stress-tensor
is found by varying the action with respect to the inverse-metric.  When
considered with the gravitational action \cite{Wald_GR}, the
stress-tensor for minimal coupling has the form
\begin{equation}
T_{\mu\nu} = (\partial_\mu\phi)(\partial_\nu\phi) - \frac{1}{2} g_{\mu\nu}\left[
g^{\alpha\beta}(\partial_\alpha\phi)(\partial_\beta\phi)- V \phi^2 \right].
\label{eq:classical stress-tensor}
\end{equation}

We now make a two choices so that the mathematics which follows is more
tractable. First, we choose to work on the the standard two-dimensional
cylinder spacetime $\reals\times S^1$.  This is done for two
reasons: a) the spactime is boundaryless so there are no boundary conditions
to consider, and b) the spectrum of the
Laplace operator on $S^1$, with and without the potential, is discreet.
We use the standard Minkowski space $(t,x)$ coordinates with the
identification of points such that $(t,x) = (t,x+L)$.  Here, $L$ is the
circumference of the spatial sections of the universe. Secondly, on this
spacetime we have a scalar, Mamev-Trunov-type potential \cite{Ma&Tr82}
given by Eq.~(\ref{eq:M-T-potential}).

The classical mode functions to the wave equation can be solved for independently
in both regions.  To determine mode functions on the whole spacetime, we take each
mode function from the IN region and require that the function and its first
derivative match across the $t=0$ Cauchy surface to a general Fourier decomposition
of the wave function in the OUT region, i.e. we require $C^1$ continuity in $t$
of the wave functions. This matching is used to determine the Fourier coefficients
for the OUT solution of the wave solution. We now present the details of this process.

\subsection{Mode Solutions on the IN Region, $t<0$}

For our spacetime, and upon substitution of the potential, the wave equation
for the IN region is
\begin{equation}
\partial_t^2 \phi - \partial_x^2 \phi + 2 \xi\,\delta(x) \phi = 0.
\end{equation}
Using the standard techniques for separation of variables, we assume a
solution of the form $\phi(x,t) = u(x) T(t)$, such that the time dependence solves
\begin{equation}
T_{tt}(t) + \lambda T(t) = 0,
\label{eq:timedepend}
\end{equation}
while the space dependence leads to the Schr\"{o}dinger-like equation
\begin{equation}
-u_{xx}(x) + 2 \xi\,\delta(x) u(x) = \lambda u(x).
\label{eq:Schrodinger}
\end{equation}
Here, $\lambda$ is the separation constant, playing a role akin to the
energy in ordinary quantum mechanics. The operator
\begin{equation}
\bm{O} = -\frac{d^2}{dx^2} + 2 \xi\,\delta(x)
\end{equation}
is Hermitian, i.e.,  $(u,\bm{O} v)_{L^2} = (\bm{O} u, v)_{L^2}$, with
respect to the standard $L^2$ inner product on $S^1$.

The spatial sections of the universe are compact, therefore the eigenvalues $\lambda$
are discrete.  Furthermore, the eigenvalues are real-valued and greater than or equal
to zero.  A convenient $L^2$-orthonormalized basis of eigenfunction to
Eq.~(\ref{eq:Schrodinger}) is given by (a) a family of antisymmetric eigenfunctions,
\begin{equation}
u^{\rm odd}(n,x)=\sqrt{\frac{2}{L}} \sin(k_n x),
\end{equation}
where $k_n={2\pi n}/{L}$, $\lambda_n = (k_n)^2$ and $n=1,2,3,\dots$, and (b)
a family of symmetric eigenfunctions,
\begin{equation}
u^{\rm even}(j,x)= \sqrt{\frac{2}{L}} A_j \left[ \cos(\kappa_j x) + \frac{\xi}{\kappa_j}
\sin(\kappa_j |x|)\right],
\end{equation}
where $\kappa_j = {2 Z_j}/{L}$, $\lambda_j = (\kappa_j)^2$, and $Z_j$ is the $j$-th
positive root of the transcendental equation
\begin{equation}
Z = \frac{\xi L}{2}\cot(Z).
\end{equation}
For any value of $j$, the value of $Z_j$ lays in the interval
between $(j-1)\pi$ and $(j-\frac{1}{2})\pi$. For $(j-1)>\xi L/2\pi$,
the values of the $Z_j$'s approach the poles of the
cotangent function from above. A fairly good approximation for $Z_j$
using the first two terms in the Taylor series of the cotangent
function is
\begin{eqnarray}
Z_j &\approx& (j-1)\pi + \frac{1}{2}\left(1+\frac{\chi}{3}\right)^{-1}
\left[\sqrt{(j-1)^2 \pi^2+4\chi\left(1+\frac{\chi}{3}\right)}-(j-1)\pi\right]\nonumber\\
&=& (j-1)\pi +2\chi\left[(j-1)\pi+\sqrt{(j-1)^2\pi^2+4\chi\left(1+\frac{\chi}{3}\right)}\right]^{-1},
\label{eqn:Z_j_approx}
\end{eqnarray}
where $\chi = \xi L/2$. Strictly speaking, the exact value of $Z_j$ is always
less that the value of the approximation above.

The normalization constant for the symmetric eigenfunctions is
\begin{equation}
A_j = \cos(Z_j) \left[ 1+\frac{\sin(Z_j)\cos(Z_j)}{Z_j}\right]^{-1/2}.
\label{eq:A_j_normalization_coeff}
\end{equation}
There do not exist any eigenfunctions with eigenvalue $\lambda=0$.

From the above $L^2$-eigenfunctions, we can define positive-frequency mode solutions
to the wave equation on the IN-region:
\begin{equation}
\phi^{\rm odd}(n,x,t) = (2 k_n)^{-1/2} u^{\rm odd}(n,x) \, e^{-i k_n t}
\label{eq:antisym_IN}
\end{equation}
and
\begin{equation}
\phi^{\rm even}(j,x,t) = (2 \kappa_j)^{-1/2}u^{\rm even}(j,x) \,
e^{-i \kappa_j t}.
\label{eq:symm_IN}
\end{equation}
The normalization for these mode solutions has been chosen such that the modes
are orthonormal with respect to the standard bilinear product used in QFT, Eq.~(\ref{eq:normalization}).
Negative-frequency mode solutions are given by the complex conjugate of
the above expressions.

\subsection{Mode Solutions on the OUT Region, $t>0$}
The OUT region is simply the spacetime $\reals\times S^1$ with no potential, i.e.,
it is the standard cylinder spacetime.
Assuming a solution of the form $\psi(x,t) = v(x) T(t)$, we find that the
time dependence  again solves Eq.~(\ref{eq:timedepend}), while the space
dependence leads to
\begin{equation}
-v_{xx}(x) = \lambda v(x).
\end{equation}
Here, $\lambda$ is again the separation constant.  The eigenvalues and
eigenfunctions to the spatial equation are well known;  There are
(a) antisymmetric eigenfunctions
\begin{equation}
v^{\rm odd}(n, x) = u^{\rm odd}(n, x),
\end{equation}
(b) symmetric eigenfunctions
\begin{equation}
v^{\rm even}(n,x) = \sqrt{\frac{2}{L}} \cos(k_n x),
\end{equation}
and (c) a zero-eigenvalue topological solution
\begin{equation}
v^{\rm top.}(x) = \frac{1}{\sqrt{L}}.
\end{equation}
Both the symmetric and antisymmetric eigenfunctions have $k_n =
2\pi n/L$ with $\lambda_n= (k_n)^2$.  A generic function on the
circle can be represented as a Fourier series in this basis as
\begin{equation}
f(x) =  c\, v^{\rm top.}(x) + \sum_{n=1}^\infty \left( a_n v^{\rm odd}(n, x)+
b_n v^{\rm even}(n,x)\right),
\end{equation}
where $c, \{a_n\}$, and $\{b_n\}$ are all Fourier coefficients.  In particular, the
Dirac $\delta$-function on $S^1$ has the representation
\begin{equation}
\delta(x-x') =  v^{\rm top.}(x) v^{\rm top.}(x') + \sum_{n=1}^\infty \left(
v^{\rm odd}(n, x) v^{\rm odd}(n, x')+ v^{\rm even}(n,x) v^{\rm even}(n,x')\right).
\label{eq:delta_fourier}
\end{equation}

The positive-frequency mode solutions to the wave equation on the OUT region for the
antisymmetric and symmetric eigenfunctions are simply
\begin{equation}
\psi^{\rm odd}(n,x,t) = (2 k_n)^{-1/2 }v^{\rm odd}(n,x) e^{-i k_n t}
\label{eq:psi_odd_out}
\end{equation}
and
\begin{equation}
\psi^{\rm even}(n,x,t) = (2 k_n)^{-1/2} v^{\rm even}(n,x) e^{-i k_n t},
\label{eq:psi_even_out}
\end{equation}
respectively. The negative-frequency solutions are given by the complex conjugate of the
above expressions.  The topological eigenfunction leads to an often neglected solution of
the wave equation,
\begin{equation}
\psi^{\rm top.}(x,t) = \sqrt{\frac{\ell}{2}}\, v^{\rm top.}(x)\left(1 - i \frac{t}{\ell}\right)
= \sqrt{\frac{\ell}{2L}}\left(1 - i \frac{t}{\ell}\right),
\label{eq:psi_top_out}
\end{equation}
where $\ell$ is an arbitrary constant that sets a length scale \cite{F&Pa89}.
Unlike the time oscillatory solutions, the topological solution is not an
eigenfunction of the energy operator $i\partial_t$. The complex conjugate
of the topological solution is also a linearly independent solution
of the wave equation.  All three types of solutions are orthonormal with
respect to the bilinear product Eq.~(\ref{eq:normalization}), i.e., they satisfy
\begin{equation}
(\psi_j,\psi_{j'})_{\rm QFT} = \delta_{jj'},  \qquad (\overline{\psi_j},
\overline{\psi_{j'}})_{\rm QFT} = -\delta_{jj'}, \qquad\mbox{and}\qquad
(\psi_j,\overline{\psi_{j'}})_{\rm QFT} = 0.
\end{equation}
where the labels $j$ and $j'$ specify both the type of mode and the value of $n$.

A generic, complex-valued, classical solution to the wave equation in the OUT
region is given by the Fourier series
\begin{equation}
\psi(x,t) = a\, \psi^{\rm top.}(x,t) + b\, \overline{\psi^{\rm top.}(x,t)}+
\sum_{n=1}^\infty
\left[ a_n \psi^{\rm odd}(n,x,t) + b_n \overline{\psi^{\rm odd}(n,x,t)} + c_n
\psi^{\rm even}(n,x,t) + d_n \overline{\psi^{\rm even}(n,x,t)} \right],
\label{eq:fourier}
\end{equation}
where $a$, $b$, $\{a_n\}$, $\{b_n\}$, $\{c_n\}$, and $\{d_n\}$ are
complex-valued constants.

\subsection{Mode Solutions on the Whole Spacetime}

Next, we determine mode solutions on the whole of the spacetime for the time-dependent
potential. Let $\phi(x,t)$ be any solution to the wave equation on the IN region.  We
know that a general solution in the OUT region is given by Eq.~(\ref{eq:fourier})
above. At the $t=0$ Cauchy surface where the potential abruptly turns off, we require
continuity of the wave function and its first time derivative, i.e.,
\begin{equation}
\phi(x,0) = \psi(x,0) \qquad\mbox{ and }\qquad \partial_t \phi(x,0) = \partial_t \psi(x,0).
\end{equation}
Upon substitution, we find
\begin{equation}
\phi(x,0) = \sqrt{\frac{L}{2}} v^{\rm top}(x) (a+b)  + \sum_{n=1}^\infty \frac{1}{\sqrt{2 k_n}}
\left[ v^{\rm odd}(n,x) (a_n + b_n) +v^{\rm even}(n,x) (c_n+d_n)\right]
\end{equation}
and
\begin{equation}
\partial_t \phi(x,0) = -\frac{i}{\sqrt{2L}} v^{\rm top}(x)(a- b) - i \sum_{n=1}^\infty \sqrt{
\frac{k_n}{2}} \left[ v^{\rm odd}(n,x)(a_n - b_n) + v^{\rm even}(n,x) (c_n-d_n)\right].
\end{equation}
Next, we apply Fourier's trick; put the above expressions into the first slot
of the $L^2$ inner product with one of the OUT basis functions in the second
slot.  Permuting through all the basis functions results in
\begin{eqnarray}
a &=& \frac{1}{\sqrt{2\ell}}\left(\phi(x,0)+i\ell\partial_t\phi(x,0),v^{\rm top.}(x)\right)_{L^2},\\
b &=& \frac{1}{\sqrt{2\ell}}\left(\phi(x,0)-i\ell\partial_t\phi(x,0),v^{\rm top.}(x)\right)_{L^2},\\
a_n &=& \sqrt{\frac{k_n}{2}}\left(\phi(x,0)-\frac{1}{ik_n}\partial_t\phi(x,0),v^{\rm odd}(n,x)\right)_{L^2},\\
b_n &=& \sqrt{\frac{k_n}{2}}\left(\phi(x,0)+\frac{1}{ik_n}\partial_t\phi(x,0),v^{\rm odd}(n,x)\right)_{L^2},\\
c_n &=& \sqrt{\frac{k_n}{2}}\left(\phi(x,0)-\frac{1}{ik_n}\partial_t\phi(x,0),v^{\rm even}(n,x)\right)_{L^2},\\
d_n &=& \sqrt{\frac{k_n}{2}}\left(\phi(x,0)+\frac{1}{ik_n}\partial_t\phi(x,0),v^{\rm even}(n,x)\right)_{L^2}.
\end{eqnarray}
We now explicitly determine these coefficients for the basis of IN mode solutions:

\vspace*{10pt}
\noindent{\bf Odd Mode Solutions}:
If $\phi(x,t) = \phi^{\rm odd}(m,x,t)$ for $t\leq0$, then $\phi(x,0) = (2 k_m)^{-1/2} v^{\rm odd}(m,x)$ and
$ \partial_t \phi(x,0) = -i k_m (2 k_m)^{-1/2} v^{\rm odd}(m,x)$.  Upon substitution into the
above expressions, we find $a=b=b_n=c_n=d_n=0$ for all $n$, and $a_n = \delta_{nm}$.  So the
antisymmetric, positive-frequency mode solutions to the wave equation on the whole
spacetime is given by
\begin{equation}
\Phi^{\rm odd}(n,x,t) = (2 k_n)^{-1/2} \, u^{\rm odd}(n,x) \, e^{-i k_n t} \qquad\mbox{ with }
\qquad
k_n = \frac{2\pi n}{L} \mbox{ and } n=1,2,3, \dots.
\end{equation}
This family of solutions, as well as its complex conjugate, are
entirely ignorant to the presence of the potential.

\vspace*{10pt}
\noindent{\bf Even Mode Solutions}:
If $\phi(x,t) = \phi^{\rm even}(j,x,t)$ for $t\leq0$, then $\phi(x,0) = (2\kappa_j)^{-1/2}
u^{\rm even}(j,x)$ and $\partial_t\phi(x,0) = -i (\kappa_j/2)^{1/2}u^{\rm even}(j,x)$.
Because both of the preceding expressions are even functions in the variable $x$,
it is immediately obvious that $a_n=b_n=0$ for all $n$.  Additionally,
\begin{equation}
a =\frac{1}{2\sqrt{\kappa_j \ell}}(1+\kappa_j \ell) Y_{j,0}
\qquad\mbox{ and }\qquad
b =\frac{1}{2\sqrt{\kappa_j \ell}}(1 - \kappa_j \ell) Y_{j,0},
\label{Eq_Fourier_coeff_a}
\end{equation}
where the coefficient
\begin{equation}
Y_{j,0} = \left(u^{\rm even}(j,x),v^{\rm top.}(x) \right)_{L^2} =
\frac{\xi L A_j}{\sqrt{2} Z_j^2}.
\end{equation}
The remaining two sets of coefficients are found to be
\begin{equation}
c_n = \frac{1}{2}\sqrt{\frac{k_n}{\kappa_j}}\left(1+\frac{\kappa_j}{k_n}\right) Y_{j,n}
\qquad\mbox{ and }\qquad
d_n = \frac{1}{2}\sqrt{\frac{k_n}{\kappa_j}}\left(1-\frac{\kappa_j}{k_n}\right) Y_{j,n},
\label{Eq_Fourier_coeff_b}
\end{equation}
where
\begin{equation}
Y_{j,n}=\left( u^{\rm even}(j,x), v^{\rm even}(n,x)\right)_{L^2} = \frac{\xi L A_j}{Z_j^2-(\pi n)^2}.
\end{equation}
Note, $Y_{j,0}$ is not the $n=0$ expression of $Y_{j,n}$; the two differ by a factor of
$\sqrt{2}$.  The $Y_{j,n}$'s turn out to be the Fourier coefficients for the Fourier
series of $u^{\rm even}(j,x)$ when written in the OUT eigenfunctions, i.e.,
\begin{equation}
u^{\rm even}(j,x) = Y_{j,0} v^{\rm top.}(x) + \sum_{n=1}^{\infty} Y_{j,n} v^{\rm even}(n,x).
\label{eq:Fourier for Ueven}
\end{equation}
From the Bogolubov identities below, Eq.~(\ref{eq:Bogolubov}), one can demonstrate
that the $Y_{j,n}$ coefficients satisfy
\begin{equation}
\sum_{j=1}^\infty Y_{j,m} Y_{j,n} = \delta_{mn},\label{eq:y_orthog}
\end{equation}
where the allowed $m$ and $n$ also include zero.

Substituting the coefficients back into the Fourier decomposition of $\psi(x,t)$,
we have that the time-evolution of an IN mode-solution into the OUT
region is
\begin{eqnarray}
\phi^{\rm even}_{\rm OUT} (j,x,t) 
&=& \sum_{n=0}^\infty\left[ c_n \psi^{\rm even}(n,x,t) +
d_n \overline{\psi^{\rm even}(n,x,t)}\right],
\label{eq:psi_odd_mode_decomposition}
\end{eqnarray}
Here we are abusing our notation a bit with $\psi^{\rm even}(0,x,t) =
\psi^{\rm top}(x,t)$, $c_0 = a$ and $d_0 = b$.  We also wish to alert
the reader that all of the Fourier coefficients given above are dependent
upon the value of $j$ for the mode in question, although we have not
explicitly written it that way.  This notational deficiency will be
rectified shortly when the Bogolubov coefficients are defined below.

On the whole of the
spacetime, we have that the symmetric mode solutions to the wave equation are
of the form
\begin{equation}
\Phi^{\rm even}(j,x,t) = \left\{ \begin{array}{cc}
\phi^{\rm even}(j,x,t)\quad & \mbox{for } t\leq 0,\\[4pt]
\phi^{\rm even}_{\rm OUT}(j,x,t)\quad & \mbox{for } t\ge0.
\end{array}\right.
\end{equation}
The symmetric modes start out as purely positive frequency, however, the
shutting off of the potential at $t=0$ causes them to develop topological
and negative frequency components.

There is one more important property of the symmetric mode solutions to
the wave equation; we prove in Appendix~\ref{Appendix_overlap} below
that
\begin{equation}
\phi^{\rm even}(j,x,t) = \phi^{\rm even}_{\rm OUT}(j,x,t)
\end{equation}
on the domain $\mathcal{D}\cup\{(0,0)\}$, where the open, bow-tie-shaped domain
\begin{equation}
\mathcal{D} \equiv\left\{ (x,t)\in \left[-\frac{L}{2},\frac{L}{2}\right] \times
\left[-\frac{L}{2},\frac{L}{2}\right]
\big|-|x| < t < |x|\right\}.
\end{equation}
In other words, we can extend the IN mode solutions to the future of
$t=0$ Cauchy surface, and likewise extend the OUT mode solutions to the
past of the same Cauchy surface.  This is because of causality in the
spacetime, i.e., the mode solutions don't alter their behavior until
information has had time to propagate outward from the location of the shutting
off of the potential.  So physically and mathematically we actually have
\begin{equation}
\Phi^{\rm even}(j,x,t) = \left\{ \begin{array}{cl}
\phi^{\rm even}(j,x,t)\quad & \mbox{for } t< |x|,\\[4pt]
\phi^{\rm even}(j,0,0)=\phi^{\rm even}_{\rm OUT}(j,0,0)\quad & \mbox{for } t=x=0,\\[4pt]
\phi^{\rm even}_{\rm OUT}(j,x,t)\quad & \mbox{for } t>-|x|.
\end{array}\right.
\end{equation}

A generic complex-valued classical solution to the wave equation on the whole
spacetime is given by the Fourier series
\begin{equation}
\Phi(x,t) = \sum_{n=1}^\infty \left[ \alpha_n \Phi^{\rm odd}(n,x,t) + \alpha_n^*
\overline{\Phi^{\rm odd}(n,x,t)} \right]+ \sum_{j=1}^\infty\left[ \beta_j
\Phi^{\rm even}(j,x,t) + \beta_j^* \overline{\Phi^{\rm even}(j,x,t)} \right]
\label{eq:fullFourdecom}
\end{equation}
where $\{\alpha_n\}$, $\{\alpha^*_n\}$, $\{\beta_n\}$, $\{\beta^*_n\}$ are
complex-valued constants.

We have seen that the odd mode solutions are unaffected by the delta-function
potential, and therefore remain monochromatic with the same positive frequency.
On the other hand, the even mode solutions change behavior when the delta-function
is turned off, so even classically, the initially monochromatic positive frequency
solution develops polychromatic positive and negative frequency components in
the OUT region. Also notice that the mode solutions contain a contribution
from the topological mode.  In the quantum treatment of this problem, we will
see that both of these give rise to particle creation at the moment the
potential turns off.

\section{Canonical Quantization}\label{sec:Quantization}

To second-quantize our system, we will follow the standard canonical quantization
scheme in literature (see, for example, Birrell and Davies \cite{Brl&Dv}).  In
this scheme, one lifts the real-valued classical field $\Phi$ to a self-adjoint
operator $ \bm{\Phi}$ on a Hilbert space of states.  The typical Hilbert space
is usually given by a Fock representation. For a Bosonic field theory, the field
operator and its conjugate momenta $\bm{\Pi}$ must also satisfy a standard set
of equal time commutation relations. This process works well for our spacetime
because it has a convenient timelike Killing vector.

\subsection{Quantization of the Field Operator on $\reals\times S^1$ with Potential}
\label{subsec:INquantization}
 For real-valued fields based on Eq.~(\ref{eq:fullFourdecom}), we must require
$\alpha_n^* = \overline{\alpha_n}$ and $\beta_n^* =\overline{\beta_n}$.  Next,
we promote the Fourier coefficients to operators on a Hilbert space, i.e.,
$a_n\mapsto\bm{a}_n$ and $b_j\mapsto\bm{b}_j$, to form a self-adjoint
field operator
\begin{equation}
\bm{\Phi}(x,t) = \sum_{n=1}^\infty \left[ \bm{a}_n \Phi^{\rm odd}(n,x,t)
+ \bm{a}_n^\dagger\overline{\Phi^{\rm odd}(n,x,t)} \right]+
\sum_{j=1}^\infty\left[ \bm{b}_j \Phi^{\rm even}(j,x,t) +
\bm{b}_j^\dagger \overline{\Phi^{\rm even}(j,x,t)} \right].
\label{eq:standard_field_operator}
\end{equation}
Here, $\dagger$ specifies the Hermitian conjugate and.
The field operator must also satisfy the equal time commutation relations
\begin{equation}
\left[ \bm{\Phi}(x,t), \bm{\Phi}(x',t) \right] = 0 =
\left[ \bm{\Pi}(x,t), \bm{\Pi}(x',t) \right]
\qquad\mbox{ and }\qquad
\left[ \bm{\Phi}(x,t), \bm{\Pi}(x',t) \right] = i \delta(x-x')\mathbb{I},
\end{equation}
where $ \bm{\Pi}(x,t)\equiv\partial_t\bm{\Phi}(x,t)$ and $\mathbb{I}$ is
the identity operator.
These commutation relationships hold if the operators $\bm{a}_n$
and $\bm{b}_j$ are required to satisfy
\begin{equation}
[\bm{a}_n,\bm{a}_m^\dagger] = \delta_{nm}\mathbb{I}
\qquad\mbox{ and }\qquad
[\bm{b}_j,\bm{b}_{j'}^\dagger] = \delta_{j j'}\mathbb{I},
\end{equation}
with all other commutators vanishing.

The vacuum state for the IN region, which we will denote by
 $|0_L\rangle$, satisfies
\begin{equation}
\bm{a}_n |0_L\rangle = 0 =\bm{b}_j|0_L\rangle\label{eq:defn_IN_vac_state}
\end{equation}
for all $n$ and $j$. One-particle states are created by acting on the
vacuum state with the creation operators $\bm{a}_n^\dagger$ and
$\bm{b}_j^\dagger$, i.e.,
\begin{equation}
|1_{L,n}\rangle = \bm{a}_n^\dagger|0_L\rangle \qquad\mbox{ and }\qquad
|1_{L,j}\rangle = \bm{b}_j^\dagger|0_L\rangle.
\end{equation}
One can construct higher number particle states by repeated
action of the creation operators.

The positive-frequency Wightman's function is the vacuum expectation value of the
point-split field-squared operator,
\begin{eqnarray}
G^+(x,t;x',t') &=&\langle 0_L |\bm{\Phi}(x,t)\bm{\Phi}(x',t')| 0_L \rangle\nonumber\\
&=&\frac{1}{2}\sum_{n=1}^\infty k_n^{-1} u^{\rm odd}(n,x) \overline{u^{\rm odd}(n,x')}
e^{-i k_n(t-t')}+ \sum_{j=1}^\infty \Phi^{\rm even}(j,x,t)\overline{\Phi^{\rm even}(j,x',t')}
\end{eqnarray}
The form of the Wightman's function varies depending on the time coordinates,
i.e., if $t$ and $t'$ are on the IN or OUT regions of the spacetime.
In particular, for the IN region, the Wightman's function has the form
\begin{equation}
G^+_{\rm IN}(x,t;x',t') = \frac{1}{2}\left[\sum_{n=1}^\infty \frac{1}{k_n}
u^{\rm odd}(n,x)\overline{u^{\rm odd}(n,x')}e^{-i k_n(t-t')}
+\sum_{j=1}^\infty\frac{1}{\kappa_j}u^{\rm even}(j,x)
\overline{u^{\rm even}(j,x')}e^{-i \kappa_j(t-t')}\right].
\label{eq:IN_Wightman}
\end{equation}
The form for the Wightman function for the OUT region will be given after
the definition of the Bogolobuv coefficients below. (See Eq.~(\ref{eq:G+out}) for
the explicit form.)

\subsection{Unitarily Equivalent Representation of the Field Operator for the OUT Region}

For the OUT region of the spacetime, we have seen above that there is a second
complete set of orthormal modes solutions to the wave equation given in terms of
the odd modes Eq.~(\ref{eq:psi_odd_out}), the even modes Eq.~(\ref{eq:psi_even_out}),
and the topological modes Eq.~(\ref{eq:psi_top_out}). As in the preceding subsection,
we can promote Eq.~(\ref{eq:fourier}) to a real-valued, self-adjoint field
operator, with
\begin{equation}
\bm{\psi}(x,t) = \widetilde{\bm{a}}\, \psi^{\rm top.}(x,t) +  \widetilde{\bm{a}}^\dagger\,
\overline{\psi^{\rm top.}(x,t)} + \sum_{n=1}^\infty \left[\widetilde{\bm{a}}_n\,
\psi^{\rm odd}(n,x,t) + \widetilde{\bm{a}}_n^\dagger\,\overline{\psi^{\rm odd}(n,x,t)} +
\widetilde{\bm{b}}_n\, \psi^{\rm even}(n,x,t) + \widetilde{\bm{b}}_n^\dagger\,
\overline{\psi^{\rm even}(n,x,t)}\right],
\label{eq:RS1_field_operator}
\end{equation}
where we assume the commutation relations \cite{F&Pa89}
\begin{equation}
[ \widetilde{\bm{a}},\widetilde{\bm{a}}^\dagger ] =\mathbb{I} \qquad\mbox{ and }\qquad
[\widetilde{\bm{a}}_n,\widetilde{\bm{a}}_m^\dagger ] = \delta_{nm}\mathbb{I}=
[\widetilde{\bm{b}}_n,\widetilde{\bm{b}}_m^\dagger ],
\end{equation}
with all other commutators vanishing.

It is straightforward to show that this yields
the correct equal-time commutation relations for the field operator and its
conjugate momenta $\bm{\pi}(x,t)\equiv\partial_t\bm{\psi}(x,t)$. Substituting,
we have
\begin{eqnarray}
\left[ \bm{\psi}(x,t),\bm{\pi}(x',t)\right] &=&
i\left[v^{\rm top.}(x) v^{\rm top.}(x') +  \sum_{n=1}^\infty \left( v^{\rm odd}(x)
v^{\rm odd}(x') +v^{\rm even}(x)v^{\rm even}(x')\right)\right].
\end{eqnarray}
By Eq.~(\ref{eq:delta_fourier}) above, this expression reduces to the standard
$\left[ \bm{\psi}(x,t),\bm{\pi}(x',t)\right] = i \delta(x-x')\mathbb{I}.$
It is also straightforward to demonstrate that
\begin{equation}
\left[\bm{\psi}(x,t),\bm{\psi}(x',t)
\right] = \left[ \bm{\pi}(x,t),\bm{\pi}(x',t)\right] = 0 \qquad\mbox{ and }\qquad
\left[ \bm{\psi}(x,t), \bm{\psi}(x',t') \right] = i  E(x,t;x',t')\, \mathbb{I},
\end{equation}
where $E(x,t;x',t')$ is the advanced-minus-retarded two point function on
$\reals\times S^1$ constructed in Appendix~\ref{sec:adv-ret Green}.

The Hilbert space on which these operators act is given by the conventional Fock
space used in QFT; the ground state with respect to this
field operator is $|\tilde{0}_L\rangle$, such that
\begin{equation}
\widetilde{\bm{a}} |\tilde{0}_L\rangle = 0 \qquad\mbox{ and }\qquad
\widetilde{\bm{a}}_n |\tilde{0}_L\rangle = 0 =\widetilde{\bm{b}}_n |\tilde{0}_L\rangle\qquad\forall\, n.
\end{equation}
The positive-frequency Wightman function  is quickly
found to be
\begin{eqnarray}
\tilde{G}^+(x,t;x',t') &=& \langle \tilde{0}_L | \bm{\psi}(x,t) \bm{\psi}(x',t') |
\tilde{0}_L \rangle \nonumber\\
&=& \frac{\ell}{2L}\left( 1-i\frac{t}{\ell}\right)\left(1+i\frac{t'}{\ell}\right) +\frac{1}{L}
\sum_{n=1}^\infty k_n^{-1} \cos[k_n(x-x')] e^{-i k_n (t-t')}
\label{eq:out wightman}\\
&=& \frac{\ell}{2L}\left( 1-i\frac{t}{\ell}\right)\left(1+i\frac{t'}{\ell}\right) -
\frac{1}{4\pi}\ln\left\{ \left[1-e^{-i2\pi(\Delta t - \Delta x)/L}\right]
\left[1-e^{-i2\pi(\Delta t + \Delta x)/L}\right]\right\}
\label{eq:out wightman_Kay}
\end{eqnarray}
where $\Delta x = x -x'$ and $\Delta t = t-t'$.

One final note before we leave this section.  With the specification and properties
of the Bogolubov coefficients below, it is a straightforward exercise to
check that for the OUT region $\langle\widetilde{0}_L|\bm{\Phi}(x,t)\bm{\Phi}(x',t')%
|\widetilde{0}_L\rangle = \widetilde{G}^+(x,t;x',t')$, as expected.

\subsection{Bogolubov Transform and Particle Creation}

For the OUT region, we have two representations for the field operator, one given in terms of
the mode solutions on the whole spacetime, Eq.(\ref{eq:standard_field_operator}), and
one given by the standard modes on $\reals\times S^1$, Eq.~(\ref{eq:RS1_field_operator}).
It is immediately obvious that the odd modes solutions are common to both representations,
i.e., $\phi^{\rm odd}(n,x,t) = \psi^{\rm odd}(n,x,t)$, therefore
$\widetilde{\bm{a}}_n = \bm{a}_n$ and $\widetilde{\bm{a}}_n^\dagger = \bm{a}_n^\dagger$,
In keeping with the notation of Birrell and Davies\cite{Brl&Dv}, one can simply read the
remaining Bogolubov coefficients from Eq.~(\ref{eq:psi_odd_mode_decomposition}).
We have
\begin{eqnarray}
\overline{\alpha_{0j}} &=& a\\
\beta_{0j} &=& -b\\
\overline{\alpha_{nj}} &=& c_n\\
\beta_{nj} &=& -d_n
\end{eqnarray}
Therefore, on the OUT region, it is possible to express the $\reals\times S^1$
annihilation and creation operators in terms of the annihilation and creation operators
on the whole spacetime, i.e,
\begin{equation}
\widetilde{\bm{a}} = \sum_{j=1}^\infty (\overline{\alpha_{0j}}\,\bm{b}_j-\overline{\beta_{0j}}\,
\bm{b}_j^\dagger )
\qquad\mbox{ and }\qquad
\widetilde{\bm{b}}_n = \sum_{j=1}^\infty (\overline{\alpha_{nj}}\,\bm{b}_j-\overline{\beta_{nj}}\,
\bm{b}_j^\dagger ).
\end{equation}
If the quantum state of the system is initially in the IN vacuum state,
$|0_L\rangle$, then observers in the OUT region will observe the creation
of field quanta with an expectation value per mode given by
\begin{equation}
\langle 0_L | \widetilde{\bm{N}}_0 | 0_L \rangle =
\langle 0_L | \widetilde{\bm{a}}^\dagger \widetilde{\bm{a}} | 0_L \rangle=
\sum_{j=1}^\infty \left|\beta_{0j}\right|^2
\end{equation}
for the topological modes, and
\begin{equation}
\langle 0_L | \widetilde{\bm{N}}_n | 0_L \rangle =
\langle 0_L | \widetilde{\bm{b}}_n^\dagger \widetilde{\bm{b}}_n | 0_L \rangle=
\sum_{j=1}^\infty \left|\beta_{nj}\right|^2
\end{equation}
for the even modes.  No quanta are created in the odd modes. By definition,
the number of quanta created is a strictly positive quantity.
Substituting the expressions for the Bogolubov coefficients and using
Eq.~(\ref{eq:y_orthog}), we find
\begin{eqnarray}
\langle 0_L | \widetilde{\bm{N}}_0 | 0_L \rangle &=& -\frac{1}{2} + \frac{1}{4}
\sum_{j=1}^\infty \left( \frac{1}{\kappa_j\ell}+\kappa_j\ell\right)Y_{j,0}^2
\nonumber\\
&=&
-\frac{1}{2}+\left(\frac{\xi L}{2}\right)^2\left(\frac{\ell}{L}\right)
\sum_{j=1}^{\infty} \left[ Z_j^2 + \left(\frac{L}{2\ell}\right)^2\right]
\frac{A_j^2}{Z_j^5}\nonumber\\
&=& -\frac{1}{2} + \frac{\ell}{L}F_3(\chi)+
\frac{1}{4} \frac{L}{\ell} F_5(\chi),
\label{eq:number_topological_excited}
\end{eqnarray}
where  the function $F_n(x)$ is defined in Appendix~\ref{sec:series}, and
\begin{equation}
\langle 0_L | \widetilde{\bm{N}}_n | 0_L \rangle =-\frac{1}{2} + \frac{1}{4}
\sum_{j=1}^\infty \left( \frac{k_n}{\kappa_j}+\frac{\kappa_j}{k_n}\right)Y_{j,n}^2
=
-\frac{1}{2}+\left(\frac{\xi L}{2}\right)^2\frac{1}{\pi n}\sum_{j=1}^{\infty}
\frac{Z_j^2+(\pi n)^2}{\left[ Z_j^2-(\pi n)^2\right]^2}\frac{A_j^2}{Z_j}.
\label{eq:number_oscillatory_excited}
\end{equation}
Both $\langle 0_L | \widetilde{\bm{N}}_0 | 0_L\rangle$ and $\langle 0_L | \widetilde{\bm{N}}_n
 | 0_L\rangle$ are absolutely convergent.  Furthermore, the sum formed from
the upper bound of these two sums is also absolutely convergent.  Therefore,
the total number of particles created at the shutting off of the potential is finite.
Numerical values found using Mathematica for the first ten coefficients are
presented in Table~\ref{mytable}. The dominant pathway for particle creation
is into the topological mode.

\renewcommand{\arraystretch}{1.4}
\setlength{\tabcolsep}{10pt}
\begin{table}
\caption{\label{mytable} Expectation value for the number of quanta
excited per mode when the potential is turned off for various values
of the coupling constant $\xi$.  The values were generated using $L=1$
and summing the first 500 terms in the series using Mathematica. The
$n=0$ values were determined with $\ell = L$.}

\begin{tabular}{|c|c|c|c|c|}\hline
&\multicolumn{4}{|c|}{$\displaystyle \langle 0_L | \widetilde{\bm{N}}_n | 0_L \rangle$}\\
\hline
$n$ & $\xi=1$ & $\xi=5$ & $\xi=10$ & $\xi=100$ \\ \hline
0 & 0.023987 & 0.255469 & 0.416834 & 1.082297\\
1 & 0.003875 & 0.024742 & 0.047086 & 0.198755\\
2 & 0.000665 & 0.005465 & 0.011781 & 0.070152\\
3 & 0.000231 & 0.002154 & 0.004975 & 0.036841\\
4 & 0.000108 & 0.001091 & 0.002639 & 0.022904\\
5 & 0.000059 & 0.000637 & 0.001594 & 0.015659\\
6 & 0.000036 & 0.000408 & 0.001048 & 0.011386\\
7 & 0.000024 & 0.000277 & 0.000731 & 0.008647\\
8 & 0.000017 & 0.000200 & 0.000533 & 0.006782\\
9 & 0.000012 & 0.000149 & 0.000402 & 0.005455\\
10 & 0.000009 & 0.000114 & 0.000312 & 0.004477\\\hline
\end{tabular}
\end{table}

With the definition of the Bogolubov coefficients completed, we now give
the expression for the Wightman's function of the IN ground state on
the OUT region of the spacetime. Making use of the series expansion of
$\phi^{\rm even}_{\rm OUT}(j,x,t)$ in terms of the conventional modes on the OUT
region, i.e.,
\begin{equation}
\phi^{\rm even}_{\rm OUT}(j,x,t) = \sum_{n=0}^\infty \left[ \overline{\alpha_{nj}}
\psi^{\rm even}(n,x,t) - \beta_{nj}\overline{\psi^{\rm even}(n,x,t)}\right],
\end{equation}
we have
\begin{eqnarray}
G^+_{\rm OUT}(x,t;x',t') &=& \sum_{n=1}^\infty
\phi^{\rm odd}(n,x,t)\overline{\phi^{\rm odd}(n,x',t')}\nonumber\\
&&+ \sum_{j=1}^\infty
\sum_{n=0}^\infty \sum_{m=0}^\infty \left[ \overline{\alpha_{nj}}\, \alpha_{mj}\,
\psi^{\rm even}(n,x,t) \overline{\psi^{\rm even}(m,x',t')}
-\overline{\alpha_{nj}}\,\overline{\beta_{mj}}\, \psi^{\rm even}(n,x,t)
\psi^{\rm even}(m,x',t')\right.\nonumber\\
&&\left.-\beta_{nj}\, \alpha_{mj}\,\overline{\psi^{\rm even}(n,x,t)}
\overline{\psi^{\rm even}(m,x',t')}+\beta_{nj}\,\overline{\beta_{mj}}\,
\overline{\psi^{\rm even}(n,x,t)} \psi^{\rm even}(m,x',t')\right].
\end{eqnarray}
Swapping the order of the $j$-summation with the $n$ and $m$ summations
and using the properties of the Bogolubov coefficients,
\begin{equation}
\sum_{j=1}^\infty \left( \alpha_{mj}\, \overline{\alpha_{nj}} - \beta_{mj}\,
\overline{\beta_{nj}}\right) = \delta_{nm} \qquad\mbox{ and }\qquad
\sum_{j=1}^\infty \left( \alpha_{mj}\, \beta_{nj} - \beta_{mj}\,
\alpha_{nj}\right) = 0,
\label{eq:Bogolubov}
\end{equation}
we can simplify the above expression to
\begin{eqnarray}
G^+_{\rm OUT}(x,t;x',t') &=& \sum_{n=1}^\infty
\psi^{\rm odd}(n,x,t)\overline{\psi^{\rm odd}(n,x',t')}+
\sum_{n=0}^\infty
\psi^{\rm even}(n,x,t)\overline{\psi^{\rm even}(n,x',t')}\nonumber\\
&&+ 2\Re \left\{\sum_{n=0}^\infty
\sum_{m=0}^\infty \left[ \left(\sum_{j=1}^\infty \overline{\beta_{nj}}\, \beta_{mj}\right)
\psi^{\rm even}(n,x,t) \overline{\psi^{\rm even}(m,x',t')}\right.\right.
\nonumber\\
&&-\left.\left.\left(\sum_{j=1}^\infty\overline{\alpha_{nj}}\,\overline{\beta_{mj}}\right)
\psi^{\rm even}(n,x,t) \psi^{\rm even}(m,x',t')\right]\right\}.
\end{eqnarray}
However, the first two summations are the definition of the positive-frequency
Wightman function for the OUT ground state given by Eq.~(\ref{eq:out wightman_Kay}),
thus
\begin{eqnarray}
G^+_{\rm OUT}(x,t;x',t')-\widetilde{G}^+(x,t;x',t') &=&
2\Re \left\{\sum_{n=0}^\infty
\sum_{m=0}^\infty \left[ \left(\sum_{j=1}^\infty \overline{\beta_{nj}}\, \beta_{mj}\right)
\psi^{\rm even}(n,x,t) \overline{\psi^{\rm even}(m,x',t')}\right.\right.
\nonumber\\
&&-\left.\left.\left(\sum_{j=1}^\infty\overline{\alpha_{nj}}\,\overline{\beta_{mj}}\right)
\psi^{\rm even}(n,x,t) \psi^{\rm even}(m,x',t')\right]\right\}.\label{eq:G+out}
\end{eqnarray}
We will determine the renormalized expectation value of the stress-energy tensor in
the OUT region using this expression in the next section. Two final notes: first,
both of the summations over $j$ in the above expression are absolutely convergent, and
second, on the domain $\mathcal{D}\cup\{(0,0)\}$ the IN and OUT Wightman functions
can be used interchangeably, i.e.,
\begin{equation}
G^+_{\rm OUT}(x,t;x',t') = G^+_{\rm IN}(x,t;x',t').
\end{equation}
This second property follows from the mode solutions being equal on the domain
$\mathcal{D}\cup\{(0,0)\}$.

\section{Stress-Energy Tensor}\label{sec:Stress-Tensor}

With the second quantization of the field now completed, we address the
expectation value of the stress-tensor.  The classical stress-tensor,
Eq.~(\ref{eq:classical stress-tensor}), is promoted to the self-adjoint
operator
\begin{equation}
\bm{T}_{\mu\nu} = \frac{1}{2}\left\{(\partial_\mu \bm{\Phi})(\partial_\nu\bm{\Phi})+
(\partial_\nu \bm{\Phi})(\partial_\mu\bm{\Phi})-
g_{\mu\nu}\left[g^{\alpha\beta} (\partial_\alpha \bm{\Phi})(\partial_\beta\bm{\Phi})
-V \bm{\Phi}\bm{\Phi}\right]\right\}.
\label{eq:Tmunu_quantum}
\end{equation}
For any normalized state $|\chi_L\rangle$ in the Fock space, it
is well know that the expectation value of the stress-tensor is divergent.
For free fields in Minkowski spacetime, the divergences are removed by the normal
ordering process, but in curved spacetimes and flat spacetimes of non-trivial
topology, we are required to employ renormalization to obtain finite
results.  For a quantum field interacting with a potential, as here, further
local renormalization counterterms are required which are dependent upon the
potential.  Mamev and Trunov discuss this in their paper and the references
therein \cite{Ma&Tr82}. Further work has been carried out by others, including
Graham, Jaffe, and colleagues \cite{Gr&Ja02, Gr&Ol03a}, working primarily
in Minkowski spacetime.

This leaves us in an awkward position. Progress has been made on the two
fronts, but we are unaware of both renormalizations being combined to fully
treat the problem at hand. To do so here would be beyond the intent of this
paper, so we follow the path of Mamev and Trunov who calculate the renormalized
stress-tensor in regions of the spacetime where the potential is zero.  For
such localized potentials, the potential-dependent counterterms are not
necessary outside of the support of the potential.  This is the same
path that Solomon takes, which gives rise to his notion of the kinetic-tensor.
Thus, outside the support of the potential, we define
\begin{equation}
\langle\chi_L| \bm{T}_{\mu\nu} |\chi_L\rangle_{Ren.} \equiv \langle\chi_L| \bm{T}_{\mu\nu} |\chi_L\rangle
-\langle0| \bm{T}_{\mu\nu} |0\rangle
\end{equation}
where $|0\rangle$ is the Minkowski vacuum state.  The rigorous mathematical interpretation
of this renormalization scheme is discussed by Kay \cite{Kay79}.

The renormalized expectation value of the stress-tensor for the OUT
 vacuum state on the OUT region of the spacetime is identical to the
determination of the Casimir effect in the standard $\reals\times S^1$
spacetime that is found in literature. (For example, see Chap. 4 of
Birrell and Davies, or Kay \cite{Kay79} and the references therein.)
With the inclusion of the topological modes \cite{F&Pa89}, we have
the simple expression
\begin{equation}
\langle \widetilde{0}_L | \bm{T}_{\mu\nu} | \widetilde{0}_L \rangle_{\rm Ren.}
= \left(\frac{1}{4\ell L}-\frac{\pi}{6L^2} \right)
\delta_{\mu\nu}.
\end{equation}
Notice that the topological modes only add a positive-constant term to the
renormalized stress-tensor. The additional term is dependent upon the arbitrary
constant $\ell$.

For the calculation to follow below, we define normal ordering
of the unrenormalized stress-tensor in any allowable state $|\chi_L\rangle$,
with respect to any other allowable state $|\rho_L\rangle$, as
\begin{equation}
\langle \chi_L| : \bm{T}_{\mu\nu} :_{\rho_L} |\chi_L \rangle
=\langle \chi_L| \bm{T}_{\mu\nu} |\chi_L \rangle
-\langle \rho_L| \bm{T}_{\mu\nu} |\rho_L \rangle.
\end{equation}
It is computationally useful to combine this with the renormalization
scheme defined above, yielding
\begin{equation}
\langle\chi_L| \bm{T}_{\mu\nu} |\chi_L\rangle_{Ren.} =
\langle\chi_L| :\bm{T}_{\mu\nu}:_{\rho_L} |\chi_L\rangle
+\langle\rho_L| \bm{T}_{\mu\nu} |\rho_L\rangle_{Ren.}.
\end{equation}
The remainder of this section is dedicated to determining expressions for each
of the terms above when $|\chi_L\rangle = |0_L\rangle$ and $|\rho_L\rangle
=|\tilde{0}_L\rangle$.
Because the mode decompositon of the field changes at the $t=0$ Cauchy surface, the
expression for the stress-tensor can be written as
\begin{equation}
\langle 0_L| \bm{T}_{\mu\nu} | 0_L\rangle_{Ren.} = \left\{\begin{array}{ll}
\langle 0_L| :\bm{T}_{\mu\nu}^{IN}:_{\widetilde{0}_L} | 0_L\rangle +
\langle \widetilde{0}_L|\bm{T}_{\mu\nu}|\widetilde{0}_L \rangle_{Ren.} &
\mbox{for } t\leq0,\\
\langle 0_L| :\bm{T}_{\mu\nu}^{OUT}:_{\widetilde{0}_L} | 0_L\rangle +
\langle \widetilde{0}_L|\bm{T}_{\mu\nu}|\widetilde{0}_L \rangle_{Ren.} &
\mbox{for } t\geq0.
\end{array}\right.
\end{equation}

\subsection{Renormalized Stress-Tensor for $|0_L\rangle$ on the OUT Region}

For the OUT region, we can make progress toward an explicit expression if we first
look at the ingoing ground state's normal-ordered, point-split, field-squared operator,
\begin{equation}
\Delta G^+(x,t;x',t')=\langle 0_L | :\bm{\Phi}(x) \bm{\Phi}(x'):_{\tilde{0}_L} | 0_L \rangle =
\langle 0_L | \bm{\Phi}(x) \bm{\Phi}(x') | 0_L \rangle
-\langle \widetilde{0}_L | \bm{\Phi}(x) \bm{\Phi}(x') | \widetilde{0}_L \rangle.
\end{equation}
However, the right-hand side is the difference of the positive-frequency
Wightman functions we determined above in Eq.~(\ref{eq:G+out}).
Recall, the $n=0$ topological mode is unique from the rest of the even modes,
so we expand the products out;
\begin{eqnarray}
\Delta G^+(x,t;x',t')
%
%
&=&2\Re \left\{\left[\left(\sum_{j=1}^\infty |\beta_{0j}|^2\right)
\psi^{\rm top.}(x,t) \overline{\psi^{\rm top.}(x',t')}
-\left(\sum_{j=1}^\infty\overline{\alpha_{0j}}\,\overline{\beta_{0j}}\right)
\psi^{\rm top.}(x,t) \psi^{\rm top.}(x',t')\right]\right.\nonumber\\
%
%
&& +\sum_{m=1}^\infty \left[ \left(\sum_{j=1}^\infty \overline{\beta_{0j}}\,\beta_{mj}
\right)\psi^{\rm top.}(x,t) \overline{\psi^{\rm even}(m,x',t')}
-\left(\sum_{j=1}^\infty\overline{\alpha_{0j}}\,\overline{\beta_{mj}}\right)
\psi^{\rm top.}(x,t) \psi^{\rm even}(m,x',t')\right]\nonumber\\
%
%
&& + \sum_{n=1}^\infty
\left[ \left(\sum_{j=1}^\infty \overline{\beta_{nj}}\, \beta_{0j}\right)
\psi^{\rm even}(n,x,t) \overline{\psi^{\rm top.}(x',t')}
-\left(\sum_{j=1}^\infty\overline{\alpha_{nj}}\,\overline{\beta_{0j}}\right)
\psi^{\rm even}(n,x,t) \psi^{\rm top.}(x',t')\right]\nonumber\\
%
%
&& + \left.\sum_{n=1}^\infty
\sum_{m=1}^\infty \left[ \left(\sum_{j=1}^\infty \overline{\beta_{nj}}\, \beta_{mj}\right)
\psi^{\rm even}(n,x,t) \overline{\psi^{\rm even}(m,x',t')}\right.\right.
\nonumber\\
&&-\left.\left.\left(\sum_{j=1}^\infty\overline{\alpha_{nj}}\,\overline{\beta_{mj}}\right)
\psi^{\rm even}(n,x,t) \psi^{\rm even}(m,x',t')\right]\right\}.\nonumber\\
\end{eqnarray}
The expectation value of the normal-ordered energy-density for the IN ground
state on the OUT region can be found from this expression by
\begin{equation}
\langle 0_L | :\bm{T}_{tt}: | 0_L \rangle = \frac{1}{2}\lim_{(t',x')\rightarrow(t,x)
} \left(\partial_t \partial_{t'} + \partial_x \partial_{x'}\right)
\Delta G^+(x,t;x',t').
\end{equation}
Evaluating the derivatives and taking the limit as the spacetime points come together yields
\begin{eqnarray}
\langle 0_L|: \bm{T}_{tt}: | 0_L \rangle &=& \frac{1}{L}\left\{\sum_{n=1}^\infty k_n \sum_{j=1}^\infty
|\beta_{nj}|^2 + \frac{1}{2\ell} \sum_{j=1}^\infty \left[ |\beta_{0j}|^2 +\Re (\alpha_{0j} \beta_{0j})
\right] \right\}+\Re \left[ \rho(t-x) + \rho(t+x) \right]
\label{eq:Energy_dens}\\
&=& - \frac{1}{4\ell L}+\frac{\mathcal{B}}{L^2}+\Re \left[ \rho(t-x) + \rho(t+x) \right],
\end{eqnarray}
where we have made use of the properties of the Bogolubov coefficients to simplify the
summation,
\begin{equation}
\sum_{j=1}^\infty \left[ |\beta_{0j}|^2 + \Re(\alpha_{0j}\beta_{0j})\right] =-\frac{1}{2} +\frac{1}{2}
\sum_{j=1}^\infty \left| \beta_{0j} + \overline{\alpha_{0j}}\right|^2,
\end{equation}
in order to define the positive constant
\begin{equation}
\mathcal{B} \equiv L \left(\sum_{n=1}^\infty k_n \sum_{j=1}^\infty
|\beta_{nj}|^2 +\frac{1}{4\ell} \sum_{j=1}^\infty |\beta_{0j}+\overline{\alpha_{0j}}|^2\right).
\end{equation}
Upon substitution of the expressions for the Bogolubov coefficients, we find
\begin{eqnarray}
\mathcal{B} &=& \frac{(\xi L)^2}{2}\sum_{n=1}^\infty \sum_{j=1}^\infty\frac{A_j^2}{Z_j(Z_j+\pi n)^2}
+\frac{(\xi L)^2}{4}\sum_{j=1}^\infty \frac{A_j^2}{Z_j^3}
\nonumber\\
&=&\frac{(\xi L)^2}{2\pi^2}\sum_{j=1}^{\infty}\frac{A_j^2}{Z_j}\left[\psi^{(1)}
\left(1+\frac{Z_j}{\pi}\right)+\frac{\pi^2}{2Z_j^2}
\right],
\label{eq_B_defn}
\end{eqnarray}
where $\psi^{(1)}(x)$ is the polygamma function of order one (p.\ 260 of \cite{Abramowitz}).
$\psi^{(1)}(x)$ is a positive, strictly decreasing function on the interval
$x\in(0,\infty)$, with a pole of order-2 at $x=0$. (We have no interest in the polygamma
function for values of $x<1$.)  Also, $\psi^{(1)}(1) = \zeta(2) = \pi^2/6$, thus
\begin{equation}
\pi^2/6 \geq \psi^{(1)}\left(1+\frac{z}{\pi}\right) > 0
\end{equation}
on the interval $z\in[0,\infty).$  One remarkable fact to note is that the constant
$\mathcal{B}$ is independent of $\ell$.

The coordinate-dependent function $\rho$ is given by
\begin{eqnarray}
\rho(z) &\equiv& \frac{1}{2L\sqrt{2\ell}} \sum_{n=1}^\infty k_{n}^{1/2} \left[ e^{-i k_n z}
\sum_{j=1}^{\infty}(\overline{\beta_{nj}}\beta_{0j} + \overline{\alpha_{nj}}\overline{\beta_{0j}}
+ \overline{\alpha_{0j}}\overline{\beta_{nj}}) + e^{i k_n z}\sum_{j=1}^\infty\overline{\beta_{0j}}
\beta_{nj}\right]\nonumber\\
&&+\frac{1}{2L} \sum_{n=1}^\infty \sum_{m=1}^\infty (k_n k_m)^{1/2} \left[ (1-\delta_{nm})e^{-i(k_n-k_m)z}
\sum_{j=1}^\infty \overline{\beta_{nj}} \beta_{mj} + e^{-i(k_n+k_m) z} \sum_{j=1}^\infty
\overline{\alpha_{nj}}\overline{\beta_{mj}} \right].
\label{eq:rho_defn}
\end{eqnarray}
For this particular case, the Bogolubov coefficients are all real valued and we find
\begin{eqnarray}
\Re\rho(z) &=& \frac{1}{2L\sqrt{2\ell}} \sum_{n=1}^\infty k_{n}^{1/2} \cos( k_n z)
\sum_{j=1}^{\infty}(2\beta_{nj}\beta_{0j} + \alpha_{nj}\beta_{0j} + \alpha_{0j}\beta_{nj})
\nonumber\\
&&+\frac{1}{2L} \sum_{n=1}^\infty \sum_{m=1}^\infty (k_n k_m)^{1/2} \left\{ (1-\delta_{nm}) \cos\left[(k_n-k_m)z\right]
\sum_{j=1}^\infty \beta_{nj}\beta_{mj} + \cos\left[(k_n+k_m) z\right] \sum_{j=1}^\infty
\alpha_{nj}\beta_{mj} \right\}.
\end{eqnarray}
The double summations of $n$ and $m$ over the above range can be reorganized to
simplify our expression.

Upon substitution for the Bogolubov coefficients, we have
\begin{eqnarray}
\Re \rho(z) &=& \frac{1}{2L} \left\{  \sum_{n=1}^\infty \cos(k_n z)\left[
\frac{1}{2\sqrt{2}} \sum_{j=1}^\infty \left(2\kappa_j-k_n -\frac{1}{\ell}
\right)Y_{j,0}Y_{j,n}\right]\right.\nonumber\\
&&+\sum_{n=1}^\infty \cos(k_n z) \left[\frac{1}{2} \sum_{m=1}^\infty
\sum_{j=1}^\infty \frac{1}{\kappa_j}(\kappa_j-k_m)(\kappa_j-k_{n+m})Y_{j,m}Y_{j,n+m}
\right]\nonumber\\
&&+\left.\sum_{n=2}^\infty \cos(k_n z) \left[ \frac{1}{4}\sum_{m=1}^{n-1}
\sum_{j=1}^\infty \frac{1}{\kappa_j}(\kappa_j+k_m)(\kappa_j-k_{n-m})
Y_{j,m} Y_{j,n-m}\right]\right\}.
\label{eq:Cdefn}
\end{eqnarray}
The first $j$-summation  may be simplified by using the orthogonality relation,
Eq.~(\ref{eq:y_orthog}), which eliminates two of the terms;
\begin{equation}
\sum_{j=1}^\infty \left(2\kappa_j-k_n -\frac{1}{\ell}\right)Y_{j,0}Y_{j,n}=
2\sum_{j=1}^\infty \kappa_j Y_{j,0}Y_{j,n}
-\left(k_n +\frac{1}{\ell}\right)\sum_{j=1}^\infty Y_{j,0}Y_{j,n}=
2\sum_{j=1}^\infty \kappa_j Y_{j,0}Y_{j,n}.
\end{equation}
Making the further substitutions to obtain an expression in terms of $Z_j$,
we find
\begin{eqnarray}
\Re \rho(z) &=& \frac{\xi^2}{2} \left\{  \sum_{n=1}^\infty \cos(k_n z)\left[
\sum_{j=1}^\infty \frac{A_j^2}{Z_j(Z_j^2-(\pi n)^2)}\right]\right.\nonumber\\
&&+\sum_{n=1}^\infty \cos(k_n z) \left[ \sum_{m=1}^\infty \sum_{j=1}^\infty
\frac{A_j^2}{Z_j(Z_j+\pi m)(Z_j+\pi(n+m))}
\right]\nonumber\\
&&+\left.\sum_{n=2}^\infty \cos(k_n z) \left[ \frac{1}{2}\sum_{m=1}^{n-1}
\sum_{j=1}^\infty \frac{A_j^2}{Z_j(Z_j-\pi m)(Z_j+\pi(n-m))}
\right]\right\}.
\end{eqnarray}

Interchanging the order of the $m$ and $j$ summations in
the second and third terms gives
\begin{eqnarray}
\Re \rho(z) &=& \frac{\xi^2}{2} \left\{  \sum_{n=1}^\infty \cos(k_n z)\left[
\sum_{j=1}^\infty \frac{A_j^2}{Z_j(Z_j^2-(\pi n)^2)}\right]\right.\nonumber\\
&&+\sum_{n=1}^\infty \cos(k_n z) \left[  \sum_{j=1}^\infty \frac{A_j^2}{Z_j} \sum_{m=1}^\infty
\frac{1}{(Z_j+\pi m)(Z_j+\pi(n+m))}
\right]\nonumber\\
&&+\left.\sum_{n=2}^\infty \cos(k_n z) \left[ \frac{1}{2}\sum_{j=1}^\infty
\frac{A_j^2}{Z_j}\sum_{m=1}^{n-1}\frac{1}{(Z_j-\pi m)(Z_j+\pi(n-m))}
\right]\right\}.
\end{eqnarray}
Next, we use two facts:
\begin{equation}
\sum_{m=1}^{\infty}\frac{1}{(Z_j + \pi m)(Z_j+\pi(n+m))}=\frac{1}{\pi n}
\sum_{m=1}^{n}\frac{1}{Z_j+\pi m}
\end{equation}
and
\begin{equation}
\frac{1}{2}\sum_{m=1}^{n-1}\frac{1}{(Z_j - \pi m)(Z_j+\pi(n-m))}=\frac{1}{\pi n}
\sum_{m=1}^{n-1}\frac{\pi m}{Z_j^2-(\pi m)^2}.
\end{equation}
Upon substitution, one finds that it is possible to combine the three summations
into a single compact expression,
\begin{eqnarray}
\Re \rho(z) &=& \frac{1}{2} \sum_{n=1}^\infty \cos(k_n z)\left[ \frac{\xi^2}{\pi n}
\sum_{j=1}^\infty A_j^2 \sum_{m=1}^n \frac{1}{Z_j^2-(\pi m)^2}\right]\nonumber\\
&=&\frac{1}{2} \sum_{n=1}^\infty \cos(k_n z)\left[ \frac{\xi^2}{\pi n}
\sum_{m=1}^n \sum_{j=1}^\infty   \frac{A_j^2}{Z_j^2-(\pi m)^2}\right]
\label{Eq:almost_there}
\end{eqnarray}

Next, consider Eq.~(\ref{eq:y_orthog}) when $m\neq0$ and $n=0$.  Substituting the
definition of the $Y_{j,n}$'s, we immediately find
\begin{equation}
\sum_{j=1}^\infty \frac{A_j^2}{Z_j^2\left[Z_j^2-(\pi m)^2 \right]}=0.
\end{equation}
The summation itself is convergent, therefore we can alternatively
write the above as
\begin{equation}
\frac{1}{(\pi m)^2}\sum_{j=1}^\infty \left[ \frac{A_j^2}{Z_j^2-(\pi m)^2}-
\frac{A_j^2}{Z_j^2}\right]=0.
\end{equation}
Both of the above terms in the summation are individually convergent, therefore
we conclude that for all $m$
\begin{equation}
\sum_{j=1}^\infty\frac{A_j^2}{Z_j^2-(\pi m)^2}=
\sum_{j=1}^\infty\frac{A_j^2}{Z_j^2}.
\end{equation}

To simplify the expression for the moving part of the energy-density, we define
the positive constant
\begin{equation}
\mathcal{C} \equiv \frac{(\xi L)^2}{2\pi} \sum_{j=1}^{\infty}\frac{A_j^2}{Z_j^2}=
\frac{(\xi L)^2}{2\pi} \sum_{j=1}^{\infty}\frac{A_j^2}{Z_j^2-(\pi m)^2}.
\label{eq_C_defn}
\end{equation}
$\mathcal{C}$ is a dependent upon the product of the coupling constant $\xi$
and the size of the universe $L$, but it is independent of the
free parameter $\ell$. It is shown in Appendix~\ref{sec:series} that the
series of the form above for even powers of the $Z_j$ in the denominator
often result in a simple analytic expression in terms of the variable
$\chi=\xi L/2$, such that
\begin{equation}
\mathcal{C} = \frac{2}{\pi} F_2(\chi) = \frac{\xi L}{2\pi}.
\end{equation}
Substituting $\mathcal{C}$ into Eq.~(\ref{Eq:almost_there}) yields
\begin{eqnarray}
\Re \rho(z) =  \frac{\mathcal{C}}{L^2} \sum_{n=1}^\infty \cos(k_n z)
= \frac{\mathcal{C}}{2L^2} \left[\delta(z/L)-1\right],
\label{eq:subtlety}
\end{eqnarray}
where we have made use of the definition of the delta-function, Eq.~(\ref{eq:delta_fourier}).
Finally, this yields a remarkably simple expression for the normal-ordered energy-density of the `IN'-vacuum state in the `OUT'-region of the spacetime,
\begin{eqnarray}
\langle 0_L|: \bm{T}_{tt}:_{\tilde{0}_L} | 0_L \rangle =- \frac{1}{4\ell L}+ \frac{\mathcal{B - C}}{L^2} +\frac{\mathcal{C}}{2L^2}
\sum_{n=-\infty}^\infty\left[\delta\left(\frac{t-x}{L}-n\right) + \delta\left(\frac{t+x}{L}-n\right)\right].
\end{eqnarray}
Although not immediately obvious, the summation over $n$ is necessary in the above
expression to account for the spacial periodicity of the spacetime%
\footnote{The expression $\sum_{n=1}^\infty\cos(k_n(t\pm x))$ is inherently
periodic in the $t$ coordinate on $\reals\times S^1$. However, the expression
$\delta[(t\pm x)/L]-1$ is not periodic at all. While Eq.~(\ref{eq:subtlety})
is certainly correct on the circle, to lift it to the cylinder spacetime requires a
restoration of the time periodicity.  This can be accomplished in a number of ways,
for example, with the modulo operator or as an infinite series as given here.}.
A similar analysis yields the expectation value of the  remaining components of the
stress-tensor.  Combining leads to the complete expression for the
expectation value of the normal-ordered stress-tensor;
\begin{eqnarray}
\langle 0_L|:\bm{T}_{\mu\nu}:_{\tilde{0}_L}| 0_L \rangle &=& \left\{-\frac{1}{4\ell L}+\frac{\mathcal{B
- C}}{L^2} + \frac{\mathcal{C}}{2L^2}\sum_{n=-\infty}^\infty
\left[\delta\left(\frac{t+x}{L}-n\right) + \delta\left(\frac{t-x}{L}-n\right)\right]\right\}
\mathbb{I}\nonumber\\
 &&+ \frac{\mathcal{C}}{2L^2}\sum_{n=-\infty}^\infty\left[\delta\left(\frac{t+x}{L}-n\right)
- \delta\left(\frac{t-x}{L}-n\right)\right] \left( \begin{array}{cc}
0 & 1\\ 1 & 0 \end{array}\right).
\label{Eq.norm_ordered_stress_tensor}
\end{eqnarray}
Adding the Casmir energy for the OUT region of the spacetime
leads directly to the renormalized stress-tensor on the OUT region;
\begin{eqnarray}
\langle 0_L|\bm{T}_{\mu\nu}| 0_L \rangle_{Ren.} &=& \left\{-\frac{\pi}{6 L^2}+\frac{\mathcal{B
- C}}{L^2} + \frac{\mathcal{C}}{2L^2}\sum_{n=-\infty}^\infty
\left[\delta\left(\frac{t+x}{L}-n\right) + \delta\left(\frac{t-x}{L}-n\right)\right]\right\}
\mathbb{I}\nonumber\\
 &&+ \frac{\mathcal{C}}{2L^2}\sum_{n=-\infty}^\infty\left[\delta\left(\frac{t+x}{L}-n\right)
- \delta\left(\frac{t-x}{L}-n\right)\right] \left( \begin{array}{cc}
0 & 1\\ 1 & 0 \end{array}\right).
\label{Eq.ren_stress_tensor}
\end{eqnarray}
The first term in the stress-tensor is the ``standard'' expression for
the Casimir energy on the spacetime.  Further, since both positive constants
$B$ and $C$ are independent of $\ell$, the expectation value of the stress-tensor
is also independent of $\ell$. The trace of this stress-tensor vanishes.

Finally, the energy-density and pressure terms for
the OUT region depend on the the difference between the constants
\begin{equation}
\mathcal{B - C} = \frac{(\xi L)^2}{2\pi^2}\sum_{j=1}^{\infty}\frac{A_j^2}{Z_j}\left[
\psi^{(1)}\left(1+\frac{Z_j}{\pi}\right)+\frac{\pi^2}{2Z_j^2} - \frac{\pi}{Z_j}\right].
\label{eq:B-C}
\end{equation}
To show that this is positive, we begin by defining the function
\begin{equation}
f(y)\equiv \psi^{(1)}(1+y) +\frac{1}{2 y^2} - \frac{1}{y}
\end{equation}
over the domain $y\in[0,\infty)$. By the recurrence formula
for polygamma functions, $f(y)$ can also be written as
\begin{equation}
f(y)\equiv \psi^{(1)}(y) -\frac{1}{2 y^2} - \frac{1}{y}.
\end{equation}
We now use two facts:  the integral definition of the polygamma function,
\begin{equation}
\psi^{(1)}(y) = \int_0^\infty \frac{t e^{-yt}}{1-e^{-t}} dt,
\end{equation}
and the relation
\begin{equation}
\frac{1}{y^n} = \frac{1}{\Gamma(n)}\int_0^\infty t^{n-1}e^{-yt} dt.
\end{equation}
Substituting both into the definition of $f(y)$ yields
\begin{equation}
f(y) = \int_0^\infty e^{-yt}\left( \frac{t}{1-e^{-t}}-\frac{1}{2}t-1\right)dt
= \int_0^\infty \frac{e^{-yt}}{1-e^{-t}}\left[ \left(\frac{1}{2}t-1\right)+
\left(\frac{1}{2}t+1\right)e^{-t}\right]dt.
\label{eq:final integral}
\end{equation}
It is straightforward to see that the integrand is positive when $t>2$.  To
show that the integrand is always positive, it is sufficient to demonstrate
that
\begin{equation}
g(t)=  \left(\frac{1}{2}t-1\right)+\left(\frac{1}{2}t+1\right)e^{-t}
\end{equation}
is positive over the remainder of the domain of integration, i.e. $t\in[0,2]$.

The real-valued function $g(t)$ is continuous on $[0,2]$, thus, by the extreme value
theorem in calculus, we know that $g(t)$ achieves both a minimum and a maximum on
$[0,2]$.  The extrema can occur at either endpoint of the interval or at
critical points of the function.  For the endpoints, we have $g(0) = 0$ and $g(2)=2e^{-2}>0$.
Checking for critical points using the first derivative test, we need to determine
the root(s) of the equation
\begin{equation}
\frac{1}{2} - \frac{1}{2}(t+1)e^{-t}=0.
\end{equation}
Rearranging, we have $t+1 = e^t$. This is only satisfied at $t=0$, therefore we conclude
the minimum of $g(t)$ on the interval $[0,2]$ occurs at $t=0$. Combining this with the
straightforward positivity for $t>2$,  we can deduce that the integrand in
Eq.~(\ref{eq:final integral}) is always positive, thus implying that $f(y)\geq0$.
So far we can conclude that
\begin{equation}
\mathcal{B - C} \geq 0.
\end{equation}

To find an upper bound on our expression, we define the function
\begin{equation}
h(t)\equiv\frac{12+6t+t^2}{12-6t+t^2} \qquad\mbox{ for }t\geq0.
\end{equation}
A straightforward calculation shows that $h(0)=1$ and
\begin{equation}
h(t)-h'(t) = \frac{t^4}{(12-6t+t^2)^2} \geq 0.
\end{equation}
This is equivalent to the set
\begin{equation}
\ln h(0)=0 \qquad\mbox{ and }\qquad \left[\ln h(t) \right]'\leq 1.
\end{equation}
Integrating this equation with respect to $t$ over the range
of $0$ to $t$ yields
\begin{equation}
h(t)\leq e^t.
\end{equation}
Substituting the definition of $h(t)$ and rearranging the terms yields
\begin{equation}
\frac{t}{1-e^{-t}}-\frac{1}{2}t-1\leq \frac{1}{12}t^2.
\end{equation}
This is an upper bound on the integrand of Eq.~(\ref{eq:final integral}),
thus
\begin{equation}
f(y) \leq \frac{1}{12}\int_0^\infty e^{-yt}t^2 dt = \frac{1}{6 y^3}.
\end{equation}
When applied to Eq.~(\ref{eq:B-C}), we have
\begin{equation}
\mathcal{B}-\mathcal{C} \leq \frac{\pi}{3} \chi^2\sum_{j=1}^\infty \frac{A_j^2}{Z_j^4} = \frac{\pi}{3}
F_4(\chi).
\end{equation}
Consulting Appendix~\ref{sec:series}, we find $F_4(\chi) = 1/2$, therefore, we can
conclude that the difference between $\mathcal{B}$ and $\mathcal{C}$ always satisfies
\begin{equation}
0\leq \left(\mathcal{B}-\mathcal{C}\right) \leq \frac{\pi}{6}.
\end{equation}

\subsection{Renormalized Stress-Tensor for $|0_L\rangle$ on the IN Region}
\label{sec:Tren._IN_Region}

For a moment, let us consider the static cylinder spacetime $\reals\times S^1$
with the potential $V(x,t) = 2\xi\delta(x).$  From the time independence
of the potential, and the symmetry of the potential along the $x$-direction, we can
expect that the renormalized vacuum expectation value of the stress-energy tensor
to be time independent, and a symmetric function in the $x$-variable
\footnote{For a free field in a static spacetime, we would also expect that the
renormalized stress-tenor
is conserved. The addition of the background potential complicates the
conservation equation; before renormalizing we have
$
\nabla^\mu \langle 0_L | \bm{T}_{\mu\nu} |0_L \rangle(x,t) = \frac{1}{2}
\left(\nabla_\nu V(x)\right) \langle 0_L|\bm{\Phi}(x,t) \bm{\Phi}(x,t) | 0_L \rangle.
$
This equation gives insight into the comment by Mamev and Trunov \cite{Ma&Tr82},
``In the presence of an external field, in addition to normal ordering it is
necessary to carry out renormalizations which depend locally on the potential
and its derivatives.''  So, to enforce conservation of the renormalized
stress-tensor everywhere, we would need to have one or more renormazation
counterterms that would cancel the right-hand side of the above equation.},
 i.e.,
\begin{equation}
\langle 0_L | \bm{T}_{\mu\nu} | 0_L\rangle_{Ren.}(x,t) =
\langle 0_L | \bm{T}_{\mu\nu} | 0_L\rangle_{Ren.}(x) =
\langle 0_L | \bm{T}_{\mu\nu} | 0_L\rangle_{Ren.}(-x).
\end{equation}

If we now consider our time dependent potential, for times $t<0$ we expect the
renormalized expectation value of the IN-state stress-tensor to be of
the same functional form as above, i.e. a time-independent, symmetric function
in $x$. Additionally, on the open, bow-tie-shaped region $\mathcal{D}$,
causality enforces the condition
\begin{equation}
\langle 0_L | \bm{T}_{\mu\nu}^{\rm IN} | 0_L\rangle_{Ren.}(x) =
\langle 0_L | \bm{T}_{\mu\nu}^{\rm OUT} | 0_L\rangle_{Ren.}(x,t) =
\left( -\frac{\pi}{6 L^2} + \frac{\mathcal{B} - \mathcal{C}}{L^2}\right)
\delta_{\mu\nu}.
\label{Eq.ren_IN_stress_tensor}
\end{equation}
The above expression is missing the moving delta-function terms of
Eq.~(\ref{Eq.ren_stress_tensor}) because they
have support on the boundary of $\mathcal{D}$, i.e., on the future and past
lightcone of the origin, and not on $\mathcal{D}$ itself.
So on $\mathcal{D}$, the expectation value of the stress-tensor
for the IN vacuum state is a position-independent constant.  However, it
is a function of the parameter $\chi$.

Because of the static nature of the $t<0$ portion of the spacetime, we can then
extend the above expression back to $t=-\infty$ for all spacetime points
except those along the line $x=0$, where the delta-function potential exists.
In other words Eq.~(\ref{Eq.ren_IN_stress_tensor}) is the renormalized
expectation-value of the stress-tensor for the IN vacuum state on the IN region,
outside of the support of the potential.

We now confirm that the above analysis yields the correct expression by
deriving the renormalized stress-tensor directly. From Sec.~\ref{subsec:INquantization}
above, we already know the IN mode-solution representation of the
IN-region Wightman function, Eq.~(\ref{eq:IN_Wightman}). One can simply
substitute the explicit form of the eigenfunctions, Eqs.~(\ref{eq:antisym_IN})
and (\ref{eq:symm_IN}) and then proceed to take the appropriate derivatives
to calculate the unrenormalized stress-tensor.  The details of this approach
are given in Appendix~\ref{sec:A_derivation}.  Unfortunately, this method is fraught with the
technical difficulties of having to explicity determine the difference
between two divergent sums to obtain a renormalized answer.
Furthermore, this approach seems to yield only one of the two terms of the
Casimir energy on the IN-region.

Instead, we proceed by first substituting the Fourier expansion of the IN
region, even-parity eigenfunctions in terms of the OUT region eigenfunctions given by
Eq.~(\ref{eq:Fourier for Ueven}).  We then define the difference between the
Wightman functions on the IN region as
\begin{eqnarray}
\Delta G_{IN}^+(x,t;x',t') & \equiv & G_{IN}^+(x,t;x',t') - \widetilde{G}^+(x,t;x',t')
\nonumber\\
&=& -\frac{\ell}{2L}\left(1-i\frac{t}{\ell}\right)
\left( 1+i\frac{t'}{\ell}\right) - \frac{1}{2}\sum_{n=1}^\infty\frac{e^{-i k_n (t-t')}}{k_n}
v^{\rm even}(n,x) v^{\rm even}(n,x')\nonumber\\
&&+\frac{1}{2}\sum_{j=1}^\infty \frac{e^{-i\kappa_j(t-t')}}{\kappa_j}
\left\{\frac{Y_{j,0}^2}{L} + \frac{Y_{j,0}}{\sqrt{L}}\sum_{n=1}^\infty
Y_{j,n}\left[ v^{\rm even}(n,x) + v^{\rm even}(n,x')\right]\right.\nonumber\\
&& \left. +\sum_{n=1}^\infty \sum_{m=1}^\infty Y_{j,n} Y_{j,m} v^{\rm even}(n,x)
v^{\rm even}(n,x')\right\}
\end{eqnarray}
Next, to determine any of the normal-ordered components of the kinetic-tensor, we act with the
appropriate point-split derivative operator on the difference of the Wightmen's functions and
then take the limit as the spacetime points come together.  For example, the $tt$-component
of the kinetic-tensor is given by
\begin{eqnarray}
\langle 0_L | : \bm{K}_{tt} :_{\widetilde{0}_L} | 0_L\rangle &=& -\frac{1}{4\ell L}
-\frac{1}{2L} \sum_{n=1}^\infty k_n + \frac{1}{4L}\sum_{j=1}^\infty \kappa_j Y_{j,0}^2
+\frac{1}{2\sqrt{L}}\sum_{n=1}^\infty v^{\rm even}(n,x) \sum_{j=1}^\infty \kappa_j
Y_{j,0} Y_{j,n}\nonumber\\
&& +\frac{1}{4} \sum_{j=1}^\infty \frac{1}{\kappa_j} \sum_{n=1}^\infty
\sum_{m=1}^\infty Y_{j,n} Y_{j,m} \left[\kappa_j^2 v^{\rm even}(n,x)
v^{\rm even}(m,x) + k_n k_m v^{\rm odd}(n,x) v^{\rm odd}(m,x)\right].
\end{eqnarray}
Our goal now is to pull out of the final triple summation a divergent piece
that exactly cancels the second term of the expression.  This is accomplished
by separating out a term proportional to
\begin{equation}
\kappa_j (k_n+k_m)  \left[v^{\rm even}(n,x) v^{\rm even}(m,x) +
v^{\rm odd}(n,x) v^{\rm odd}(m,x)\right],
\end{equation}
such that the final triple summation can be rewritten as
\begin{eqnarray}
\langle 0_L | : \bm{K}_{tt} :_{\widetilde{0}_L} | 0_L\rangle &=& -\frac{1}{4\ell L}
-\frac{1}{2L} \sum_{n=1}^\infty k_n + \frac{1}{4L}\sum_{j=1}^\infty \kappa_j Y_{j,0}^2
+\frac{1}{2\sqrt{L}}\sum_{n=1}^\infty v^{\rm even}(n,x) \sum_{j=1}^\infty \kappa_j
Y_{j,0} Y_{j,n}\nonumber\\
&& +\frac{1}{8} \sum_{j=1}^\infty  \sum_{n=1}^\infty
\sum_{m=1}^\infty Y_{j,n} Y_{j,m} (2\kappa_j - k_n -k_m) v^{\rm even}(n,x)
v^{\rm even}(m,x) \nonumber\\
&& +\frac{1}{8} \sum_{j=1}^\infty  \sum_{n=1}^\infty
\sum_{m=1}^\infty Y_{j,n} Y_{j,m} \left(\frac{2k_n k_m}{\kappa_j}-k_n-k_m\right) v^{\rm odd}(n,x) v^{\rm odd}(m,x)
\nonumber\\
&& +\frac{1}{8} \sum_{j=1}^\infty \sum_{n=1}^\infty
\sum_{m=1}^\infty Y_{j,n} Y_{j,m} (k_n+k_m) \left[v^{\rm even}(n,x) v^{\rm even}(m,x) +
v^{\rm odd}(n,x) v^{\rm odd}(m,x)\right].
\end{eqnarray}
Notice, in the final triple-summation term that the only $j$ dependence is
in the $Y_{j,n}$, thus, we can use Eq.~(\ref{eq:y_orthog}) to eliminate both the
$j$-summation and the $m$-summation.  Recalling the property of the modes that
\begin{equation}
v^{\rm even}(n,x)^2 + v^{\rm odd}(n,x)^2 = \frac{2}{L},
\end{equation}
we find that this final term exactly cancels the second term of the kinetic
energy-density, yielding a fully regularized expression,
\begin{eqnarray}
\langle 0_L | : \bm{K}_{tt} :_{\widetilde{0}_L} | 0_L\rangle &=& -\frac{1}{4\ell L}
+\frac{1}{4L} \sum_{j=1}^\infty \kappa_j Y_{j,0}^2 + \frac{1}{\sqrt{2}L}
\sum_{n=1}^\infty \cos(k_n x)\sum_{j=1}^\infty \kappa_j Y_{j,0} Y_{j,n}\nonumber\\
&&+\frac{1}{4}\sum_{n=1}^\infty \sum_{m=1}^\infty \cos(k_n x)\cos(k_m x)
\sum_{J=1}^\infty (2\kappa_j - k_n - k_m) Y_{j,n} Y_{j,m}\nonumber\\
&&+\frac{1}{4}\sum_{n=1}^\infty \sum_{m=1}^\infty \sin(k_n x)\sin(k_m x)
\sum_{J=1}^\infty \left(\frac{2 k_n k_m}{\kappa_j} - k_n - k_m\right)
Y_{j,n} Y_{j,m}
\end{eqnarray}
The next step is use the product rule for sine and cosine to rewrite the
summations;
\begin{eqnarray}
\langle 0_L | : \bm{K}_{tt} :_{\widetilde{0}_L} | 0_L\rangle &=& -\frac{1}{4\ell L}
+\frac{1}{4L} \sum_{j=1}^\infty \kappa_j Y_{j,0}^2 + \frac{1}{\sqrt{2}L}
\sum_{n=1}^\infty \cos(k_n x)\sum_{j=1}^\infty \kappa_j Y_{j,0} Y_{j,n}
+\frac{1}{4L}\sum_{n=1}^\infty \sum_{j=1}^\infty \frac{(\kappa_j-k_n)^2}{\kappa_j}
Y_{j,n}^2\nonumber\\
&&+\frac{1}{4L}\sum_{n=2}^\infty \cos(k_n x)\sum_{m=1}^{n-1}\sum_{j=1}^\infty
\left[ \frac{1}{\kappa_j}(\kappa_j +  k_m) (\kappa_j - k_{n-m}) - k_m + k_{n-m} \right] Y_{j,m} Y_{j,n-m}
\nonumber\\
&&+\frac{1}{2L}\sum_{n=1}^\infty \cos(k_n x)\sum_{m=1}^\infty\sum_{j=1}^\infty
\frac{1}{\kappa_j}(\kappa_j  - k_m)(\kappa_j - k_{n+m})  Y_{j,m} Y_{j,n+m}.
\end{eqnarray}
Because of Eq.~(\ref{eq:y_orthog}), we note that
\begin{equation}
\sum_{m=1}^{n-1}\sum_{j=1}^\infty (- k_m + k_{n-m} ) Y_{j,m} Y_{j,n-m} = 0,
\end{equation}
because the only value where the Kronecker delta is nonzero occurs when $m=n/2$,
but for this value of $m$ the $k_{n-m} - k_m=0$. Reorganizing our terms leads to
\begin{eqnarray}
\langle 0_L | : \bm{K}_{tt} :_{\widetilde{0}_L} | 0_L\rangle &=& -\frac{1}{4\ell L}
+\frac{1}{4L}\left[ \sum_{j=1}^\infty \kappa_j Y_{j,0}^2
+\sum_{n=1}^\infty \sum_{j=1}^\infty \frac{(\kappa_j-k_n)^2}{\kappa_j}
Y_{j,n}^2\right]\nonumber\\
&&+ \frac{1}{L}\left\{
\sum_{n=1}^\infty \cos(k_n x)\left[\frac{1}{\sqrt{2}}\sum_{j=1}^\infty \kappa_j Y_{j,0} Y_{j,n}
\right]\right.\nonumber\\
&&+\sum_{n=2}^\infty \cos(k_n x)\left[\frac{1}{4}\sum_{m=1}^{n-1}\sum_{j=1}^\infty
\frac{1}{\kappa_j}(\kappa_j +  k_m) (\kappa_j - k_{n-m}) Y_{j,m} Y_{j,n-m}\right]
\nonumber\\
&&\left.+\sum_{n=1}^\infty \cos(k_n x)\left[\frac{1}{2}\sum_{m=1}^\infty\sum_{j=1}^\infty
\frac{1}{\kappa_j}(\kappa_j  - k_m)(\kappa_j - k_{n+m})  Y_{j,m} Y_{j,n+m}\right]\right\}.
\end{eqnarray}
Comparing with Eqs.~(\ref{eq:Cdefn}) from the preceding subsection, we can immediately
identify the three cosine series terms with $2 \Re \rho(x)$, thus
\begin{eqnarray}
\langle 0_L | : \bm{K}_{tt} :_{\widetilde{0}_L} | 0_L\rangle &=& -\frac{1}{4\ell L}
+\frac{1}{4L}\left[ \sum_{j=1}^\infty \kappa_j Y_{j,0}^2
+\sum_{n=1}^\infty \sum_{j=1}^\infty \frac{(\kappa_j-k_n)^2}{\kappa_j}
Y_{j,n}^2\right]+2\Re\rho(x).
\end{eqnarray}
After substituting the explicit form $\kappa_j$, $k_n$, and the Fourier
coefficients into the remaining summations and comparing with Eq.~(\ref{eq_B_defn}),
we find
\begin{eqnarray}
\langle 0_L | : \bm{K}_{tt} :_{\widetilde{0}_L} | 0_L\rangle &=& -\frac{1}{4\ell L}
+\frac{\mathcal{B} - \mathcal{C}}{L^2} +\frac{\mathcal{C}}{L^2}\delta(x/L).
\end{eqnarray}
From the definition of the kinetic tensor and the fact that we are working in
two-dimension, we have that
\begin{equation}
\langle 0_L | : \bm{K}_{xx} :_{\widetilde{0}_L}
| 0_L\rangle=\langle 0_L | : \bm{K}_{tt} :_{\widetilde{0}_L} | 0_L\rangle
\end{equation}
A similar calculations can be performed for the remaining components of the
kinetic tensor.

Finally, recalling that the Mamev-Trunov potential $V(x,t)$ only has support at
$x=0$ for times $t<0$ and adding the Casimir energy for the cylinder
spacetime yields the {\it almost-everywhere} renormalized stress-tensor
on the IN region;
\begin{eqnarray}
\langle 0_L |  \bm{T}_{\mu\nu} | 0_L\rangle_{\rm Ren.} &=& \left(-\frac{\pi}{6 L^2}
+\frac{\mathcal{B} - \mathcal{C}}{L^2}\right) \delta_{\mu\nu}.
\label{Eq:Tren_IN}
\end{eqnarray}
This expression does not hold along the half line where the potential is
non-zero.

\section{Energy Conditions on the OUT Region}\label{sec:Energy Cond}

On the covering space of this spacetime, a timelike geodesic can be
parameterized as
\begin{equation}
\gamma^\mu(\tau) = \frac{1}{\sqrt{1-v^2}}(\tau, v\tau) + (t_0,x_0),
\label{eq.timelike geodesic}
\end{equation}
where $v$ is the speed of the observer, $\gamma = (1-v^2)^{-1/2}$, and $(t_0,x_0)$
is the location in spacetime of the geodesic at proper time $\tau=0$.  At every point along
the geodesic, we have the tangent
\begin{equation}
u^\mu(\tau) = \frac{1}{\sqrt{1-v^2}}(1, v)
\label{eq.timelike vector}
\end{equation}
and the orthogonal spacelike vector
\begin{equation}
r^\mu = \frac{1}{\sqrt{1-v^2}}(v,1).
\end{equation}
Both of these vectors can be extended to vector fields on the whole of the manifold, which
we will denote as $v_0^\mu = u^\mu$ and $v_1^\nu = r^\mu$.  For a given $(t_0,x_0)$,
the geodesic is contained within the IN region for $\tau\in(-\infty,-\gamma^{-1}t_0)$,
on the $t=0$ Cauchy surface when $\tau=-\gamma^{-1}t_0$, and contained within the OUT
region for $\tau\in(-\gamma^{-1}t_0,\infty)$.

The renormalized expectation value of the energy-density along the
worldline of any timelike geodesic observer on the OUT region is
\begin{eqnarray}
\langle 0_L|\bm{\rho}|0_L \rangle_{Ren.}(\tau) &=& \langle 0_L|\bm{T}_{\mu\nu}| 0_L \rangle_{Ren.} u^\mu u^\nu
\nonumber\\
&=& \frac{1+v^2}{1-v^2} \langle 0_L|\bm{T}_{tt}| 0_L \rangle_{Ren.}+\frac{2v}{1-v^2}
\langle 0_L|\bm{T}_{tx}| 0_L \rangle_{Ren.}\nonumber\\
&=&\frac{1+v^2}{1-v^2}\left(-\frac{\pi}{6L^2}+\frac{\mathcal{B-C}}{L^2} \right) +\frac{\mathcal{C}}{2L^2}
\sum_{n=-\infty}^{\infty} \left[ \frac{1+v}{1-v} \delta\left(\frac{t_0+x_0+(1+v)\gamma\tau}{L}
-n\right)\right.\nonumber\\ && \left.+ \frac{1-v}{1+v}\delta \left(\frac{t_0-x_0+(1-v)
\gamma\tau}{L}-n\right)\right].
\label{eq:Timelike_contracted}
\end{eqnarray}

The interpretation of this expression is straight forward.  The geodesic observer
``measures'' that the universe is filled with (a) a static, uniform cloud of
negative energy-density given by
\begin{equation}
\langle \rho_{\rm cloud} \rangle_{\rm Ren.} = \frac{1+v^2}{1-v^2}
\left(-\frac{\pi}{6L^2}+\frac{\mathcal{B-C}}{L^2} \right)
\end{equation}
and (b) two Dirac-delta-function pulses of particles that were created by the
shutting off of the potential which circle
around the universe, one moving in the $+x$-direction and the other moving in the
$-x$-direction, that repeatedly cross with the observers worldline with fixed periods of
\begin{equation}
T_{\rm right} = \sqrt{\frac{1+v}{1-v}}L \qquad\mbox{ and }\qquad
T_{\rm left} = \sqrt{\frac{1-v}{1+v}}L,
\end{equation}
respectively.  Both pulses have positive energy-density.  Similarly, the
renormalized expectation value of the momentum density in the $r^\nu$-direction
\cite{Wald_GR} is
\begin{eqnarray}
\langle 0_L|\bm{p}|0_L\rangle_{Ren.} (\tau)&=& -\langle 0_L|\bm{T}_{\mu\nu}| 0_L \rangle_{Ren.}
u^\mu r^\nu\nonumber\\
&=& -\frac{2v}{1-v^2}\left(-\frac{\pi}{6L^2}+\frac{\mathcal{B- C}}{L^2} \right) +\frac{\mathcal{C}}{2L^2}
\sum_{n=-\infty}^{\infty} \left[ -\frac{1+v}{1-v} \delta\left(\frac{t_0+x_0+(1+v)\gamma\tau}{L}
-n\right)\right.\nonumber\\ && \left.+ \frac{1-v}{1+v}\delta \left(\frac{t_0-x_0+(1-v)
\gamma\tau}{L}-n\right)\right].
\end{eqnarray}

For a right-going (+) or left-going (-) null geodesic parameterized by the
variable $\lambda$, such that
\begin{equation}
\eta_\pm^\mu(\lambda) = (\lambda, \pm\lambda) + (t_0,x_0)\qquad\mbox{ with }\qquad
K_\pm^\mu = \frac{d}{d\lambda} \eta_\pm^\mu(\lambda) = (1,\pm 1),
\end{equation}
we determine that the renormalized energy-density along the worldline of a null
geodesic observer on the OUT region of the spacetime to be
\begin{eqnarray}
\langle 0_L |\bm{T}_{\mu\nu}|0_L \rangle_{Ren.} K_\pm^\mu K_\pm^\nu (\tau) &=&
2 \langle 0_L |\bm{T}_{tt}|0_L \rangle_{Ren.}\pm 2\langle 0_L |\bm{T}_{tx}|0_L \rangle_{Ren.}
\nonumber\\
&=& 2\left(-\frac{\pi}{6L^2}+\frac{\mathcal{B}-\mathcal{C}}{L^2} \right) + 2
\frac{\mathcal{C}}{L^2} \sum_{n=-\infty}^\infty \delta\left(\frac{2\lambda+
t_0\pm x_0}{L}-n\right).
\label{eq:Null_contracted}
\end{eqnarray}
Notice that the null observer only picks up a contribution from the positive-energy
delta-function pulse that is moving in the opposite direction to that of the observer.
The co-moving delta-function pulse never crosses the null observer's worldline.

\subsection{Classical Energy Conditions on the OUT Region}

With the expressions for the renormalized energy-density along the worldline of both
a timelike and null observer, we can now evaluate whether the stress-tensor for our
scalar quantum field obeys or violates each of the point-wise classical energy
conditions of general relativity on the OUT region of the spacetime:

\begin{itemize}

\item {\bf Null Energy Condition}  A stress-tensor is said to satisfy the null energy
condition from general relativity if it obeys
\begin{equation}
T_{\mu\nu} K^\mu K^\nu \geq 0
\end{equation}
at all points in the spacetime.  From Eq.~(\ref{eq:Null_contracted}),
the renormalized expectation value of the stress-energy tensor for
the IN vacuum state on the OUT region {\bf fails} to
satisfy the NEC for all values of the $\chi$ because
$\mathcal{B}-\mathcal{C} < \pi/6$.

\item {\bf Weak Energy Condition}  A stress-tensor is said to satisfy the weak energy
condition from general relativity if it obeys
\begin{equation}
T_{\mu\nu} u^\mu u^\nu \geq 0
\end{equation}
at all points in the spacetime. From Eq.~(\ref{eq:Timelike_contracted}),
the renormalized expectation value of the stress-tensor for the IN
vacuum state on the OUT region {\bf fails} to satisfy the WEC for
all values of $\chi$.  This was the same situation as for the NEC.

\item {\bf Strong Energy Condition}  A stress-tensor is said to satisfy the strong energy
condition from general relativity if it obeys
\begin{equation}
\left(T_{\mu\nu}-\frac{1}{2}T\, g_{\mu\nu}\right) u^\mu u^\nu \geq 0
\end{equation}
at all points in the spacetime. The renormalized expectation value of
the stress-tensor for the IN vacuum state on the OUT region,
Eq.~(\ref{Eq.ren_stress_tensor}), is traceless, thus, the SEC is equivalent
to the WEC for our problem and will {\bf fail} under the same circumstances.

\item {\bf Dominant Energy Condition} A stress-tensor is said to satisfy the dominant energy
condition from general relativity if, for every future-pointing
causal vector field $Y^\mu$ (timelike and null), the vector
\begin{equation}
V^\mu \equiv {T^\mu}_\nu Y^\nu
\end{equation}
is also a future-pointing and causal.
Using the above definition of $V^\mu$ and the expectation value of our
normalized stress-tensor, the future-pointing condition is
\begin{equation}
\langle 0_L | {\bm T}_{tt} | 0_L \rangle_{\rm Ren.} Y^t +
\langle 0_L | {\bm T}_{tx} | 0_L \rangle_{\rm Ren.} Y^x > 0
\end{equation}
and the causal condition is
\begin{equation}
\left( \langle 0_L | {\bm T}_{tt} | 0_L \rangle_{\rm Ren.}^2 -
\langle 0_L | {\bm T}_{tx} | 0_L \rangle_{\rm Ren.}^2\right)
\left[\left(Y^t\right)^2-\left(Y^x\right)^2\right] \geq 0.
\end{equation}
If $Y^\mu$ is everywhere null, then the causal condition is
satisfied. As for the future-pointing condition, setting $Y^\mu = K_\pm^\mu$
reduces the condition to
\begin{equation}
-\frac{\pi}{6L^2} + \frac{\mathcal{B-C}}{L^2} + \frac{\mathcal{C}}{L^2}
\sum_{n=-\infty}^\infty \delta\left( \frac{t\pm x}{L}-n\right) > 0.
\end{equation}
However, $\mathcal{B-C} \leq \pi/6$, so the above inequality fails on large
regions of the spacetime. We can therefore conclude that the DEC is also
violated under the same condition as all the other energy conditions.
\end{itemize}

\subsection{Total Energy in a Constant-Time Hypersurface}

Let $t_s >0 $, determine a constant time Cauchy surface on the OUT
region of the spacetime.  The unit normal to the Cauchy surface is
given by $n^\mu=(1,0)$.  Contracting Eq.~(\ref{Eq.ren_stress_tensor})
with the unit normal twice yields the energy-density contained in the
Cauchy surface,
\begin{eqnarray}
\rho_L(x,t_s) &=& -\langle 0_L| \bm{T}_{\mu\nu} |0_L\rangle_{Ren.}n^\mu n^\nu\nonumber\\
&=& -\frac{\pi}{6L^2} + \frac{\mathcal{B}-\mathcal{C}}{L^2} +
\frac{\mathcal{C}}{2L^2} \sum_{n=-\infty}^\infty\left[ \delta\left(\frac{t_s-x}{L}-n\right)
 + \delta\left(\frac{t_s+x}{L}-n\right)\right].
\end{eqnarray}
To determine the ``total'' energy contained in the Cauchy surface, we integrate the
above expression over the spatial direction, thus
\begin{equation}
\mathcal{E}(t_s) = -\frac{\pi}{6L} + \frac{\mathcal{B}-\mathcal{C}}{L} +
\frac{\mathcal{C}}{2L}\int_{0}^{L} \sum_{n=-\infty}^\infty\left[
\delta\left(t_s-x-nL\right)+ \delta\left(t_s+x-nL\right)\right] dx.
\end{equation}
Only two of the delta-functions in the infinite sum give nontrivial contributions
to the integral. They occur when $n=\mbox{IntegerPart}(t_s/L)$, and the positions
$x = \mbox{Remainder}(t_s/L)$ for the right moving pulse and $x = L -
\mbox{Remainder}(t_s/L)$ for the left moving pulse.  The result for the energy
in the Cauchy surface is
\begin{equation}
\mathcal{E}(t_s) = -\frac{\pi}{6L} + \frac{\mathcal{B}}{L}, \qquad\mbox{for}\qquad
t_s>0.
\end{equation}
This expression is a time-independent constant, thus energy is conserved
on the OUT region of the spacetime by the scalar field. From the numerical
simulations of $\mathcal{B}$ as a function of $\chi$, the total energy in
the Cauchy surface is negative for values of $\chi\leq0.82$, positive for
values of $\chi\geq0.83$, and it passes through zero somewhere in the
range $0.82<\chi<0.83$.

\subsection{Quantum Weak Energy Inequality on the OUT Region}

In Appendix~\ref{sec:QWEI for cylinder}, we derive a QWEI for the
quantized scalar field on the cylinder spacetime with no potential that
includes the contributions from the topological modes.  For a timelike
geodesic observer moving through the spacetime, the QWEI is
\begin{eqnarray}
\int_\reals d\tau\, \langle\omega| : \bm{\rho} :_{\tilde{0}_L}  |\omega\rangle (\tau) \left[g(\tau)\right]^2 &\geq&
\frac{1+v^2}{1-v^2}\left(-\frac{1}{4\ell L}\right) \int_\reals d\tau\, \left[g(\tau)\right]^2
-\frac{1}{2L}\sum_{n=1}^\infty k_n \left[\frac{1+v}{1-v} \int_0^\infty
\frac{d\alpha}{\pi} \left| \hat{g}\left(\alpha + k_n\sqrt{\frac{1+v}{1-v}}\right)\right|^2
\right.
\nonumber\\
&&\left. +\frac{1-v}{1+v} \int_0^\infty
\frac{d\alpha}{\pi} \left| \hat{g}\left(\alpha + k_n\sqrt{\frac{1-v}{1+v}}\right)\right|^2
\right],
\label{eq:QWEI specific}
\end{eqnarray}
where $|\omega\rangle$ is any Hadamard state on the spacetime $\reals\times S^1$,
$g(\tau)$ is a smooth, real-valued, compactly-supported test function on the real
line, $\hat{g}(\alpha)$ is the Fourier transform of $g(\tau)$, and normal ordering
is done with respect to the OUT ground state $|\widetilde{0}_L\rangle$. We emphasize
that the above form of the QWEI is
a difference inequality.  To convert the above inequality into one
for the renormalized energy-density, we must include the energy-density due
to the Casimir effect, which adds to both sides of the inequality a term of
the form
\begin{equation}
\int_\reals d\tau \langle \tilde{0}_L | \bm{\rho} |\tilde{0}_L \rangle_{\rm Ren.}(\tau)
\,\left[ g(\tau)\right]^2 =
\frac{1+v^2}{1-v^2}\left(\frac{1}{4\ell L} - \frac{\pi}{6 L^2}\right)
\int_\reals d\tau \,\left[ g(\tau)\right]^2.
\end{equation}
Thus, the {\em absolute} QWEI for the scalar field on the cylinder spacetime with
no potential is
\begin{eqnarray}
\int_\reals d\tau\, \langle\omega|  \bm{\rho}   |\omega\rangle_{\rm Ren.} (\tau) \left[g(\tau)\right]^2 &\geq&
\frac{1+v^2}{1-v^2}\left(-\frac{\pi}{6 L^2}\right) \int_\reals d\tau\, \left[g(\tau)\right]^2
-\frac{1}{2L}\sum_{n=1}^\infty k_n \left[\frac{1+v}{1-v} \int_0^\infty
\frac{d\alpha}{\pi} \left| \hat{g}\left(\alpha + k_n\sqrt{\frac{1+v}{1-v}}\right)\right|^2
\right.
\nonumber\\
&&\left. +\frac{1-v}{1+v} \int_0^\infty
\frac{d\alpha}{\pi} \left| \hat{g}\left(\alpha + k_n\sqrt{\frac{1-v}{1+v}}\right)\right|^2
\right].
\label{eq:QWEI absolute}
\end{eqnarray}
All the terms on the right-hand side of our QWEI are negative.

To apply this QWEI to states of the quantized scalar field living on the
OUT region of the cylinder spacetime with potential, we appeal to the
causal isometric embedding arguments of Fewster and Pfenning \cite{Pfen06}.
The OUT region of our spacetime is causally isometric to the $t\geq0$
portion of the the cylinder spacetime without potential. (It is assumed we
maintain the time orientiation in the isometry.)  By the principle
of local causality, an observer who performs local experiments in the $t>0$ portion
of either of these spacetimes should not be able to discern which spacetime
they actually inhabit.  Basically, an observer whose experiments do not
extend back in time beyond the $t=0$ Cauchy surface is not able to determine
that the stress-tensor they are measuring is due to the IN vacuum state
of a field that used to interact with a potential, or just some very highly
prepared state of a quantum field that never interacted with the potential.
So by this locality argument, quantum inequalities on the OUT region of
our spacetime should be the same as those on the quantum inequalities on
the $t>0$ portion of the standard cylinder spacetime.
Thus, to apply the  QWEI above to states on the OUT region of our spacetime,
all we have to do is restrict the space of allowable test functions to
only those which have support to the future of the $t=0$ Cauchy surface,
i.e., our test function space for $g(\tau)$ is a subspace of the full test
function space on $\reals\times S^1$.

Next, we evaluate the left-hand side of the QWEI when the state of interest is
the IN ground state $|0_L\rangle$ with $g(\tau)$ being any test function from the
restricted space of test functions. Substituting the expression for the renormalized
energy-density on the OUT region, Eq.~(\ref{eq:Timelike_contracted}), we find
\begin{eqnarray}
L.H.S &=& \int d\tau \, \langle 0_L | \bm{\rho}  | 0_L \rangle_{\rm Ren.}(\tau)
[g(\tau)]^2\nonumber\\
&=& \frac{1+v^2}{1-v^2}\left(-\frac{\pi}{6 L^2}\right) \int d\tau\, [g(\tau)]^2
+\frac{1+v^2}{1-v^2}\left(\frac{\mathcal{B - C}}{L^2}\right) \int d\tau [g(\tau)]^2\nonumber\\
&&+\frac{\mathcal{C}}{2L^2} \sum_{n=-\infty}^{\infty} \left[ \frac{1+v}{1-v} \int d\tau [g(\tau)]^2
\delta\left(\frac{t_0+x_0+(1+v)\gamma\tau}{L}-n\right)\right.\nonumber\\
&&\left.+\frac{1-v}{1+v} \int d\tau [g(\tau)]^2
\delta\left(\frac{t_0-x_0+(1-v)\gamma\tau}{L}-n\right)\right].
\end{eqnarray}
Only the first term of this expression is negative and it is identical
to the first term on the right-hand side of the absolute QWEI above.
As we pointed out above, all the remaining terms on the right-hand
side of the QWEI are negative, therefore, we conclude the IN state
$|0_L\rangle$ obeys the QWEI on the OUT region of the spacetime for
all $g(\tau)$ in the restricted space of test functions.

\section{Conclusions}\label{sec:Conclusion}

In this paper, we studied the behavior of a quantized scalar field coupled to
an external, time-dependent, Mamev-Trunov potential on the cylinder spacetime
$\reals\times S^1$. We found for a quantum field that begins in the
IN vacuum state that the shutting off of the potential at time
$t=0$ causes the field to respond with the creation of particles out of the vacuum
on the OUT region of the spacetime.  We determined analytic expressions for the number
of particles
created and showed that the number of particles in each mode is finite, and that
the total number of particles is also finite.  We then determined the renormalized
stress-tensor on both the IN and OUT regions of the spacetime. For the IN region,
we found the {\it almost-everywhere} expression, Eq.~(\ref{Eq:Tren_IN}), consisted
of the standard Casimir effect of $-\pi/6L^2$ and an additional term of
$(\mathcal{B}(\chi)-\mathcal{C}(\chi))/L^2$ that is due to the potential. This
result was valid on the IN region away from the location of the potential.
For the OUT region, we found that the stress-tensor, Eq.~(\ref{Eq.ren_stress_tensor}),
consisted of the same two parts as the IN region, plus additional terms that
describe the positive energy-density and flux of the particles created out
of the vacuum. We went on to show that all of the point-wise energy conditions
of general relativity are violated by this stress-tensor. However, we also found
that stress-tensor for the IN vacuum state satisfies a quantum inequality for all
timelike geodesic observers on the OUT region of the spacetime, with the constraint
that the compactly supported test functions have support only to the future of the $t=0$
Cauchy surface.  The quantum inequality was satisfied because of the positive-energy
contributions to the stress-tensor from the particles created out of the vacuum.

With regard to Solomon's claims of violations of the quantum inequalities for the
double delta-function potential of Mamev and Trunov, we see from the analysis of
this paper that the particle creation and their resulting positive-energy contributions
to the renormalized stress-tensor cannot be ignored.  In all likelihood, if these
contributions could be determined and added to the partial results of Solomon,
we would find that the quantum inequalities hold.  This is a topic we will
return to in the future.

Finally,  a great deal of the research work of this paper was directed toward
determining the behavior of infinite series over the positive solutions of
the transcendental equation $Z=\chi\cot Z$.  This includes $\mathcal{B}(\chi)$
and $\mathcal{C}(\chi)$ in the main body, and $F_p(\chi)$ and $\mathcal{A}(\chi)$ in
Appendices~\ref{sec:series} and \ref{sec:A_derivation}, respectively. From
Eq.~(\ref{eq:Fp zeta}) below, the functions $F_p(\chi)$ look remarkably like
the derivative of a some form of generalized Riemann zeta function. This probably
explains why it was possible to determine analytic expressions for $F_p(\chi)$
when $p>1$ was an even integer.  We have two conjectures about the functions
$\mathcal{A}(\chi)$ and $\mathcal{B}(\chi)$ which are based on the numerical
simulation of each in Mathematica:  that $\mathcal{A}(\chi)=\mathcal{B}(\chi)$
and that $\mathcal{B}$ is a hyperbola of the form $\mathcal{B}(\chi) = (\pi)^{-1}
\sqrt{\chi (\chi-2b)}$ where $0<b\leq\pi^2/6$.  The Mathematica plots seem to
indicate that $b\approx 12/\pi^2$ is a good fit.


\begin{acknowledgments}
I would like to thank J.C.\ Loftin, L.E.\ Harrell, P.\ Fekete, K.\ Ingold, and
D.O.\ Kashinski for many useful discussions.  I also want to thank L.H.\ Ford,
K.D. Olum, and C.J.\ Fewster for their hospitality and various discussions during
the preparation of this paper.
I would also like to acknowledge The Department of Physics and Nuclear Engineering,
The Office of The Dean, and The Photonics Research Center of Excellence, all at
the United States Military Academy, for their financial and logistical support
in the completion of this work.\\

The views expressed herein are those of the author and do not reflect the
position of the United States Military Academy, the Department of the Army,
or the Department of Defense.

\end{acknowledgments}

\appendix
\section{Equivalence of ``IN'-Mode Functions with the Fourier Time
Evolution on the Bow-Tie Shaped Region $-|x|\leq t \leq |x|$}
\label{Appendix_overlap}

In this appendix, we show that
\begin{equation}
\phi^{\rm even}(j,x,t) = \phi^{\rm even}_{\rm OUT}(j,x,t)
\end{equation}
on the the region $\mathcal{D} \cup \{(0,0)\}$, where the open, bow-tie-shaped domain
\begin{equation}
\mathcal{D} \equiv\left\{ (x,t)\in \left[-\frac{L}{2},\frac{L}{2}\right] \times
\left[-\frac{L}{2},\frac{L}{2}\right]
\big|-|x| < t < |x| \right\}.
\end{equation}
In other words, the explicit form of the IN-region mode functions can
be used to the future of the  $t=0$ Cauchy surface, i.e, to the portion of the
IN-region where $0\leq t < |x|$.  Similarly, the explicit form of the
OUT-regions solutions can be used to the past of the $t=0$ Cauchy surface,
i.e. to the portion of the IN-region where $-|x|< t \leq 0$.  On this
domain, one can use the expressions for the IN and OUT forms
of the mode solutions interchangeably.

Substituting the definitions of the mode functions and Fourier coefficients
into the above expression, we need to show that
\begin{equation}
\left[\cos(\kappa_j x) + \frac{\xi}{\kappa_j} \sin(\kappa_j |x|)\right]
 e^{-i\kappa_j t}
= \frac{\xi L}{2}\left\{ \frac{1-i\kappa_j t}{Z_j^2} + 2
\sum_{n=1}^\infty \frac{\cos(k_n x)}{Z_j^2-(\pi n)^2}
\left[ \cos(k_n t) - i \frac{\kappa_j}{k_n} \sin(k_n t)\right]
\right\}\label{Eq:Complex_expression}
\end{equation}
on the specified region. It easier to handle the real and imaginary
parts separately, thus, the above expression breaks into two conditions
which we must prove:
\begin{equation}
\left[\cos(\kappa_j x) + \frac{\xi}{\kappa_j} \sin(\kappa_j |x|)\right]
\cos(\kappa_j t) = \frac{\xi L}{2}\left[ \frac{1}{Z_j^2} + 2
\sum_{n=1}^\infty \frac{\cos(k_n x)\cos(k_n t)}{Z_j^2-(\pi n)^2}
\right]
\label{Eq:Real_part}
\end{equation}
for the real part, and
\begin{equation}
-\left[\cos(\kappa_j x) + \frac{\xi}{\kappa_j} \sin(\kappa_j |x|)\right]
\sin(\kappa_j t) =
-\frac{\xi L}{2}\left[ \frac{\kappa_j t}{Z_j^2} + 2
\sum_{n=1}^\infty \left(\frac{\kappa_j}{k_n}\right) \frac{\cos(k_n x)\sin(k_n t)}
{Z_j^2-(\pi n)^2}\right]
\label{Eq:Imaginary_part}
\end{equation}
for the imaginary part.

To begin, it is a straightforward exercise of Fourier analysis on the circle of
circumference $L$ to show that
\begin{equation}
f(x)\equiv\cos(\kappa_j x) + \frac{\xi}{\kappa_j} \sin(\kappa_j |x|) =\frac{\xi L}{2}
\left[ \frac{1}{Z_j^2} + 2
\sum_{n=1}^\infty \frac{\cos(k_n x)}{Z_j^2-(\pi n)^2}
\right].\label{eq:f_definition}
\end{equation}
From this, it is easy to see that Eq.~(\ref{Eq:Real_part}) holds when $t=0$,
while Eq.~(\ref{Eq:Imaginary_part}) is trivially true along the same line.
Next, using the product identities for cosines, it is possible to
rewrite the right-hand side of the real-part equation as
\begin{equation}
f(x) \cos(\kappa_j t) \stackrel{?}{=} \frac{1}{2}\left( \frac{\xi L}{2}
\left\{ \frac{1}{Z_j^2} + 2 \sum_{n=1}^\infty \frac{\cos[k_n (x-t)]}{Z_j^2-(\pi n)^2}
\right\}+\frac{\xi L}{2}
\left\{ \frac{1}{Z_j^2} + 2 \sum_{n=1}^\infty \frac{\cos[k_n (x+t)]}{Z_j^2-(\pi n)^2}
\right\}\right),
\end{equation}
or more simply,
\begin{equation}
f(x) \cos(\kappa_j t) \stackrel{?}{=} \frac{1}{2}\left[f(x-t)+f(x+t)\right].
\end{equation}
Using the product identities of the trigonometric functions on the left-hand side
of this equation, substituting the compact definition of $f(x)$ on the right-hand
side,  and simplifying, results in us having to determine the domain of validity
of the equation
\begin{equation}
\sin[\kappa_j(|x|-t)]+ \sin[\kappa_j(|x|+t)] \stackrel{?}{=} \sin(\kappa_j|x-t|)
+\sin(\kappa_j|x+t|).
\end{equation}
This equation is satisfied if we can meet either of the following conditions:
\begin{equation}
\mbox{a) } |x|-t = |x-t| \mbox{ and } |x|+t = |x+t|,
\end{equation}
or
\begin{equation}
\mbox{b) } |x|-t = |x+t| \mbox{ and } |x|+t = |x-t|.
\end{equation}
However, for both cases, it is always true that $|x\pm t|\geq 0$,
which implies that we would simultaneously need $|x|-t\geq0$ and
$|x|+t\geq0$.  These are compatible conditions which hold on the
non-trivial domain $-|x| \leq t \leq |x|$.  Therefore, this implies
that the real part, Eq.~(\ref{Eq:Real_part}), holds on the domain
$\mathcal{D}$, and on the boundary of the domain.

We now turn our attention to the proving that on the domain $\mathcal{D}$
the imaginary part, Eq.~(\ref{Eq:Imaginary_part}), is true. We begin with
Eq.~(\ref{eq:f_definition}) and integrate it in $x$ from $x-t$ to $x+t$,
i.e.,
\begin{eqnarray}
\int_{x-t}^{x+t}\left[\cos(\kappa_j x') + \frac{\xi}{\kappa_j} \sin(\kappa_j |x'|)
\right]dx' &=& \frac{\xi L}{2}\int_{x-t}^{x+t}\left[ \frac{1}{Z_j^2} + 2
\sum_{n=1}^\infty \frac{\cos(k_n x')}{Z_j^2-(\pi n)^2}\right]dx',
\end{eqnarray}
resulting in
\begin{eqnarray}\frac{2}{\kappa_j}\left[\cos(\kappa_j x) +
\left(\frac{\xi}{\kappa_j}\right) \Theta^+(x,t)
\sin(\kappa_j x)\right]\sin(\kappa_j t)
-\frac{2\xi}{\kappa_j^2}\Theta^-(x,t)\left[\cos(\kappa_j x)
\cos(\kappa_j t)-1 \right]\nonumber\\
= 2\left( \frac{\xi L}{2} \right)\left[ \frac{t}{Z_j^2}+2\sum_{n=1}^\infty
\frac{1}{k_n} \frac{\cos(k_n x) \sin(k_n t)}{Z_j^2 - (\pi n)^2}\right],
\label{eq:imag_int}
\end{eqnarray}
where we have simplified using trigonometric identities and the
definition
\begin{equation}
\Theta^{\pm}(x,t) \equiv \frac{1}{2}\left[ \sign(x+t) \pm \sign(x-t)\right]
\end{equation}
with the $\sign$-function being the numerical sign of the argument, i.e., it equals $+1$
for positive arguments, $-1$ for negative arguments, and undefined at zero.
Notice, the right-hand side of Eq.~(\ref{eq:imag_int}) is, up to a factor
of $\kappa_j/2$, identical to the right-hand side of Eq.~(\ref{Eq:Imaginary_part}).

The final step is to determine if there exist any regions where
\begin{equation}
\Theta^+(x,t) = \sign(x)\qquad\mbox{ and }\qquad\Theta^-(x,t) = 0.
\end{equation}
The second of these two conditions is equivalent to $\sign(x+t) = \sign(x-t)$,
which is satisfied within the bow tie region $\mathcal{D}$.  Furthermore, for
all $x<0$ inside of $\mathcal{D}$ we have $\Theta^+(x,t)=-1$, and for all
$x>0$ inside of $\mathcal{D}$ we have $\Theta^+(x,t) = 1$.  Therefore, we have
$\Theta^+(x,t) = \sign(x)$ on $\mathcal{D}$.  Because both relations hold
within $\mathcal{D}$, we have that
\begin{equation}\frac{2}{\kappa_j}\left[\cos(\kappa_j x) +
\left(\frac{\xi}{\kappa_j}\right)
\sin(\kappa_j |x|)\right]\sin(\kappa_j t)
= 2\left( \frac{\xi L}{2} \right)\left[ \frac{t}{Z_j^2}+2\sum_{n=1}^\infty
\frac{1}{k_n} \frac{\cos(k_n x) \sin(k_n t)}{Z_j^2 - (\pi n)^2}\right]
\end{equation}
holds within $\mathcal{D}$.  Multiply both sides by of this equation by $\kappa_j/2$,
we can conclude that Eq.~(\ref{Eq:Imaginary_part}) indeed holds on
$\mathcal{D}$.  Recall from above that Eq.~(\ref{Eq:Real_part}) and
Eq.~(\ref{Eq:Imaginary_part}) also hold for all values of $x$ when $t=0$,
which includes the origin point $(0,0)$.
Because the real and imaginary parts hold on $\mathcal{D}\cup\{(0,0)\}$,
we can finally conclude that $\phi^{\rm even}(j,x,t) = \phi^{\rm even}_{\rm OUT}(j,x,t)$
on this region.


\section{Construction of the advanced-minus-retarded Green's function
on $M \simeq \reals \times S^1$}\label{sec:adv-ret Green}

In this appendix, we derive the advanced-minus-retarded Green's function
for the scalar Klein-Gordon-Fock equation on $\stM \approx \mathbb{R}\times S^1$
without a potential. We use the conventions of Fulling \cite{Fulling}.
Let $\J \in C_0^\infty (\stM;\reals)$ be a smooth, compactly-supported function
on $\stM$, then by spectral theory, the
advanced-minus-retarded operator $E: C_0^\infty(M)\rightarrow C^\infty(M)$
is given by
\begin{equation}
(E\J)(x,t) = -\int dt' \sum_{j=0}^\infty \frac{(-\hat{K})^j}{(2j+1)!} (t-t')^{2j+1} \J(x,t'),
\end{equation}
where the operator $\hat{K} = -\partial_x^2$ is Hermitian under integration
on the circle. The completeness theorem for functions on $S^1$ tells us
\begin{equation}
\J(x,t') =({v^{top.}}\J)(t')\,v_{top.}(x) + \sum_{n=1}^\infty
\left[({v^{odd}_n}\J)(t')\, v^{odd}(n,x)  + ({v^{even}_n}\J)(t')
\, v^{even}(n,x) \right]
\end{equation}
where we define
\begin{eqnarray}
({v^{top.}}\J)(t') &\equiv& \int_{S^1} {v^{top.}(x')}\, \J(x',t')\, dx',\\
({v^{odd}_n}\J)(t') &\equiv& \int_{S^1} {v^{odd}(n,x')}\, \J(x',t')\, dx',\\
({v^{even}_n}\J)(t') &\equiv& \int_{S^1} {v^{even}(n,x')}\, \J(x',t')\, dx'.
\end{eqnarray}
The advanced-minus-retarded Green's function smeared in both slots by
$\J_1, \J_2 \in C_0^\infty(M;\reals)$ is defined as
\begin{equation}
E(\J_1, \J_2) =  \int dt \int_{S^1} dx\, \J_1(x,t) (E\J_2)(x,t).
\end{equation}
Substituting into this expression yields
\begin{eqnarray}
E(\J_1,\J_2) &=& -\int dt\int dt' (t-t') (v^{top.}\J_1)(t) ({v^{top.}}\J_2)(t')\nonumber\\
&&-\int dt\int dt' \sum_{n=1}^\infty \frac{\sin[k_n(t-t')]}{k_n} \left[ (v^{odd}_n \J_1)(t)
({v^{odd}_n}\J_2)(t')+(v^{even}_n \J_1)(t)({v^{even}_n}\J_2)(t')\right].
\end{eqnarray}
The kernel of this expression is easily seen to be
\begin{eqnarray}
E(x,t; x',t') &=& -(t-t')\, v^{top.}(x) \,{v^{top.}(x')}\nonumber\\
&& -\sum_{n=1}^\infty \frac{\sin[k_n(t-t')]}{k_n} \left[ v^{odd}(n,x)\,
{v^{odd}(n,x')}+v^{even}(n,x)\,{v^{even}(n,x')}\right],\nonumber\\
&=& -\frac{(t-t')}{L} - \frac{1}{L} \sum_{n=1}^{\infty} \frac{1}{k_n}
\left( \sin\left\{ k_n\left[(t-t')-(x-x')\right]\right\}
+\sin\left\{ k_n\left[(t-t')+(x-x')\right]\right\}\right).
\label{eq:Adv-Ret_Kernal}
\end{eqnarray}
It is clear the kernel consists of a smooth topological part, a purely
``right''-moving part and a purely ``left''-moving part, both moving parts
propagating at the speed of light.   Also, from the
above expression we see that the kernel is antisymmetric under the interchange
of the coordinates, i.e., $E(x',t';x,t) = -E(x,t;x',t')$, which in turn implies
$E(\J_2,\J_1) = -E(\J_1,\J_2)$.

By elementary Fourier analysis, it is straightforward to show that the
function below, constructed from the modulo operation, has the Fourier
representation
\begin{equation}
\frac{L}{4}\left[ 1-\frac{2}{L}\left(x\bmod L\right)\right] = \sum_{n=1}^\infty \frac{1}{k_n}
\sin(k_n x),
\end{equation}
whereby, we may alternatively express the kernel of the advanced-minus-retarded
Green's function as
\begin{equation}
E(x,t; x',t')=-\frac{(t-t')}{L}-\frac{1}{2} +\frac{1}{2L}\left\{
\left[(t-t')-(x-x')\right]\bmod L +\left[(t-t')+(x-x')\right]\bmod L\right\}.
\end{equation}

For the Cauchy problem, where the initial data is given by $\phi(x,0) = f(x)$ and
$\partial_t\phi(x,0) = g(x)$, the unique classical solution is given by
\begin{equation}
\phi(x,t) = -\int_{S^1} \left[ \left( \partial_t E(x,t;x',0)\right) f(x')
+ E(x,t;x',0) g(x') \right]dx'.
\end{equation}


\section{Convergence of Series}\label{sec:series}

In this appendix, we are interested in the properties of series of the
form
\begin{equation}
F_p(x) \equiv x^2\sum_{j=1}^\infty \frac{A_j^2}{Z_j^p},
\label{Eq:F_defn}
\end{equation}
where $x\in [0,\infty)$,  $p$ is a positive real number greater than one,
$Z_j$ is the $j$-th positive root of the transcendental equation
\begin{equation}
Z = x \cot(Z),
\end{equation}
and
\begin{equation}
A_j^2 = \frac{\cos^2Z_j}{1+\frac{\sin Z_j\cos Z_j}{Z_j}} = \frac{Z_j^2}{Z_j^2 + x^2 +x}.
\label{eq:A_defn}
\end{equation}
We remind the reader that $Z_j$ is an implict function of $x$.  The transcendental
equation implies the trigonometric relations
\begin{equation}
\sin Z_j = (-1)^{j-1} \frac{x}{\sqrt{Z_j^2 + x^2}} \qquad\mbox{ and }\qquad
\cos Z_j = (-1)^{j-1} \frac{Z_j}{\sqrt{Z_j^2 + x^2}},
\end{equation}
which were used to obtain the final equality of Eq.~(\ref{eq:A_defn}).
For the main body of the paper, we had $x = \xi L/2$.  Also, the Bogolubov
relations imply, via Eq.~(\ref{eq:y_orthog}) with $m=n=0$, that $F_4(x) = 1/2$.
(Numerical simulation in Mathematica seems to confirm this fact.)

We wish to demonstrate that such sums are
convergent for $p>1$. First, note that every term in the summation
above is positive, therefore the summation is bounded below by zero.
Next, we determine an upper bound.  Using the definition of $A_j$ and
separating out the first term in the summation, we have
\begin{equation}
\sum_{j=1}^\infty \frac{A_j^2}{Z_j^p} =\frac{\cos^2(Z_1)}{Z_1^{p-1}
\left[Z_1 + \sin(Z_1)\cos(Z_1) \right]} + \sum_{j=2}^\infty \frac{\cos^2(Z_j)}{Z_j^{p-1}
\left[Z_j + \sin(Z_j)\cos(Z_j) \right]}.
\end{equation}
Recall from above that the $Z_j$'s always satisfy
\begin{equation}
(j-1)\pi \leq Z_j \leq \left(j-\frac{1}{2}\right)\pi.
\end{equation}
On each of these intervals, the product of the sine and cosine functions is
a positive number, thus
\begin{equation}
\frac{\cos^2(Z_j)}{Z_j^{p-1}\left[Z_j + \sin(Z_j)\cos(Z_j) \right]}\leq
\frac{1}{Z_j^{p-1}\left[Z_j + \sin(Z_j)\cos(Z_j) \right]}\leq
\frac{1}{Z_j^{p}} \leq \frac{1}{(j-1)^p\pi^p}.
\end{equation}
Therefore, we have an upper bound
\begin{equation}
\sum_{j=1}^\infty \frac{A_j^2}{Z_j^p} \leq \frac{\cos^2(Z_1)}{Z_1^{p-1}
\left[Z_1 + \sin(Z_1)\cos(Z_1) \right]} + \frac{1}{\pi^p}\sum_{j=2}^\infty
\frac{1}{(j-1)^p}.
\end{equation}
The summation over $j$ is the series definition of the Riemann zeta function,
thus
\begin{equation}
\sum_{j=1}^\infty \frac{A_j^2}{Z_j^p} \leq \frac{\cos^2(Z_1)}{Z_1^{p-1}
\left[Z_1 + \sin(Z_1)\cos(Z_1) \right]} + \frac{\zeta(p)}{\pi^p}.
\end{equation}
The zeta function is convergent for all $p>1$, and the value of $Z_1$,
which may be very small as the product $\xi L$ tends to 0, is strictly
greater than zero, thus, our summation is convergent for all $p>1$.

We now wish to show that the functions $F_p(x)$ and $F_{p+2}(x)$ are related to
one and other. Beginning with definition~(\ref{Eq:F_defn}), we have
\begin{equation}
F_{p+2}(x) = x^2 \sum_{j=1}^\infty \frac{1}{Z_j^{p}} \left(\frac{A_j^2}{Z_j^2}
\right).
\end{equation}
It is a simple matter of algebra to demonstrate that
\begin{equation}
(x^2+x)\frac{A_j^2}{Z_j^2} = 1 - A_j^2,
\end{equation}
whereby,
\begin{equation}
(x^2+x) F_{p+2}(x) = x^2\sum_{j=1}^\infty \left( \frac{1}{Z_j^p} - \frac{A_j^2}{Z_j^p}\right).
\end{equation}
Each of the terms under the summation are individually convergent for $p>1$, therefore
\begin{equation}
(x^2+x) F_{p+2}(x) + F_p(x) = x^2 \sum_{j=1}^\infty \frac{1}{Z_j^p}.
\label{eq:F_recurrence}
\end{equation}
Next, take the derivative of this expression with respect to $x$;
\begin{equation}
(x^2+x) F_{p+2}'(x) + (2x+1) F_{p+2}(x) + F_p'(x) = 2x \sum_{j=1}^\infty \frac{1}{Z_j^p}
-p x^2 \sum_{j=1}^\infty \frac{1}{Z_j^{p+1}}\left( \frac{dZ_j}{dx}\right).
\end{equation}
Similarly, differentiation of the transcendental equation with respect to $x$ yields
\begin{equation}
\frac{dZ_j}{dx} = \frac{A_j^2}{Z_j}.
\end{equation}
Substituting, we finally arrive at
\begin{equation}
(x^2+x) F_{p+2}'(x) + (p-1) F_{p+2}(x) + F_p'(x) - \frac{2}{x} F_p(x) = 0.
\end{equation}

We already know that $F_4(x) = 1/2$, therefore, let us set $p=2$ in the above
expression, which yields the ordinary differential equation
\begin{equation}
x F_2'(x) - 2 F_2(x) = -\frac{1}{2}x,
\end{equation}
whose general solution is
\begin{equation}
F_2(x) = \frac{1}{2} x + c x^2,
\end{equation}
where $c$ is a constant of integration.  Because all the terms of the series
form of $F_2(x)$ are positive for all values of the allowed range of $x$, we
immediately have the constraint $c\geq0$.

We obtain an upper bound on $c$ by returning to Eq.~(\ref{eq:F_recurrence})
and separating the first term out of the summation on the right-hand side,
\begin{equation}
(x^2+x) F_{p+2}(x) + F_p(x) = x^2\left(\frac{1}{Z_1^p} + \sum_{j=2}^\infty \frac{1}{Z_j^p}
\right).
\end{equation}
Next, we recall that $Z_j \geq (j-1)\pi$, thus, employing the series definition of
the Riemann zeta function $\zeta(p)$, we find
\begin{equation}
(x^2+x) F_{p+2}(x) + F_p(x)\leq
x^2\left(\frac{1}{Z_1^p} + \frac{\zeta(p)}{\pi^p}\right).
\end{equation}
Setting $p=2$ and substituting the expressions for $F_2(x)$ and $F_4(x)$ leads to
the upper bound
\begin{equation}
c\leq \frac{1}{Z_1^2}-\left(\frac{1}{x}+\frac{1}{3}\right),
\end{equation}
which must hold for all positive values of $x$.  The strongest bound occurs when
$x\rightarrow 0$; a condition under which $Z_1$ is also going to zero, but they
approach zero at different rates.  We may use the transcendental
equation to put the entire expression in terms of $Z_1$, and then use the series
expansion for $\cot Z_1$ about zero to obtain
\begin{equation}
c\leq \left(\frac{Z_1^2}{45}+\frac{2 Z_1^4}{945}+\mathcal{O}(Z_1^6)+\dots\right).
\end{equation}
In the limit of $Z_1\rightarrow0$, we find that $c\leq0$.  Combining this bound with
the lower bound implies that $c=0$.

So far, we have found
\begin{equation}
F_2(x) \equiv x^2 \sum_{j=1}^\infty \frac{A_j^2}{Z_j^2}= \frac{1}{2}x
\end{equation}
and
\begin{equation}
F_4(x)\equiv x^2 \sum_{j=1}^\infty \frac{A_j^2}{Z_j^4} = \frac{1}{2}.
\end{equation}
An identical analysis can be use to determine
\begin{equation}
F_6(x)\equiv x^2 \sum_{j=1}^\infty \frac{A_j^2}{Z_j^6} = \frac{1}{2}\left(\frac{1}{x}+\frac{1}{3}\right).
\end{equation}

Substituting back into Eq.~(\ref{eq:F_recurrence}), we also find
\begin{equation}
\sum_{j=1}^\infty \frac{1}{Z_j^2} = \frac{1}{2}+\frac{1}{x}
\qquad\mbox{ and }\qquad
\sum_{j=1}^\infty \frac{1}{Z_j^4} = \frac{1}{6}\left(1+\frac{4}{x}+\frac{6}{x^2}\right).
\end{equation}
Numerical simulations in Mathematica appears to confirm (B23) through (B26) over a wide range
of $x$.  Finally, for $p>1$ we note the relationship
\begin{equation}
F_{p+2}(x) = -\frac{x^2}{p} \frac{d}{dx} \sum_{j=1}^\infty \frac{1}{Z_j^p}.
\label{eq:Fp zeta}
\end{equation}


\section{A flawed derivation of the IN-region stress-tensor}
\label{sec:A_derivation}

In Sec.~\ref{sec:Tren._IN_Region}, we determined the renormalized stress-tensor
for the state $|0_L\rangle$ on the IN region of the spacetime, with a final
result of Eq.~(\ref{Eq:Tren_IN}).  The derivation used the Fourier representation
of the IN-region even-parity eigenfunctions in terms of the OUT-region eigenfunctions,
given by Eq.~(\ref{eq:Fourier for Ueven}). The strength of this approach is
that one can easily cancel all of the divergent terms in the OUT-region mode
expansion of the normal-ordered kinetic-tensor early in the calculations.
It is natural to ask if the same result is found by using the explicit form
of the IN-region eigenfunctions $u^{\rm even}(j,x)$, instead of its Fourier
representation. We explore this approach in this appendix.

For the IN vacuum state $|0_L\rangle$,
we know from Eq.~(\ref{eq:Tmunu_quantum}) that the unrenormalized expectation
value of the components of the stress-tensor on the IN region is given by
\begin{eqnarray}
\langle 0_L | \bm{T}_{\mu\nu} | 0_L \rangle &=& \frac{1}{2} \left\{
\sum_{n=1}^\infty \left[  \frac{k_n}{2}\left(u^{\rm odd}(n,x)\right)^2 +
\frac{1}{2 k_n}\left(\partial_x u^{\rm odd}(n,x)\right)^2  \right]+
\sum_{j=1}^\infty \left[  \frac{\kappa_j}{2}\left(u^{\rm even}(j,x)\right)^2 +
\frac{1}{2 \kappa_j}\left(\partial_x u^{\rm even}(j,x)\right)^2\right]\right\}
\delta_{\mu\nu}\nonumber\\
&&+ \frac{\xi}{2} \delta(x)  g_{\mu\nu} \sum_{j=1}^\infty\frac{1}{\kappa_j}\left(u^{\rm even}(j,0)\right)^2.
\end{eqnarray}
The derivative in $x$ can be evaluated everywhere except at $x=0$, where it is
indeterminate for the even modes.  Looking at the kinetic part separate from the
 potential part, and substituting the modes, we have that the unrenormalized
expectation value of the kinetic-tensor is
\begin{eqnarray}
\langle 0_L | \bm{K}_{\mu\nu} | 0_L \rangle &=&\frac{1}{2L} \left\{
\sum_{n=1}^\infty k_n + \sum_{j=1}^\infty \kappa_j A_j^2\left[ 1+ \left(\frac{\xi}{\kappa_j}
\right)^2\right]\right\}\delta_{\mu\nu},
\end{eqnarray}
which has support everywhere on the IN region except at the support of the potential.
The unrenormalized potential-tensor, which only has support on the support
of the potential, is
\begin{equation}
\langle 0_L | \bm{U}_{\mu\nu} | 0_L \rangle = \frac{\xi}{2} \delta(x)  g_{\mu\nu}
\sum_{j=1}^\infty\frac{A_j^2}{Z_j}.
\end{equation}
Similarly, the unrenormalized expectation value of the stress-tensor for the OUT ground
state on the OUT region is given by
\begin{equation}
\langle \widetilde{0}_L | \bm{T}_{\mu\nu} | \widetilde{0}_L \rangle
= \left(\frac{1}{4\ell L}+\frac{1}{L}\sum_{n=1}^\infty k_n\right)\delta_{\mu\nu}.
\end{equation}
By subtracting this expression from the kinetic-tensor, and then adding the
renormalized OUT ground state expectation value, we find
\begin{equation}
\langle 0_L|\bm{K}_{\mu\nu}|0_L \rangle_{Ren.} = \left(-\frac{\pi}{6L^2}+\frac{\mathcal{A}}{L^2}
\right) \delta_{\mu\nu},
\label{Eq_almost_everywhere}
\end{equation}
where
\begin{equation}
\mathcal{A} = \sum_{j=1}^\infty Z_j A_j^2 \left( 1+\frac{\chi^2}{Z_j^2}\right)
-\sum_{j=1}^\infty (j-1)\pi
\end{equation}
is implicitly a function of the product $\chi \equiv \xi L/2$.

A bit of algebra
and substituting the transcendental equation allows us to rewrite the above
equation as
\begin{equation}
\mathcal{A} = \sum_{j=1}^\infty \left[ Z_j -\tan Z_j+(\chi+1)\chi^2\frac{A_j^2}{Z_j^3}
\right]-\sum_{j=1}^\infty (j-1)\pi.
\end{equation}
Next, let us define $\epsilon_j$, by the relation
\begin{equation}
\epsilon_j \equiv Z_j - (j-1)\pi ,
\end{equation}
thus, $\epsilon_j$ satisfies the transcendental equation
\begin{equation}
(j-1)\pi +\epsilon_j = \chi \cot\epsilon_j
\end{equation}
and its value lies in the interval from 0 to $\pi/2$ for all $j$.  Substituting
into the first two terms, we find
\begin{equation}
\mathcal{A} = \sum_{j=1}^\infty \left[ (j-1)\pi + \epsilon_j -\tan \epsilon_j
+ (\chi+1)\chi^2\frac{A_j^2}{Z_j^3} \right]-\sum_{j=1}^\infty (j-1)\pi.
\end{equation}

We now have to make sense of this expression.  Let $p$ be any integer that is
much greater than $\chi/\pi$, and define the partial sum
\begin{equation}
\mathcal{A}_p = \sum_{j=1}^p \left[ (j-1)\pi + \epsilon_j -\tan \epsilon_j
+ (\chi+1)\chi^2\frac{A_j^2}{Z_j^3} \right]-\sum_{j=1}^p (j-1)\pi,
\end{equation}
such that $\mathcal{A} = \lim_{p\rightarrow\infty}\mathcal{A}_p$.
Because partial sums are convergent, we can alternately write this as
\begin{equation}
\mathcal{A}_p = \sum_{j=1}^p(\epsilon_j -\tan \epsilon_j)
+ (\chi+1)\chi^2\sum_{j=1}^p\frac{A_j^2}{Z_j^3}
\end{equation}
which we can interpret as a mode-by-mode difference of the first $p$ energies.
Notice, the second summation is proportional to the partial sum of the
function $F_3(\chi)$ as defined in Appendix~\ref{sec:series}, and where we show that it
is convergent.  Considering the limit of $p\rightarrow\infty$, we have
\begin{equation}
\mathcal{A} = \lim_{p\rightarrow\infty}\sum_{j=1}^p(\epsilon_j -\tan \epsilon_j)
+ (\chi+1)F_3(\chi).
\label{eq:A_renormalized}
\end{equation}

We now show that the remaining summation in $\epsilon_j$ is convergent.
First, the tangent function is greater than or equal to it argument
on the interval $(0,\pi/2)$. Every term in the summation is negative,
so the sum is bounded above by zero.  Next we show that the sum is also
bounded from below by recalling the Taylor series for the tangent function,
from which we find
\begin{equation}
\tan\epsilon_j - \epsilon_j = \frac{\epsilon_j^3}{3} + \frac{2\epsilon_j^5}{15}+
\frac{17\epsilon_j^7}{315}+\cdots.
\end{equation}
Next, there exists a $j^*$ such that $\epsilon_j<1$ for all $j\geq j^*$.  For
these $\epsilon_j$'s, we know $\epsilon_j^3\geq\epsilon_j^5\geq\epsilon_j^7\geq
\dots$, therefore
\begin{equation}
\tan\epsilon_j - \epsilon_j \leq \left( \tan1 - 1\right)\epsilon_j^3
\end{equation}
which results in
\begin{eqnarray}
\lim_{p\rightarrow\infty}\sum_{j=1}^p(\epsilon_j -\tan \epsilon_j) &=&
\sum_{j=1}^{j^*-1}(\epsilon_j -\tan \epsilon_j)+
\lim_{p\rightarrow\infty}\sum_{j=j*}^p(\epsilon_j -\tan \epsilon_j)\nonumber\\
&\geq& \sum_{j=1}^{j^*-1}(\epsilon_j -\tan \epsilon_j)+ (1-\tan 1)
\lim_{p\rightarrow\infty}\sum_{j=j*}^p\epsilon_j^3.
\end{eqnarray}
From the text following Eq.~(\ref{eqn:Z_j_approx}), we know
\begin{equation}
\epsilon_j < 2\chi\left[(j-1)\pi+\sqrt{(j-1)^2\pi^2+4\chi\left(
1+\frac{\chi}{3}\right)}\right]^{-1}
<\frac{\chi}{(j-1)\pi}.
\end{equation}
Upon substitution, we find
\begin{equation}
\lim_{p\rightarrow\infty}\sum_{j=1}^p(\epsilon_j -\tan \epsilon_j) >
\sum_{j=1}^{j^*-1}(\epsilon_j -\tan \epsilon_j)+ (1-\tan 1)\left(
\frac{\chi}{\pi}\right)^3
\lim_{p\rightarrow\infty}\sum_{j=j*}^p \frac{1}{(j-1)^3}.
\end{equation}
The first sum of the above expression is bounded because there are
a finite number of terms.  The second sum we recognize
as the majority of the series for the Riemann zeta function, thus
we find the bounds
\begin{equation}
0 > \lim_{p\rightarrow\infty}\sum_{j=1}^p(\epsilon_j -\tan \epsilon_j) >
\sum_{j=1}^{j^*-1}(\epsilon_j -\tan \epsilon_j)+ (1-\tan 1)\left(
\frac{\chi}{\pi}\right)^3 \zeta(3).
\end{equation}
Therefore, we conclude that $\mathcal{A}$ is convergent.

A plot of $\mathcal{A}$ as a function of $\chi$ is given in
Figure~\ref{Fig ABC Plots}.  The plot was generated in
Mathematica by calculating the the first $2^{15}$ terms in the
partial sum of $\mathcal{A}$ for each value of $\chi$.  Comparing
the plot of $\mathcal{A}$ with the plot of $\mathcal{B}$, we
conjecture that the two function are the same, i.e.,
$\mathcal{A} = \mathcal{B}$.  Presently, we have no way of proving
this assertion.

Unfortunately, Eq.~(\ref{Eq_almost_everywhere}) does not yield the
the complete expression for the stress-tensor on the IN region of the
spacetime.  In Sec.~\ref{sec:Tren._IN_Region} above, we saw that the
Fourier representation of the even-parity modes allowed for the
derivative to be taken at the point $x=0$, which gave rise to the
delta-function contributions in the renormalized stress-tensor.
On the circle, the Fourier-representation of the delta-function,
Eq.~(\ref{eq:delta_fourier}), has a constant term, which results
in an additional constant term  of $-\mathcal{C}/{L^2}$ in the
renormalized stress-tensor.
The approach used in this appendix does not allow the derivative
to be evaluated at $x=$.  Therefore, it fails to give rise to
a delta-function portion of the stress-tensor and its associated
constant term.



\section{Quantum Weak Energy Inequality on $\stM \simeq \reals \times S^1$}
\label{sec:QWEI for cylinder}

In this appendix, we sketch the proof of a QWEI that we can use on the
OUT region of our spacetime.  We will be brief, as most of the
technical details for the rigorous derivation have been worked out
by Fewster \cite{Fews00} . All we are seeking here
is the contribution to the QWEI due to the topological modes.

Consider the complete cylinder spacetime $\stM \simeq \reals \times S^1$ with no
potential. On this spacetime we have the space of smooth compactly supported
complex valued test functions, which we denote by $C_0^\infty (\stM)$.
For $f\in C_0^\infty (\stM)$, we define the smeared quantum field operator by
\begin{equation}
\bm{\psi}(f) = \int_\stM \bm{\psi}(\bm{x}) f(\bm{x})\, d{\rm{Vol}_g},
\end{equation}
where $\bm{\psi}(\bm{x})$ is given by Eq.~(\ref{eq:RS1_field_operator}).
By the properties describe in the main body of the paper, we have that
the smeared quantum field operator satisfies the following relations:
\begin{enumerate}
\item Linearity, $\bm{\psi}(c_1 f_1 + c_2 f_2) = c_1\bm{\psi}(f_1)+
c_2 \bm{\psi}(f_2)$ for all $c_i\in\complex$ and $f_i\in C_0^\infty (\stM)$,
\item Hermiticity, $\bm{\psi}(f)^\dagger = \bm{\psi}(\overline f)$ for
all $f \in C_0^\infty (\stM)$,
\item Field equation, $\bm{\psi}(\Box f) = 0$ for all $f \in C_0^\infty (\stM)$, and
\item Canonical commutation relations, $[\bm{\psi}(f_1),\bm{\psi}(f_2)] =
i E(f_1,f_2) \mathbb{I}$ for all $f_i\in C_0^\infty (\stM)$,
\end{enumerate}
where the smeared advanced-minus-retarded Green's function $E(f_1,f_2)$ is defined
in Appendix~\ref{sec:adv-ret Green}.

These four properties look identical to the relations used by Fewster to quotient
a free, unital, $*$-algebra in the framework of algebraic QFT.
The first three relations are indeed the same.  The fourth relation looks identical,
but it has a subtle difference;  the advanced-minus-retarded Green's function
used by Fewster does not include the topological modes, thus
the resulting $*$-algebra has a trivial center.  The advanced-minus-retarded
Green's function above does include a contribution from the topological modes,
and thus the resulting $*$-algebra has a nontrivial center.  The quantization
of the topological modes within an algebraic field theory and the resulting
nontrivial center has been discussed by Dappiaggi and Lang \cite{D&Lang}.

  In Sect.~\ref{sec:Energy Cond}, we saw that a timelike geodesic can be
parameterized by Eq.~(\ref{eq.timelike geodesic}) and that there exist
vector fields $v_0^\mu$ and $v_1^\mu$ of unit length, which are parallel and
perpendicular, respectively, to the tangent vector of the geodesic.  On $M\times M$,
we define the unrenormalized, point-split, energy-density operator as
\begin{equation}
{\bm\rho}(t,x;t',x') \equiv \frac{1}{2} \left[ (v_0^\mu \partial_\mu)(v_0^{\nu'}
\partial_{\nu'}) + (v_1^\mu \partial_\mu)(v_1^{\nu'} \partial_{\nu'})\right]
{\bm\psi}(x,t){\bm\psi}(x',t'),
\end{equation}
where $\partial_{\nu'}$ is understood as taking the derivative with respect
to the primed variables and $v_i^{\nu'}$ is also in the primed variables.
Is has been shown \cite{Fews00} that the energy-density along the geodesic of
an observer is given by the pulled-back of the above expression onto
the observer's geodesic $\gamma(\tau)$, i.e.
\begin{equation}
{\bm\rho}(\tau) = \bm{\rho}(\gamma(\tau);\gamma(\tau)).
\end{equation}

Let $\omega_L$ and $\omega_0$ be any Hadamard state on the $*$-algebra
for our spacetime, and $g\in C_0^\infty(\reals)$ be a smooth, real-valued,
compactly-supported test function on the real line.  Then, we have the smeared,
normal-ordered energy-density along the worldline is given by
\begin{equation}
\int_\reals d\tau \langle\omega_L|:\bm{\rho}(\tau):_{\omega_0}|\omega_L\rangle\,
g(\tau)^2.
\end{equation}
The derivation of the quantum inequality on this expression now follows the
steps found in Fewster.  In fact, the entire derivation is identical, including
his Theorem 4.1 which yields the quantum inequality
\begin{equation}
\int_\reals d\tau \, \langle \omega_L |:_{\omega_0} \bm{\rho}(\tau) :|\omega_L \rangle
g(\tau^2) \geq -\frac{1}{\pi}\int_0^\infty d\alpha
\langle \omega_0 | \bm{\rho} |\omega_0 \rangle (\overline{g_\alpha}\otimes
g_\alpha),
\label{eq:QWEI}
\end{equation}
where $g_\alpha(\tau) = g(\tau)e^{i\alpha\tau}$ and the energy-density operator
on the right-hand side is still point split along the proper time.
All of this comes about because the topological modes only contribute a smooth
function piece to the two-point functions of the Hadamard states.  This contribution
can be seen directly as the $(t'-t)/L$ term in the advanced-minus-retarded Green's
function above.  Therefore, the topological modes do nothing to alter the wavefront
set of the any of the distributions we work with in the derivation of the
quantum inequality.

We now wish to evaluate the right-hand side of this expression for our spacetime.
Upon substitution of the explicit form of the vector fields $v_0^\mu$ and $v_1^\mu$,
the expectation value of the point-split energy-density operator is
\begin{equation}
\langle\omega_0| \bm{\rho} |\omega_0\rangle (t,x;t',x') = \frac{1}{2} \left[
\frac{1+v^2}{1-v^2} (\partial_t \partial_{t'}+\partial_x\partial_{x'} )
+\frac{2v}{1-v^2}(\partial_t \partial_{x'} +\partial_x\partial_{t'})\right]
\langle\omega_0| \bm{\Psi}(x,t)\bm{\Psi}(x',t') |\omega_0\rangle.
\end{equation}
Next, we choose our reference state for normal ordering to be the OUT
vacuum state $|\widetilde{0}_L\rangle$, such that
\begin{equation}
\langle \widetilde{0}_L| \bm{\Psi}(x,t)\bm{\Psi}(x',t') |\widetilde{0}_L\rangle = \widetilde{G}^+(x,t;x',t')
\end{equation}
is the positive-frequency Wightman function, whose series representation
is given by Eq.~(\ref{eq:out wightman}) above.  Upon substitution, we find
\begin{eqnarray}
\langle\widetilde{0}_L| \bm{\rho} |\widetilde{0}_L\rangle (t,x;t',x') &=& \frac{1}{2L}
\left( \frac{1+v^2}{1-v^2}\left\{\frac{1}{2\ell}+2\sum_{n=1}^\infty k_n
\cos[k_n(x-x')]e^{-ik_n(t-t')}\right\}\right. \nonumber\\
&&\left.-\frac{4iv}{1-v^2}\sum_{n=1}^\infty k_n \sin[k_n(x-x')]e^{-ik_n(t-t')} \right).
\end{eqnarray}
For the above point-split energy-density, the topological modes simply contributes
a constant of $1/2\ell$.

Next, pulling back onto the worldline of the observer, still point split in
the proper time used to parameterize the geodesic, we find
\begin{eqnarray}
\langle\widetilde{0}_L| \bm{\rho} |\widetilde{0}_L\rangle (\gamma(\tau));\gamma(\tau'))&=&
\frac{1}{2L} \left\{ \frac{1+v^2}{1-v^2}\left(\frac{1}{2\ell}\right)+\sum_{n=1}^\infty
k_n\left[\frac{1-v}{1+v} e^{-i(\tau-\tau')k_n\sqrt{\frac{1-v}{1+v}}}+
\frac{1+v}{1-v}e^{-i(\tau-\tau')k_n\sqrt{\frac{1+v}{1-v}}} \right]\right\}.
\end{eqnarray}
Therefore, the right-hand side of the QWEI, Eq.~(\ref{eq:QWEI}), becomes
\begin{eqnarray}
R.H.S &=& -\frac{1}{\pi} \left[ \frac{1+v^2}{1-v^2}\left(\frac{1}{4\ell L}\right)
\int_0^\infty d\alpha\, \hat{g}(-\alpha)\hat{g}(\alpha)\right.\nonumber\\
&&+\frac{1-v}{1+v}\left(\frac{1}{2L}\right)\int_0^\infty d\alpha\,
\sum_{n=1}^\infty k_n\,\hat{g}\left(-\alpha-k_n\sqrt{\frac{1-v}{1+v}}\right)
\hat{g}\left(\alpha+k_n\sqrt{\frac{1-v}{1+v}}\right)\nonumber\\
&&\left.+\frac{1+v}{1-v}\left(\frac{1}{2L}\right)\int_0^\infty d\alpha\,
\sum_{n=1}^\infty k_n\,\hat{g}\left(-\alpha-k_n\sqrt{\frac{1+v}{1-v}}\right)
\hat{g}\left(\alpha+k_n\sqrt{\frac{1+v}{1-v}}\right)\right].
\end{eqnarray}
This expression can be further simplified by recalling that for real-valued
test functions, the Fourier transforms satisfies $\hat{g}(-\alpha) = %
\overline{\hat{g}(\alpha)}$, whereby, we can then uses Parseval's theorem on
the first term.  Finally swapping the order of summation and integration, we
arrive at Eq.~(\ref{eq:QWEI specific}).


%

\newpage
\noindent
\begin{center}
\begin{figure}
\begin{center}
    \includegraphics[width=90mm]{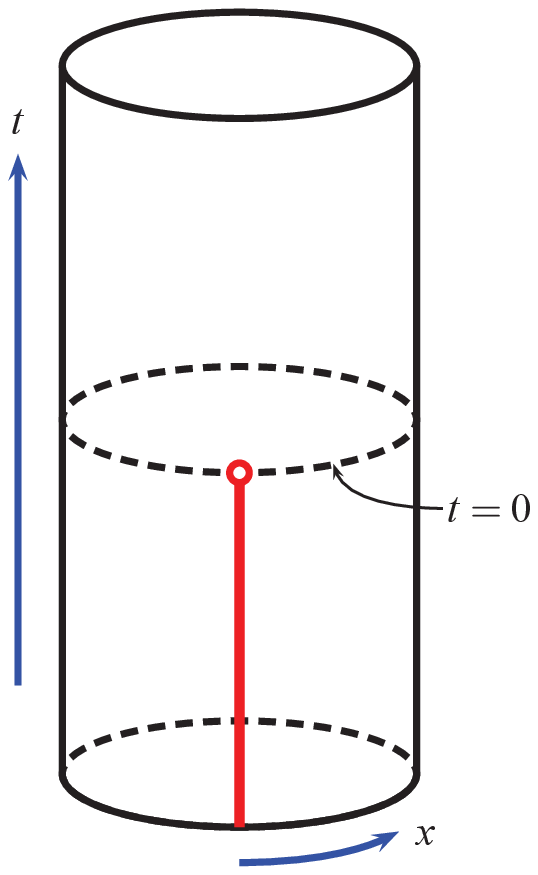}
\end{center}
\caption{A graphical representation of the spacetime $\reals\times S^1$, with
time increasing in the vertical direction.  The dashed circle midway up the
cylinder shows the $t=0$ Cauchy surface.  The red line shows the location of
the delta-function potential that is turned off at $t=0$.}
\end{figure}
\end{center}

\vfill
\noindent
\begin{center}
\begin{figure}
\begin{center}
    \includegraphics[width=6.75in]{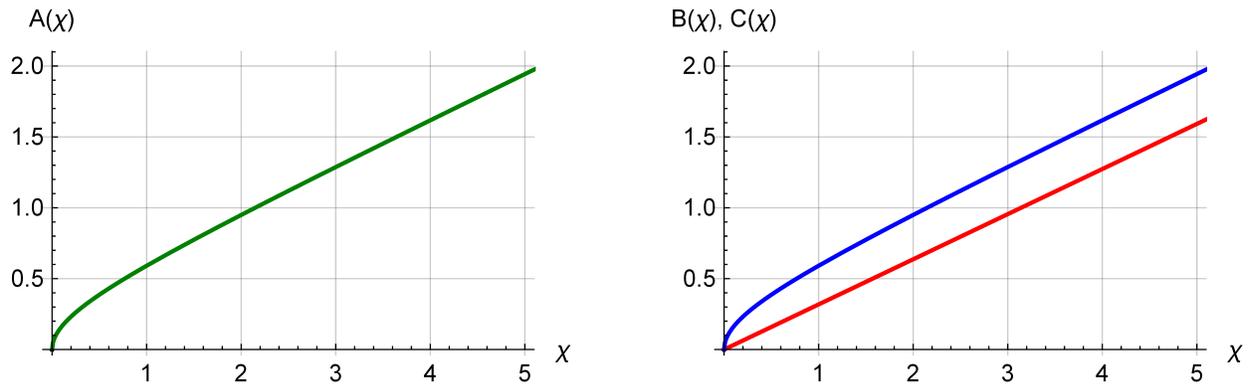}
\end{center}
\caption{Plots of the coefficients $\mathcal{A}$ (green), $\mathcal{B}$ (blue),
and $\mathcal{C}$ (red) as a function of the dimensionless variable $\chi$.
The plots were generated in Mathematica by calculating the partial sum
for the first $2^{15}$ terms of each infinite series.  The plots seem to
indicate that $\mathcal{A}=\mathcal{B}$. The analysis in Appendix~\ref{sec:series}
shows that $\mathcal{C} = \chi/\pi$, i.e., it is a straight line.}
\label{Fig ABC Plots}
\end{figure}
\end{center}

\end{document}